\documentclass{aastex}          %% The manuscript based on AASTeX v5.x
\usepackage{spr-astr-addons}    %% mimicing ASTR journal style
\usepackage{fancyhdr}           %% testing   

\newcommand{\emaila}{Baerbel.Koribalski@csiro.au}
\newcommand{\HI}{H\,{\sc i}}
\newcommand{\HII}{H\,{\sc ii}}
\newcommand{\wal}{{\sc wallaby}}
\newcommand{\kms}{\,km\,s$^{-1}$}

\newcommand{\Ha}{H$\alpha$}
\newcommand{\Lya}{Ly$\alpha$}
\newcommand{\vrot}{$V_{\rm rot}$}
\newcommand{\FHI}{$F_{\rm HI}$}

\newcommand{\MHI}{$M_{\rm HI}$}

\newcommand{\Mdyn}{$M_{\rm dyn}$}
\newcommand{\MMsun}{M$_{\odot}$}
\newcommand{\Msun}{\,M$_{\odot}$}

\newcommand{\Tsys}{$T_{\rm sys}$}
\newcommand{\Tspin}{$T_{\rm spin}$}
\newcommand{\Ho}{H$_{\rm 0}$}

\newcommand{\dingo}{{\sc dingo}}
\newcommand{\alfalfa}{{\sc alfalfa}}
\newcommand{\hipass}{{\sc hipass}}
\newcommand{\lvhis}{{\sc lvhis}}
\newcommand{\taipan}{{\sc taipan}}

\begin{document}

%% Article title
\title{WALLABY -- An SKA Pathfinder H\,{\sc i} Survey}
\email{\emaila}

%% Running heads
\shorttitle{WALLABY -- An SKA Pathfinder H\,{\sc i} Survey}
\shortauthors{B. S. Koribalski et al.}

%% Author and Affilations
% \fullcollaborationName{    Corresponding author email: ---}  \\
% \fullcollaborationName{    --- Baerbel.Koribalski @ csiro.au ---} 
\notetoeditor{accepted in ApSS \\}
\author{B\"arbel S. Koribalski,\altaffilmark{1,2,3} ~~ L. Staveley-Smith,\altaffilmark{4,3} T. Westmeier,\altaffilmark{4,3} P. Serra,\altaffilmark{5} K. Spekkens,\altaffilmark{6} O.I. Wong,\altaffilmark{4,3} K. Lee-Waddell,\altaffilmark{1} C.D.P. Lagos,\altaffilmark{4,3,36} D. Obreschkow,\altaffilmark{4,3} E.V. Ryan-Weber,\altaffilmark{7,3} M. Zwaan,\altaffilmark{8} V. Kilborn,\altaffilmark{7,3} G. Bekiaris,\altaffilmark{1} K. Bekki,\altaffilmark{4} F. Bigiel,\altaffilmark{9} A. Boselli,\altaffilmark{10} A. Bosma,\altaffilmark{10} B. Catinella,\altaffilmark{4,3} G. Chauhan,\altaffilmark{4,3} M.E. Cluver,\altaffilmark{7,11} M. Colless,\altaffilmark{12,3} H.M. Courtois,\altaffilmark{13} R.A. Crain,\altaffilmark{14}  W.J.G. de Blok,\altaffilmark{15,16,17} H. D\'enes,\altaffilmark{15} A.R. Duffy,\altaffilmark{7,3} A. Elagali,\altaffilmark{4,1,3} C.J. Fluke,\altaffilmark{7,37} B.-Q. For,\altaffilmark{4,3} G. Heald,\altaffilmark{18,3}  P.A. Henning,\altaffilmark{19} K.M. Hess,\altaffilmark{15,17} B.W. Holwerda,\altaffilmark{20} C. Howlett,\altaffilmark{21} T. Jarrett,\altaffilmark{16} D.H. Jones,\altaffilmark{22} M.G. Jones,\altaffilmark{23} G.I.G. J\'ozsa,\altaffilmark{24,25,9} R. Jurek,\altaffilmark{21} E. J\"utte,\altaffilmark{26} P. Kamphuis,\altaffilmark{26} I. Karachentsev,\altaffilmark{27} J. Kerp,\altaffilmark{9} D. Kleiner,\altaffilmark{1,5} R.C. Kraan-Korteweg,\altaffilmark{16} \'A.R. L\'opez-S\'anchez,\altaffilmark{28,29,3} J. Madrid,\altaffilmark{1} M. Meyer,\altaffilmark{4,3} J. Mould,\altaffilmark{7} C. Murugeshan,\altaffilmark{7,3} R.P. Norris,\altaffilmark{2,1} S.-H. Oh,\altaffilmark{30} T.A. Oosterloo,\altaffilmark{15,17} A. Popping,\altaffilmark{4} M. Putman,\altaffilmark{31} T.N. Reynolds,\altaffilmark{4,1,3} J. Rhee,\altaffilmark{4,3} A.S.G. Robotham,\altaffilmark{4,3} S. Ryder,\altaffilmark{29} A.C. Schr\"oder,\altaffilmark{32} Li Shao,\altaffilmark{1,33} A.R.H. Stevens,\altaffilmark{4,3} E.N. Taylor,\altaffilmark{7} J.M. van der Hulst,\altaffilmark{17} L. Verdes-Montenegro,\altaffilmark{23} B.P. Wakker,\altaffilmark{34} J. Wang,\altaffilmark{1,33} M. Whiting,\altaffilmark{1} B. Winkel\altaffilmark{35} and C. Wolf\altaffilmark{12}
}
\altaffiltext{1}{Australia Telescope National Facility, CSIRO Astronomy and Space Science, P.O. Box 76, Epping, NSW 1710, Australia. --- Corresponding author: Baerbel.Koribalski @ csiro.au} % BSK, many
\altaffiltext{2}{Western Sydney University, Locked Bag 1797, Penrith South, NSW 1797, Australia} % BSK, R. Norris
\altaffiltext{3}{ARC Centre of Excellence for All Sky Astrophysics in 3 Dimensions (ASTRO 3D)} % many
\altaffiltext{4}{International Centre for Radio Astronomy Research, University of Western Australia, 35 Stirling Hwy, Perth, WA 6009, Australia} % many
\altaffiltext{5}{INAF -- Osservatorio Astronomico di Cagliari, Via della Scienzia 5, I-09047 Selargius, Italy} % P. Serra, D. Kleiner
\altaffiltext{6}{Department of Physics and Space Science, Royal Military College of Canada, P.O. Box 17000, Kingston, Ontario K7L 3N6, Canada} % K. Spekkens
\altaffiltext{7}{Centre for Astrophysics and Supercomputing, Swinburne University of Technology, P.O. Box 218, Hawthorn, VIC 3122, Australia} % E. Ryan-Weber, V. Kilborn, A. Duffy, C. Fluke, E. Taylor
\altaffiltext{8}{European Southern Observatory, Karl-Schwarzschild Strasse 2, D-85748 Garching near Munich, Germany} % M. Zwaan
\altaffiltext{9}{Argelander-Institut f\"ur Astronomie, Auf dem H\"ugel 71, D-53121 Bonn, Germany} % F. Bigiel, J. Kerp, G. Jozsa
\altaffiltext{10}{Aix Marseille University, CNRS, CNES, LAM, Marseille, France}   % A. Boselli, A. Bosma
\altaffiltext{11}{Department of Physics and Astronomy, University of the Western Cape, Robert Sobukwe Road, Bellville, 7535, South Africa} % M. Cluver
\altaffiltext{12}{Research School of Astronomy and Astrophysics, Australian National University, Canberra, ACT 2611, Australia} % M. Colless, C. Wolf
\altaffiltext{13}{University of Lyon, UCB Lyon 1, CNRS/IN2P3, IUF, IP2I Lyon, 69622 Villeurbanne, France} % H. Courtois
\altaffiltext{14}{Astrophysics Research Institute, Liverpool John Moores University, 146 Brownlow Hill, Liverpool L3 5RF, UK} % R. Crain
\altaffiltext{15}{Netherlands Institute for Radio Astronomy (ASTRON), Oude Hoogeveensedijk 4, 7991 PD, Dwingeloo, The Netherlands} % E. de Blok, T. Oosterloo
\altaffiltext{16}{Department of Astronomy, University of Cape Town, Private Bag X3, Rondebosch 7701, South Africa} % E. de Blok, Renee KK
\altaffiltext{17}{Kapteyn Astronomical Institute, University of Groningen, Postbus 800, 9700 AV Groningen, The Netherlands} % de Blok, van der Hulst, Oosterloo
\altaffiltext{18}{CSIRO Astronomy and Space Science, P.O. Box 1130, Bentley WA 6102, Australia} % G. Heald
\altaffiltext{19}{Department of Physics and Astronomy, University of New Mexico, Albuquerque, NM 87131, USA} % P.A. Henning
\altaffiltext{20}{Department of Physics and Astronomy, University of Louisville, Louisville, KY 40292, USA} % B. Holwerda
\altaffiltext{21}{School of Mathematics and Physics, The University of Queensland, Brisbane, QLD 4072, Australia} % R. Jurek, C. Howlett 
\altaffiltext{22}{English Language and Foundation Studies Centre, University of Newcastle, Callaghan, NSW 2308, Australia} % Heath Jones
\altaffiltext{23}{Instituto de Astrof\'{i}sica de Andaluc\'{i}a (CSIC), Glorieta de Astronom\'{i}a s/n, 18008 Granada, Spain} % Mike Jones, Lourdes Verdes-Montenegro
\altaffiltext{24}{South African Radio Astronomy Observatory, 2 Fir Street, Black River Park, Observatory, Cape Town, 7925, South Africa} % SARAO - Gyula Jozsa
\altaffiltext{25}{Department of Physics and Electronics, Rhodes University, P.O. Box 94, Makhanda, 6140, South Africa} % G. Jozsa
\altaffiltext{26}{Ruhr-University Bochum, Faculty of Physics and Astronomy, Astronomical Institute, 44780 Bochum, Germany}  % P. Kamphuis, E. Juette
\altaffiltext{27}{Special Astrophysical Observatory, Russian Academy of Sciences, N. Arkhyz 369167, Russia} % Igor Karachentsev
\altaffiltext{28}{Australian Astronomical Optics, 105 Delhi Rd, North Ryde, NSW 2113, Australia} % Angel LS
\altaffiltext{29}{Department of Physics and Astronomy, Macquarie University, NSW 2109, Australia} % Angel LS, Stuart Ryder
\altaffiltext{30}{Department of Physics and Astronomy, Sejong University, 209 Neungdong-ro, Gwangjin-gu, Seoul, Republic of Korea} % Se-Heon Oh
\altaffiltext{31}{Department of Astronomy, Columbia University, New York, NY 10027, USA} % Mary Putman
\altaffiltext{32}{South African Astronomical Observatory, P.O. Box 9, Observatory 7935, Cape Town, South Africa} % SAAO - Anja Schroeder
\altaffiltext{33}{Kavli Institute for Astronomy and Astrophysics, Peking University, Beijing 100871, China} % J. Wang, L. Shao
\altaffiltext{34}{Department of Astronomy, University of Wisconsin-Madison, 475 North Charter Street, Madison, WI 53706, USA}  % B. Wakker 
\altaffiltext{35}{Max Planck Institut f\"ur Radioastronomie, Auf dem H\"ugel 69, D-53121 Bonn, Germany} % B. Winkel
\altaffiltext{36}{Cosmic Dawn Center (DAWN), Denmark} % C. Lagos
\altaffiltext{37}{Advanced Visualisation Laboratory, Swinburne University of Technology, P.O. Box 218, Hawthorn, Victoria, 3122, Australia} % C. Fluke

% \email{Baerbel.Koribalski@csiro.au} %% non-output
\email{\emaila}

\clearpage

%% Abstract
\begin{abstract}
The Widefield ASKAP L-band Legacy All-sky Blind surveY (\wal) is a next-generation survey of neutral hydrogen (\HI) in the Local Universe. It uses the widefield, high-resolution capability of the Australian Square Kilometer Array Pathfinder (ASKAP), a radio interferometer consisting of $36 \times 12$-m dishes equipped with Phased-Array Feeds (PAFs), located in an extremely radio-quiet zone in Western Australia. \wal\ aims to survey three-quarters of the sky ($-90\degr < \delta < +30\degr$) to a redshift of $z \lesssim 0.26$, and generate spectral line image cubes at $\sim$30~arcsec resolution and $\sim$1.6 mJy\,beam$^{-1}$ per 4\kms\ channel sensitivity. ASKAP's instantaneous field of view at 1.4 GHz, delivered by the PAF's 36 beams, is about 30 sq deg. At an integrated signal-to-noise ratio of five, \wal\ is expected to detect around half a million galaxies with a mean redshift of $z \sim 0.05$ ($\sim$200~Mpc).
The scientific goals of \wal\ include: (a) a census of gas-rich galaxies in the vicinity of the Local Group; (b) a study of the \HI\ properties of galaxies, groups and clusters, in particular the influence of the environment on galaxy evolution; and (c) the refinement of cosmological parameters using the spatial and redshift distribution of low-bias gas-rich galaxies. For context we provide an overview of recent and planned large-scale \HI\ surveys. Combined with existing and new multi-wavelength sky surveys, \wal\ will enable an exciting new generation of panchromatic studies of the Local Universe. --- First results from the \wal\ pilot survey are revealed, with initial data products publicly available in the CSIRO ASKAP Science Data Archive (CASDA). 
\end{abstract}

%% Keywords

\section{Introduction} % Section 1
%\label{s:introduction}

Hydrogen is the lightest chemical element in the periodic table and the most abundant in the Universe. While hydrogen makes up about 75\% of the baryonic mass, it constitutes only a tiny fraction of the overall mass-energy budget. The latter is thought to be dominated by dark matter (26.2\%) and dark energy (68.9\%) with baryons a mere 4.9\% (Planck Collaboration et al. 2018). \\

Galactic and extragalactic hydrogen is found in three primary phases: neutral molecular (H$_2$), neutral atomic (\HI) and ionized (\HII). As the Universe evolved, hydrogen underwent two major phase transitions, first from ionized to neutral (recombination) and then from neutral back to ionized (reionization). At present, the cosmological mass density of \HI, $\Omega_{\rm HI}$, is only $\sim$0.04\% (Zwaan et al. 2003, 2005). Here we focus on the \HI\ spectral line, highlight its importance for galaxy studies and its interplay with other hydrogen phases, before introducing the \wal\ observing parameters and science goals. \\

The 21-cm line of hydrogen (\HI) provides the means to study the formation and evolution of galaxies and large-scale structures, unobscured by dust and foreground stars, from the `Dark Ages' to the Local Universe. This hyperfine line results from the electron spin-flip in the electronic ground state of the hydrogen atom; its rest frequency is 1420.40575177 MHz, corresponding to a wavelength of 21.106 cm. The line was predicted by H.C. van de Hulst in 1944 (see van de Hulst 1945), first observed by Ewen \& Purcell (1951) and confirmed by Muller \& Oort (1951). Kerr et al. (1954) made the first extragalactic \HI\ detection of the Magellanic Clouds. \\

While all forms of hydrogen are typically detected in the inner discs and spiral arms of galaxies in the Local Universe, most often it is the \HI\ 21-cm line that reveals the outermost regions of galaxy discs (e.g., Warren et al. 2004; Begum et al. 2005; Sancisi et al. 2008; Heald et al. 2011; Koribalski et al. 2018). Occasionally \HI\ is found in filaments, plumes and/or bridges, tracing the gravitational interactions with neighbouring galaxies (e.g., Yun et al. 1993, 1994; Koribalski et al. 2003; Koribalski \& Dickey 2004; Pearson et al. 2016) and in groups/clusters (e.g., Verdes-Montenegro et al. 2001; Oosterloo \& van Gorkom 2005; Chung et al. 2009; English et al. 2010; Lee-Waddell et al. 2012; Hess et al. 2017; Scott et al. 2018; Saponara et al. 2018). Beyond the inner galaxy disk in which molecular hydrogen is typically the dominant gas component (Leroy et al. 2008; Bigiel \& Blitz 2012), \HI\ is also an excellent tracer of star formation in the outer discs of galaxies (e.g., Battaglia et al. 2006; Koribalski \& L\'opez-S\'anchez 2009; For et al. 2012; Koribalski 2017). The large extent of \HI\ discs in many late- and early-type galaxies, compared to their stellar discs (e.g., Meurer et al. 1996; Begum et al. 2005; Serra et al. 2012a; Bosma 2017), makes them highly susceptible to external forces such as tidal interactions, gas accretion, and ram pressure stripping, while providing fuel for star formation. As a consequence, the amount and extent of \HI\ in galaxies varies significantly with environment (e.g., Chung et al. 2009; Cortese et al. 2011; Catinella et al. 2013; D\'enes et al. 2014; Stevens \& Brown 2017; Jones et al. 2018a). Furthermore, the \HI\ kinematics of galaxies, which are typically measured with high accuracy (several\kms), can be used to derive their rotation curves as well as non-circular motions, if present. The shape, amplitude and extent of galaxy rotation curves allow detailed modelling of a galaxy's overall mass distribution as a function of radius (e.g., van Albada et al. 1985; Carignan \& Freeman 1985; de Blok et al. 2008; Oh et al. 2008, 2018).  \\

One of the highest redshift detections in \HI\ {\bf emission} from an individual galaxy, obtained in the COSMOS \HI\ Large Extragalactic Survey (CHILES), is at $z = 0.376$ (Fernandez et al. 2016), well surpassing previous detections at $z \sim 0.2$ (Zwaan et al. 2001; Verheijen et al. 2007; Catinella et al. 2008; Catinella \& Cortese 2015). Recently, \HI\ emission from a strongly lensed galaxy at $z = 0.407$ has been marginally detected by Blecher et al. (2019). The dearth of detections at high redshift is mainly due to the intrinsic weakness of the 21-cm emission line, but is also affected by the available frequency ranges (e.g., Hess et al. 2019), the presence of radio frequency interference (RFI), and limited observing time on current radio telescopes. To overcome these limitations, some pioneering research was conducted using \HI\ spectral stacking (e.g., Chengalur et al. 2001) and \HI\ intensity mapping techniques (e.g., Pen et al. 2009; Anderson et al. 2018) which are now routinely used for $z > 0.3$ \HI\ galaxy studies. The first detection (4$\sigma$) of \HI\ in emission at $z \approx 0.8$ was obtained by cross-correlating \HI\ intensity maps with optical density maps generated using $\sim$10\,000 optically identified galaxies with known, accurate redshifts (Chang et al. 2010; see also Khandai et al. 2011). \HI\ {\bf absorption} is another way to probe the neutral component of the interstellar medium (ISM) and the intergalactic medium (IGM) towards bright radio sources in the background. Rhee et al. (2018) nicely summarize the \HI\ gas density measurements obtained from \HI\ and Damped \Lya\ (DLA) measurements. \\

Although cold {\bf molecular hydrogen} gas dominates the molecular ISM of galaxies, its lack of a dipole combined with the energy requirements of its rotational and vibrational transitions, make it essentially invisible in emission for temperatures below 100~K. Carbon monoxide (CO), the next most common molecule, which is easily excited by the $UV$ radiation field coursing through the ISM, is the primary tracer of molecular gas mass. However, CO only acts as a proxy for molecular content and large uncertainties remain in the CO-to-H$_2$ conversion factor (e.g., Cormier et al. 2018). Only the densest pockets of the ISM, where the bulk of molecular gas resides, contain both H$_2$ and CO. Their immediate surroundings have a more complex composition, which is deeply influenced by ultraviolet radiation from nearby stars. Molecular hydrogen at temperatures above 100~K -- excited by high-speed shocks, for example -- is detectable in the near- and mid-infrared (Wong et al. 2014). Intergalactic \HI\ gas in galaxy groups may even be transformed into molecular hydrogen during a collision with an intruder galaxy, as shown by Cluver et al. (2010) in Stephan's Quintet. By using Herschel observations of the 158~$\mu$m line of ionized carbon ([C\,{\sc ii}]) Pineda et al. (2013) show that CO-dark H$_2$ accounts for $\sim$30\% of the Milky Way's entire reservoir of molecular gas.   \\

\begin{figure}[tb] % Figure 1 - created by BSK
\begin{center}
  \includegraphics[height=7cm,angle=-90]{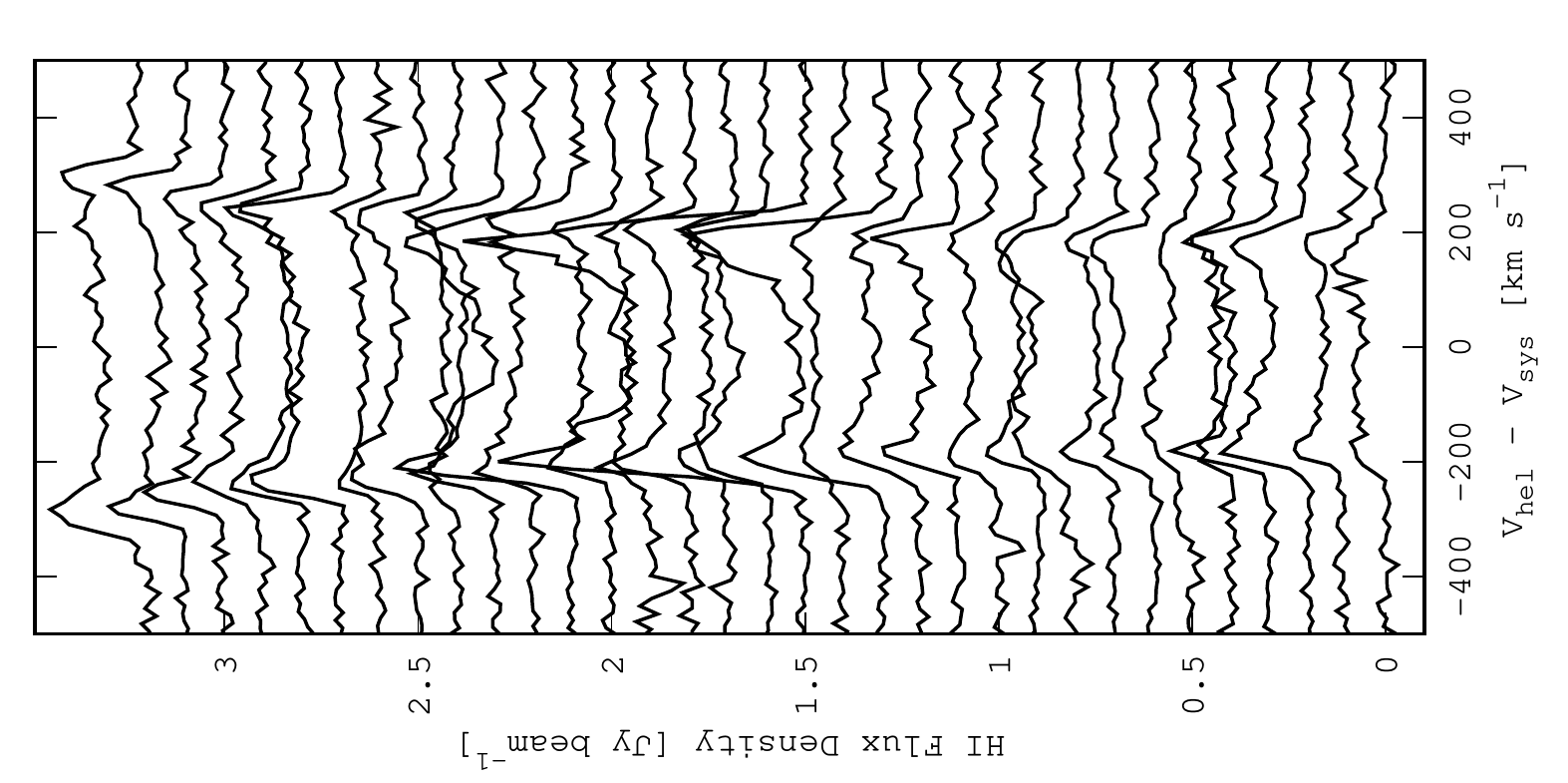}
\caption{Waterfall plot of \hipass\ spectra, offset by 0.1 Jy\,beam$^{-1}$ on the y-axis. Only the widest \HI\ spectra from the \hipass\ Bright Galaxy Catalog (Koribalski et al. 2004) are shown here, typically corresponding to fast-rotating, edge-on spiral galaxies.}
\label{fig:hipass-spectra}
\end{center}
\end{figure}

The most common way of tracing the {\bf ionized hydrogen} gas (\HII) is by observing the \Ha\ emission line. Large-scale \Ha\ surveys of nearby galaxies (e.g., Rossa \& Dettmar 2003; James et al. 2004; Parker et al. 2005; Oey et al. 2007; Kennicutt et al. 2008) reveal that extraplanar diffuse ionized gas is correlated with the level of star formation activity. In galaxy clusters, \Ha\ emission can be attributed to both star formation and shock-ionized gas (e.g., Sun et al. 2007; Kenney et al. 2008; Boselli et al. 2018a,b; Fossati et al. 2018); see also J\'achym et al. (2014) and references therein. \\

The Widefield ASKAP L-band Legacy All-sky Blind surveY (\wal) is expected to detect around half a million galaxies in \HI\ by mapping the whole southern sky and part of the northern sky ($\delta < +30\degr$), complemented by high-resolution 20-cm radio continuum maps. \wal\ Early Science results (Reynolds et al. 2019; Lee-Waddell et al. 2019; Elagali et al. 2019; Kleiner et al. 2019; For et al. 2019), include the discovery of new dwarf galaxies and \HI\ debris in galaxy groups. Full \wal\ will increase the number of catalogued \HI\ galaxies by more than an order of magnitude and provide well-resolved \HI\ maps for $\sim$5\,000 nearby galaxies. It will be the first, untargeted large-area \HI\ survey at sub-arcminute resolution. \\

\wal\ builds on the long and rich history of blind \HI\ sky surveys and is an important step towards future \HI\ surveys with the Square Kilometer Array (SKA). One of its main advantages, compared to previous large-scale \HI\ surveys, is the ability to spatially resolve thousands of nearby galaxies allowing us to study the physical processes shaping their discs and halos. \wal\ will make many discoveries --- expected and unexpected --- and, together with other SKA precursors and pathfinders, commence a new era of widefield and deep \HI\ surveys. In Section~2 we give an overview of recent and planned large-scale \HI\ surveys, including those on the Australian SKA Pathfinder (ASKAP). The ASKAP telescope and data processing pipeline are briefly described in Section~3, followed by the \wal\ parameters in Section~4, and the \wal\ science case in Section~5. The \wal\ Early Science program and first results are highlighted in Section~6. The last section contains our conclusions and future outlook.

\section{\HI\ in the Local Universe} % Section 2 

\HI\ is an ubiquitous tracer of large-scale structure in the Universe (e.g., Koribalski et al. 2004; Meyer et al. 2004; Springob et al. 2005; Papastergis et al. 2013; Kleiner et al. 2017; Hong et al. 2019), unimpeded by foreground dust and stars (Staveley-Smith et al. 2016). It is also an excellent tracer of the total mass in rotating disk galaxies (e.g., Bosma 1981a,b; de Blok et al. 2008; Ott et al. 2012; Oh et al. 2018), the gas content of galaxies, and the gaseous structures in between galaxies. \HI\ is an excellent probe of galaxy environments, allowing us to investigate the faint outer discs of galaxies, extended tidal features (such as tails, bridges, filaments) and dwarf companions (Hibbard et al. 2001; Sancisi et al. 2008; Heald et al. 2011; Serra et al. 2012a; Bosma 2017; Koribalski et al. 2018). Furthermore, \HI\ allows us to study the physical and dynamical processes within galaxies, including the kinematic properties of structures such as bars, rings, spiral arms and warps (e.g., Verdes-Montenegro et al. 2002; J\'ozsa et al. 2007; Spekkens \& Sellwood 2007; Kamphuis et al. 2015; Di Teodoro \& Fraternali 2015). Each galaxy's \HI\ spectrum provides a large set of galaxy properties, such as the systemic velocity, the integrated flux and the velocity width (e.g., Koribalski et al. 2004). These are used to derive the galaxy distance, gas mass, and total dynamical mass, respectively. The shape of the integrated \HI\ spectrum also reflects the overall galaxy symmetry, flatness of the rotation curve and presence of other \HI\ components. \\

The evolution of \HI\ is of fundamental importance for understanding the build-up of both stellar and gas mass within galaxies as well as the method by which galaxies accrete their material (e.g., Athanassoula et al. 2016). It is a crucial window into galaxy formation over time to be explored with the next generation of large-scale \HI\ surveys, spanning a significant fraction of cosmic look-back time. The latter will be carried out with new and upgraded radio interferometers, such as ASKAP (Johnston et al. 2007, 2008), providing high spatial and spectral resolution as well as good point source and surface brightness sensitivity. ASKAP's frequency coverage (700 -- 1800~MHz) allows for \HI\ studies out to redshift of $z \sim 1$. \\

The \HI\ mass of a galaxy, \MHI, is estimated using 
\begin{equation}
  M_{\rm HI}~~[M_{\odot}] = \frac{2.356}{(1+z)} \times 10^5~~ D^2~~ F_{\rm HI}~~,
\end{equation}
where \FHI\ is the integrated \HI\ flux in Jy\kms, $D$ the luminosity distance in Mpc and $z$ the redshift. The total dynamical mass of a galaxy, \Mdyn, is calculated using

\begin{equation}
  M_{\rm dyn}~~[M_{\odot}] = 2.31 \times 10^5~~ V_{\rm rot}^2~~ R_{\rm HI}~~,
\end{equation}
where \vrot\ is the inclination-corrected rotation velocity in\kms\ and $R_{\rm HI}$ the \HI\ radius in kpc. At low spatial resolution, when only integrated \HI\ spectra (see Fig.~\ref{fig:hipass-spectra}) are available, accurate estimates of \Mdyn\ depend on reliable measurement of the velocity width and knowledge of the galaxy inclination angle, while the \HI\ radius can be derived using the well-established \HI\ size-mass relation, \MHI\ $\approx 10^7 \times R^2_{\rm HI}$ (Broeils \& Rhee 1997; Wang et al. 2016). 

\begin{figure}[tb] % Figure 2
\begin{center}
 \includegraphics[width=8cm]{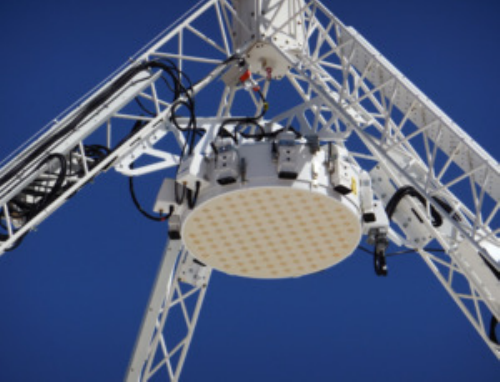}
\caption{Second-generation ASKAP Phased Array Feed (Chippendale et al. 2014), used to form 36 beams providing a $\sim$30 sq deg field-of-view at 1.4~GHz, ideal for widefield 21-cm sky surveys. --- Photo Credit: CSIRO.}
\label{fig:askap-paf}
\end{center}
\end{figure}

\subsection{Recent widefield \HI\ surveys} % Section 2.1

Multibeam receivers on large single-dish radio telescopes were built to advance studies of galaxy evolution by enabling large-area 21-cm surveys. First up was the 64-m Parkes Telescope, which in 1997 was equipped with a 13-beam receiver system, now the second largest multibeam system worldwide (second only to the recently commissioned FAST 19-beam receiver; Li et al. 2018). In 1999 the 76-m Lovell Telescope at Jodrell Bank commenced \HI\ surveys with their new 4-beam system (Lang et al. 2003), followed by installation of a 7-beam receiver on the 305-m Arecibo telescope in 2004 (Bird \& Cort\'es-Medell\'in 2003). In 2006, the 100-m Effelsberg Telescope was equipped with a 7-beam receiver system (Winkel et al. 2010). Future widefield \HI\ surveys are likely to use 21-cm Phased Array Feeds (PAFs, see Fig.~\ref{fig:askap-paf}) instead of multibeam systems, due to their increased field of view and higher survey speed. Not only have PAFs been shown to significantly reduce the standing waves between dish and receiver (Reynolds et al. 2017), they also provide greater flexibility in forming a large number of beams optimised for specific science goals, with the potential to use some beams for RFI mitigation (e.g., towards artificial satellites).

\subsubsection{Parkes HI surveys} % Section 2.1.1

The development of the innovative Parkes 13-beam receiver system ($T_{\rm sys} \approx 20$~K), coupled to a versatile correlator, instigated the era of large-scale 21-cm surveys of our Galaxy and the Local Universe. Among these, the `\HI\ Parkes All Sky Survey' (\hipass) is the largest and most prominent, covering the whole southern sky and the northern sky to a declination limit of $\delta \approx 25\degr$ over a velocity range from --1280 to 12\,700\kms. The Parkes gridded beam is $\sim$15.5 arcmin, the velocity resolution is 18\kms\ (with 4~arcmin pixel size and 13.2\kms\ channel width) and the rms noise is $\sim$13 mJy\,beam$^{-1}$ per channel for a typical integration time of eight minutes. A detailed description of the \hipass\ observations, calibration and imaging techniques is given in Barnes et al. (2001).

Current galaxy catalogues include the `\hipass\ Bright Galaxy Catalog' ({\sc hipass bgc}; Koribalski et al. 2004; see Fig.~1), the southern \hipass\ catalogue (Meyer et al. 2004) and the northern \hipass\ catalogue (Wong et al. 2006). Together these catalogues, which are highly reliable (Zwaan et al. 2004), contain more than 5\,500 \HI-rich galaxies and a few extragalactic \HI\ clouds. Detailed studies of the \HI\ mass function are presented by Zwaan et al. (2003, 2005). In addition, deeper Parkes \HI\ multibeam surveys of the Zone of Avoidance ({\sc hizoa}) have catalogued close to 1000 galaxies in the latitude range $|b| < 5\degr$; see Staveley-Smith et al. (2016) and references therein.

The large majority of catalogued \HI\ sources have been identified with optical counterparts (Koribalski et al. 2004; Doyle et al. 2005; Wong et al. 2009). About 15\% of galaxies are obscured by foreground dust and high star density, making identification in optical or infrared surveys difficult (e.g., Ryan-Weber et al. 2002). The {\sc hipass bgc} contains at least one extragalactic \HI\ cloud (HIPASS J0731--69) without an optical counterpart, located 200~kpc north-west of the asymmetric galaxy NGC~2442, which is a member of the NGC~2434 galaxy group (Ryder et al. 2001; Bekki et al. 2005a,b). Kilborn et al. (2000) and later Koribalski et al. (2004) also discuss star-less \HI\ clouds at velocities around +400\kms, e.g., HIPASS J1712--64 and HIPASS J1718--59, which are likely of Magellanic origin. Detailed \HI\ studies of the Magellanic Stream (Putman et al. 2003; For et al. 2014) shed new light on the interactions between the Magellanic Clouds and the Milky Way (see also Westmeier \& Koribalski 2008). Compact and extended populations of Galactic high-velocity clouds (HVCs) were catalogued by, for example, Putman et al. (2002). 

With a cryogenic 21-cm PAF, which is likely to replace the multibeam system in the next few years, the \HI\ survey speed might increase by a factor of four. A new Parkes \HI\ survey would be ideal to combine with \wal\ for enhanced studies of the extended and diffuse \HI\ emission in and around galaxies (see also Section~2.2.4).

\subsubsection{Jodrell HI surveys} % Section 2.1.2

The `\HI\ Jodrell All Sky Survey' ({\sc hijass}; Lang et al. 2003) was designed to match \hipass\ (in sensitivity, velocity coverage and velocity resolution) and complete \HI\ mapping of the northern sky. The gridded Lovell beam is $\sim$12~arcmin. Using the new 4-beam receiver system, the 76-m Lovell Telescope surveyed three areas, covering the North Galactic Cap region (1\,115 sq deg; Lang et al. 2003), the M\,81 group (180 sq deg; Boyce et al. 2001), and the Ursa Major region (480 sq deg; Wolfinger 2013, 2016). A total of 396 \HI\ sources were published in these three areas, including $\sim$30 newly discovered galaxies. 

\subsubsection{Effelsberg HI surveys} % Section 2.1.3

The `Effelsberg-Bonn \HI\ Survey' ({\sc ebhis}; Winkel et al. 2010, 2016; Kerp et al. 2011), conducted with the 7-beam receiver on the 100-m Effelsberg telescope, covers the full northern hemisphere ($\delta > -5\degr$, i.e. 21\,400 sq deg). Using a bandwidth of 100~MHz and 16\,384 channels, it maps the \HI\ line of the Milky Way (MW) galaxy and the Local Universe to $z < 0.07$ (270~Mpc) with an effective angular resolution of 10.8~arcmin. The first full-sky coverage was completed in 2013, and the first Milky Way \HI\ data cubes  ($|v_{\rm LSR}| \leq 600$\kms) were released in 2016 (Winkel et al. 2016) after careful removal of stray-radiation (Kalberla et al. 2005). Data acquisition with a high time resolution of 0.5\,s (Fl\"oer \& Winkel 2012) enabled flagging of time-variable RFI. While the compilation of the full extragalactic catalogue is awaiting completion of the final survey scans (Fl\"oer, Winkel \& Kerp 2014), a catalogue from the shallow survey of the northern Zone of Avoidance ($\delta >-5\degr, |b| \le 6\degr$) is already available by Schr\"oder et al. (2019). 

Data from {\sc ebhis} and the Parkes Galactic \HI\ Survey (McClure-Griffiths et al. 2009; Kalberla et al. 2010; Kalberla \& Haud 2015) were combined to form a homogeneous all-sky survey referred to as the HI4PI survey (HI4PI Collaboration 2016) with an angular resolution of 16.2~arcmin and a sensitivity of about 43~mK rms. The HI4PI survey covers the Milky Way and its system of intermediate and high-velocity clouds (Westmeier 2018) as well as several Local Group galaxies, most notably the Magellanic Clouds and the Andromeda Galaxy (Kerp et al. 2016). The 100-m Effelsberg telescope was recently equipped with a suitably-modified ASKAP PAF, which is now in operation. 

\subsubsection{Arecibo HI surveys} % Section 2.1.4

Pioneering blind surveys such as the `Arecibo \HI\ Strip Survey' (Zwaan et al. 1997) and the `Arecibo Dual Beam Survey' (Rosenberg \& Schneider 2000) used a single pixel 21-cm receiver. The installation of the 7-beam Arecibo L-band Feed Array (ALFA) in 2004 enabled the next-generation suite of blind \HI\ surveys with different trade-offs between sky coverage and depth. The most prominent among the extragalactic \HI\ surveys is the `Arecibo Legacy Fast ALFA Survey' (\alfalfa; Giovanelli et al. 2005), completed in Oct 2012, requiring $\sim$4700~h of telescope time. \alfalfa\ covers an area of 7\,000 sq deg over a redshift range of $-1600 < cz < 18\,000$\kms, divided into 5\kms\ channels. Two equatorial strips were observed in drift scan mode. \alfalfa\ has an rms noise of $\sim$2~mJy per 4~arcmin beam after Hanning smoothing to 10\kms\ channels. The final \alfalfa\ \HI\ catalogue (Haynes et al. 2018, Jones et al. 2018b) contains $\sim$31\,500 sources, including some HVCs. Several velocity ranges were wiped out by local radar interference. 

The `Arecibo Galaxy Environment Survey' (Auld et al. 2006, Cortese et al. 2008) covers a smaller sky area ($\sim$200 sq deg) compared to \alfalfa, but reaches a lower rms noise of 0.75 mJy per 10\kms\ channel. Its main goal is to investigate the variation of the \HI\ mass function with environment, by targeting selected fields spanning a large range of galaxy density, from voids to the Virgo cluster. The `ALFA Ultra Deep Survey' (Freudling et al. 2011) targets two fields with a combined sky area of 1.35 sq deg to a redshift of $z = 0.16$. Hoppmann et al. (2015) detected 102 galaxies in 60\% of the total survey with an rms of 80~$\mu$Jy (at 10\kms) in the RFI-free regions. The `ALFA Zone of Avoidance Survey' (Henning et al. 2010; Sanchez-Barrantes et al. 2019) covers the low-Galactic-latitude region with the goal to uncover heavily obscured or hidden galaxies out to 12\,000\kms\ and trace large-scale structure behind the Milky Way. 

The Galactic ALFA \HI\ survey covers 13\,146 sq deg and a velocity range of $\pm$650\kms\ with 150 mK rms noise per 1\kms\ velocity channel (Peek et al. 2018). 
Two Local Volume dwarf galaxies, Pisces\,A and B, were discovered (Tollerud et al. 2015; Carignan et al. 2016), while Saul et al. (2012) present a catalogue of compact \HI\ clouds, including HVCs and galaxy candidates.

\subsubsection{Targeted HI galaxy surveys}

A large range of targeted \HI\ surveys have been carried out with existing single-dish telescopes and radio interferometers. For a recent summary of \HI\ synthesis observations of nearby galaxies see Koribalski et al. (2018). Furthermore, Wang et al. (2016) gathered interferometric \HI\ data for over 500 nearby galaxies from 15 projects to re-examine the remarkably tight \HI\ size-mass relation of galaxies and the shape of their \HI\ surface brightness profiles, covering five orders of magnitude in \MHI. Among the single-dish \HI\ galaxy surveys, which we refer to later in this paper, are the 2MASS Tully-Fisher (2MTF) project, which gathered \HI\ spectra for 2062 nearby spiral galaxies (Springob et al. 2016; Howlett et al. 2017b; Hong et al. 2019), the `Nan\c{c}ay Interstellar Baryons Legacy Extragalactic Survey' (NIBLES) which contains 2600 \HI\ spectra of Sloan Digital Sky Survey (SDSS) selected galaxies (van Driel et al. 2016), the extended GALEX Arecibo SDSS Survey (xGASS), which contains $\sim$1200 \HI\ spectra of stellar mass and redshift selected SDSS galaxies (Catinella et al. 2018), and the {\sc amiga} project (Verdes-Montenegro et al. 2005), which contains \HI\ spectra for more than 800 isolated galaxies (Jones et al. 2018a).

\subsection{Planned large-scale \HI\ surveys} % Section 2.2

Several new radio telescopes capable of large-scale \HI\ surveys are currently being commissioned. These include ASKAP (Johnston et al. 2007, 2008; see Figs.~2 \& 4), the South African Meer-Karoo Array Telescope (MeerKAT; Jonas et al. 2016), and the Chinese Five-hundred metre Aperture Spherical Telescope (FAST; Li et al. 2018). Furthermore, several existing radio telescopes are undergoing major upgrades. We particularly note the new 1.4~GHz PAFs (Apertif) on the Westerbork Synthesis Radio Telescope (WSRT), located in the Netherlands, which has been transformed into a 21-cm survey facility (Oosterloo et al. 2010). The upgraded Jansky Very Large Array (JVLA; Perley et al. 2011) provides much increased bandwidth for deep \HI\ studies of individual galaxies, galaxy groups and clusters (e.g., CHILES, Fernandez et al. 2016).

\subsubsection{ASKAP HI Surveys} % Section 2.2.1

An overview of ASKAP is given in Section~3. Four large \HI\ projects are planned, comprising large-scale, shallow \HI\ emission and \HI\ absorption surveys of the sky ($\delta < +30\degr$), deep \HI\ pointings and a Galactic/Magellanic \HI\ survey. The top-ranked\footnote{The {\sc wallaby} and {\sc emu} proposals were both top-ranked by the ASKAP survey evaluation committee.} \wal\ project is the subject of this paper. Originally it was to be conducted jointly with {\sc emu} (Evolutionary Map of the Universe) and {\sc possum} (POlarisation Sky Survey of the Universe's Magnetism), the planned radio continuum and polarization surveys, respectively. ASKAP's strength lies in the large instantaneous field-of view ($\sim$30 sq degr) delivered by the 36 PAF-generated beams, making it a fast 21-cm survey machine.

\begin{figure}[tb] % Figure 3 - created by Aaron
\begin{center}
  \includegraphics[width=1.0\linewidth]{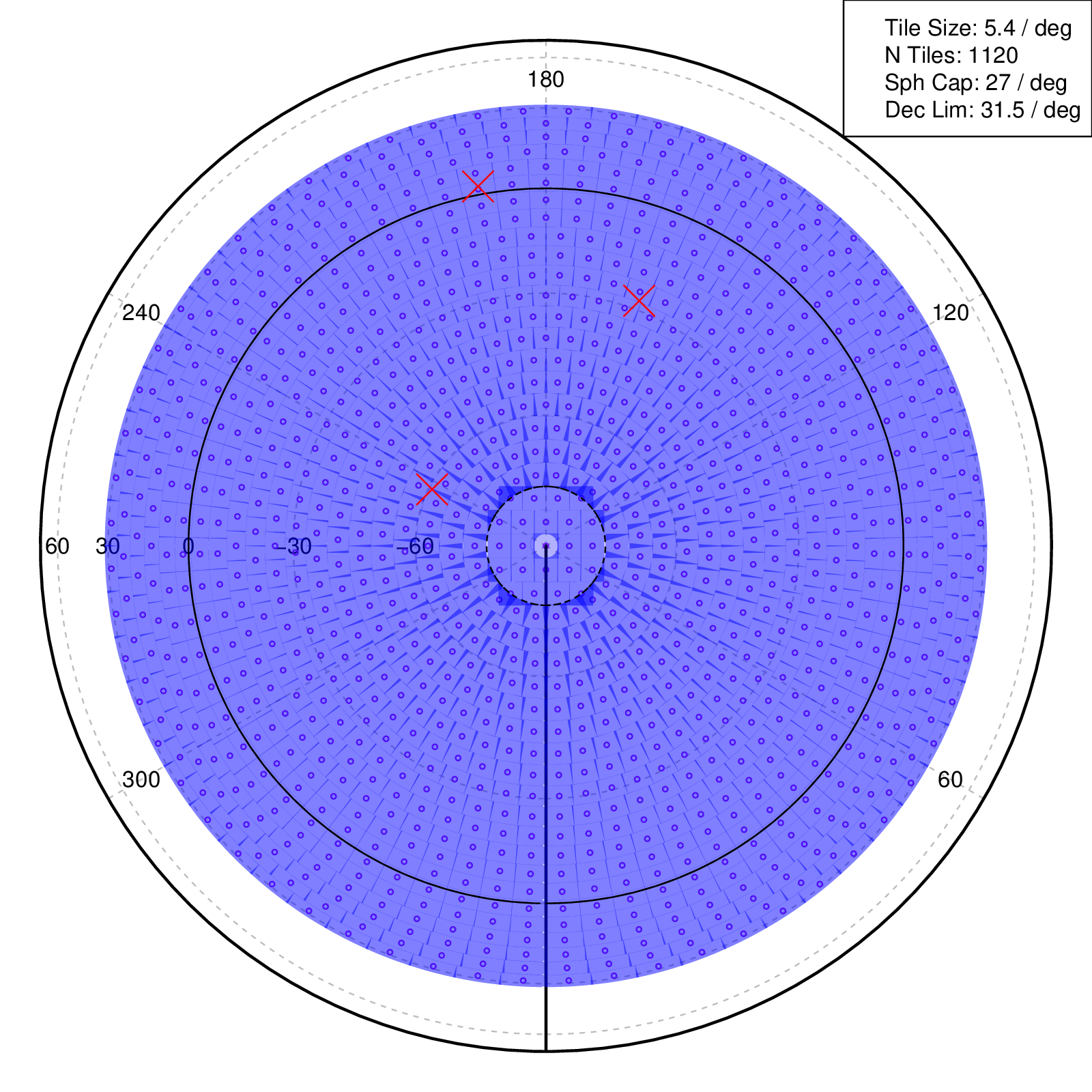}
\caption{\wal\ sky coverage (polar projection; $\delta < +30\degr$) and proposed ASKAP PAF tessellation pattern, consisting of 1120 tiles/pointings. Three red crosses mark the locations of the \wal\ pilot fields (see Section~6).}
\label{fig:wallaby-coverage}
\end{center}
\end{figure}

\begin{itemize}
\item {\bf WALLABY} will observe the whole southern sky and the northern sky up to $\delta = +30\degr$ (see Fig.~\ref{fig:wallaby-coverage}), to a redshift of $z \lesssim 0.26$. It is expected to detect around half a million galaxies in \HI\ (Johnston et al. 2008; Duffy et al. 2012). Technical details are given in Section~4 and Table~1, the main science goals are described in Section~5, and the \wal\ Early Science and pilot programs are summarized in Section~6.
\item {\bf DINGO} (`Deep Investigation of Neutral Gas Origins'), is a small-area \HI\ survey, consisting of two phases, which differ in area and depth. The deep survey aims to target five fields, i.e., 150 sq deg in total, out to $z \lesssim 0.26$. While the redshift range is the same as for \wal, the planned integration time per pointing ($\sim$500~h) is much longer. The second phase is proposed to consist of two ultra-deep fields, i.e., 60 sq deg in total, over the redshift range $z = 0.1 - 0.43$. The ultra-deep \dingo\ fields will probe the evolution of \HI\ over the last five billion years of cosmic time with a predicted total of $\sim$10\,000 galaxies (Johnston et al. 2008; Duffy et al. 2012). The target fields were selected to overlap with the `Galaxy and Mass Assembly' ({\sc gama}; Driver et al. 2011, 2016) survey.
\item {\bf FLASH} (`First Large Absorption Survey in \HI') has two components: one is commensal with \wal, exploring associated and intervening \HI\ absorption at low redshifts ($z \lesssim 0.26$), the other is at redshifts $0.4 - 1.0$ ($\sim$2~h integration time per field at 0.7 -- 1.0~GHz), expecting to detect and parametrize several hundred \HI\ absorption lines towards bright radio continuum sources (Allison et al. 2015).
\item {\bf GASKAP} (`Galactic ASKAP \HI\ Survey') plans to study \HI\ and OH lines in the Galactic Plane (for $|b| < 10\degr$), the Magellanic Clouds and the Magellanic Stream at a velocity resolution of 0.25\kms\ (Dickey et al. 2013). Within the proposed 7.3~MHz band, centered on the Galactic \HI\ line, {\sc gaskap} will also detect nearby, gas-rich galaxies. Its high velocity resolution benefits detailed kinematical modelling of nearby low-mass dwarf galaxies.
\end{itemize}

\begin{figure*}[tb] % Figure 4
\begin{center}
  \includegraphics[width=15cm]{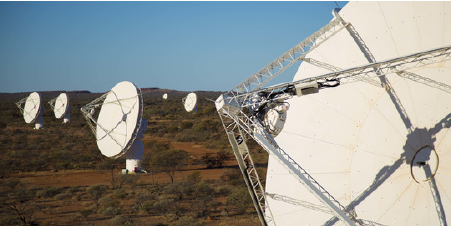}
\caption{ASKAP antennas equipped with Mk\,II Phased Array Feeds, 2017. --- Photo credit: CSIRO.}
\label{fig:askap-antennas}
\end{center}
\end{figure*}

\subsubsection{MeerKAT HI surveys} % Section 2.2.2

Five large \HI\ projects are planned for MeerKAT, which is an array of 64 $\times$ 13.5-m offset-Gregorian dishes with a primary beam FWHM of $\sim$1~deg at 1.4~GHz. MeerKAT's strength is the excellent \HI\ sensitivity thanks to the large number of antennas and low-noise single-pixel receivers, ideal for deep \HI\ surveys of small target fields.

\begin{itemize}
\item {\bf LADUMA} (`Looking at the Distant Universe with MeerKAT Array') is a very deep \HI\ survey (3424~h; single pointing) to study the evolution of galaxies out to $z \sim 1.4$ (Blyth et al. 2016). The target area includes the well-studied `Extended Chandra Deep Field South' at $\alpha,\delta$(J2000) = $3.5^{\rm h}, -28\degr$. 
\item {\bf MIGHTEE-HI} (\HI\ component of the `MeerKAT International GHz Tiered Extragalactic Exploration' survey) provides a 20 sq deg survey (16~h per pointing) out to $z \sim 0.5$ (Jarvis et al. 2016).
\item The {\bf `MeerKAT Fornax Survey'} will cover an area of 12 sq deg with a total integration of $\sim$1000~h (Serra et al. 2016). The Fornax cluster region is also one of the four \wal\ Early Science fields and was recently observed with the Australia Telescope Compact Array (ATCA); see, for example, Lee-Waddell et al. (2018).
\item {\bf MHONGOOSE} (`MeerKAT \HI\ Observations of Nearby Galactic Objects: Observing Southern Emitters') is a medium deep, pointed \HI\ survey (55~h per field) to study the low-column density environment around 30 nearby spiral galaxies (de Blok et al. 2016; Sorgho et al. 2019). 
\item {\bf MALS} (`MeerKAT Absorption Line Survey') will target 1100 bright radio continuum sources to search for intervening and associated \HI\ and OH absorbers out to $z = 2$. With roughly 3$\times$ higher sensitivity than \wal, it will -- as a by-prroduct -- be able to detect MW-like galaxies in \HI\ emission out to $z = 0.1 - 0.2$ (Gupta et al. 2016).
\end{itemize} 

\subsubsection{WSRT Apertif HI surveys} % Section 2.2.3

Twelve of the fourteen 25-m WSRT dishes have been equipped with Vivaldi PAFs, known as Apertif, providing 8 sq deg field-of-view and an effective system temperature of $\sim$70~K at 1.4~GHz (Oosterloo et al. 2010); the maximum angular resolution will be 15~arcsec. Survey observations started on 1 July 2019. Several science projects were recently approved and focus on the evolution of the \HI\ content in galaxies, covering a range of galaxy types and environments. 

The \HI\ science case for Apertif is very similar to that of ASKAP and a number of projects are planned with similar aims as those planned for ASKAP. In particular, the search for very low \HI\ mass galaxies in the nearby Universe, the role of the environment in galaxy evolution, and the \HI\ properties of active galactic nuclei (AGN) are the main focus areas. Some differences in survey strategy exist compared to ASKAP, however. 

Similar to ASKAP, a large-area, shallow Apertif \HI\ survey is carried out. With 12~h per pointing it aims to map a large part ($\sim$3\,500 sq degr) of the northern sky ($\delta > +27\degr$), and together with \wal, will provide an all-sky survey (Koribalski 2012a). Both cover the same frequency range and can provide similar resolution and sensitivity. 

In addition, a medium-deep \HI\ survey will be done to push the \HI\ column density well below 10$^{20}$ cm$^{-2}$. This survey targets a smaller area (350 sq deg) to greater depth ($10 \times 12$~h per pointing). The main scientific drivers are to explore the \HI\ mass function down to a minimum \HI\ mass of $2 \times 10^5$\Msun, to analyse in detail the extended morphologies and kinematics of the \HI\ in and around galaxies as a function of environment, and to determine the cosmic evolution of the gas content of galaxies over the past 3~Gyr. 

Current planning is for Apertif to operate until the end of 2021, with the possibility of an extension of survey time beyond that.

\subsubsection{FAST HI surveys} % Section 2.2.4
Li et al. (2018) describe the current FAST commissioning and science projects, conducted with the recently installed 19-beam horn-based array. Zhang et al. (2019) outline the `Commensal Radio Astronomy FasT Survey' (CRAFTS), aiming to cover  more than 20\,000 sq deg. They estimate that $>$600\,000  galaxies will be detected in \HI\ out to $z < 0.35$ for an effective integration time of 60s per beam. In areas where FAST \HI\ surveys overlap with \wal, we are looking forward to combine the data products to enable the detection of low-surface brightness \HI\ structures between galaxies in groups and clusters (see Section~6).

\section{ASKAP Overview} % Section 3

ASKAP\footnote{\url{https://www.atnf.csiro.au/projects/askap/news.html}} consists of $36 \times 12$-m antennas, located near Boolardy station in the Murchison Shire of Western Australia. For a detailed description of this new radio interferometer (see Fig.~\ref{fig:askap-antennas}) and key survey science areas see Johnston et al. (2007, 2008). All ASKAP antennas are equipped with widefield Phased Array Feeds (PAFs; see Fig.~\ref{fig:askap-paf}) delivering a field-of view of 30~sq deg, making ASKAP a fast 21-cm survey machine. From March 2014 to early 2016, six ASKAP antennas with first-generation PAFs were working together, leading to the first widefield continuum imaging and spectral line studies (Serra et al. 2015a, Heywood et al. 2016). This configuration is known as BETA, the Boolardy Engineering Test Array (Hotan et al. 2014, McConnell et al. 2016). In Oct 2016 the \wal\ team officially commenced ASKAP Early Science, initially using an array of 12 PAF-equipped antennas (ASKAP-12) with baselines extending out to $\sim$2~km and an observing bandwidth of 48 MHz (Reynolds et al. 2019; Lee-Waddell et al. 2019; Kleiner et al. 2019) and later 240 MHz (Elagali et al. 2019; For et al. 2019). In mid 2018, the first ASKAP observations with an array of 18 PAF-equipped antennas (ASKAP-18) and 240 MHz bandwidth were taken, expanding to an array of 28 PAF-equipped antennas in Oct 2018 (ASKAP-28), and the full 36-antenna array (ASKAP-36) in Mar 2019. The first round of ASKAP pilot surveys was conducted between Aug 2019 and May 2020.

\begin{figure*}[tb] % Figure 5 - created by Tobias
\begin{center}
 \includegraphics[height=7cm]{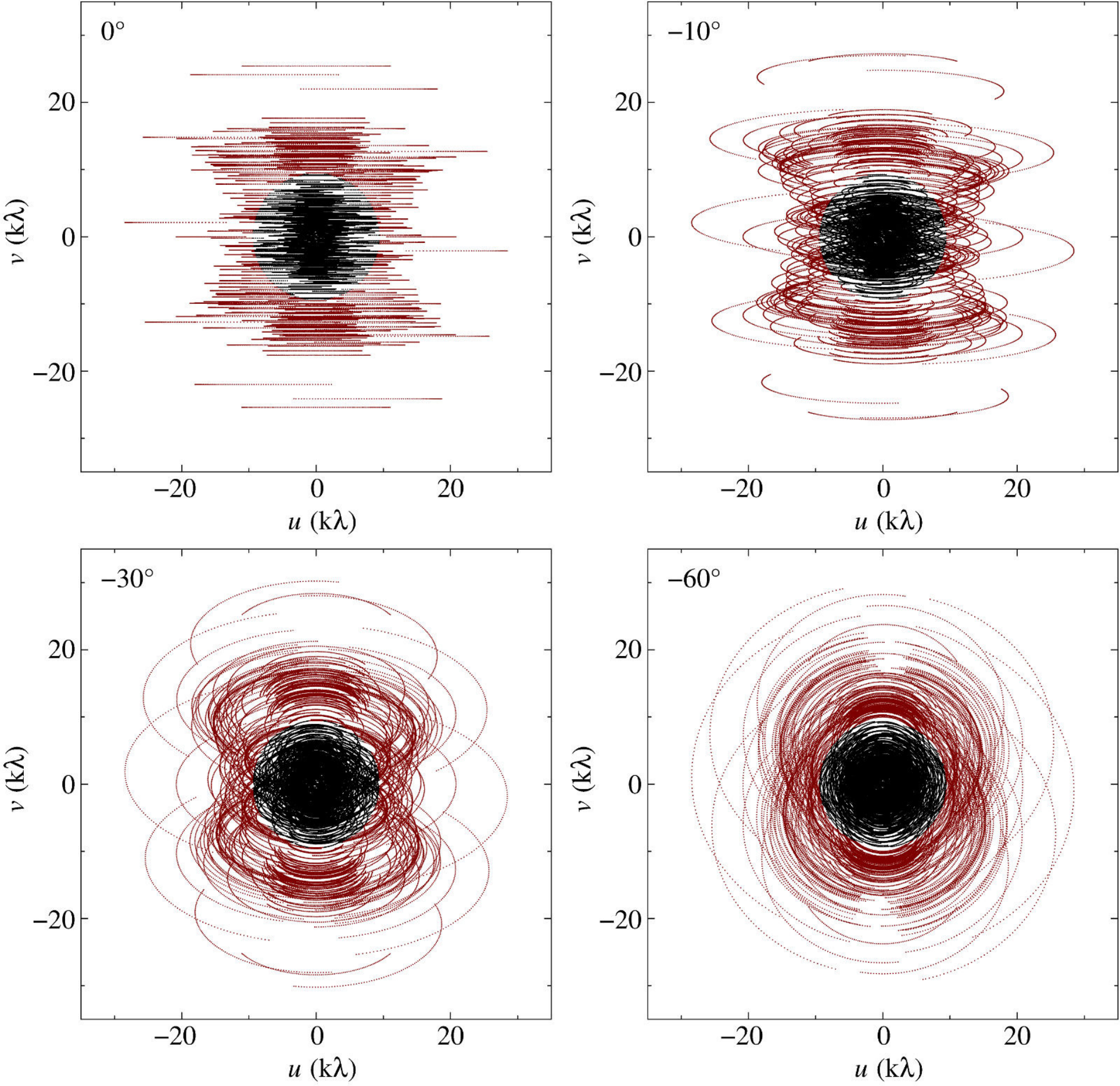} 
 \includegraphics[height=7cm]{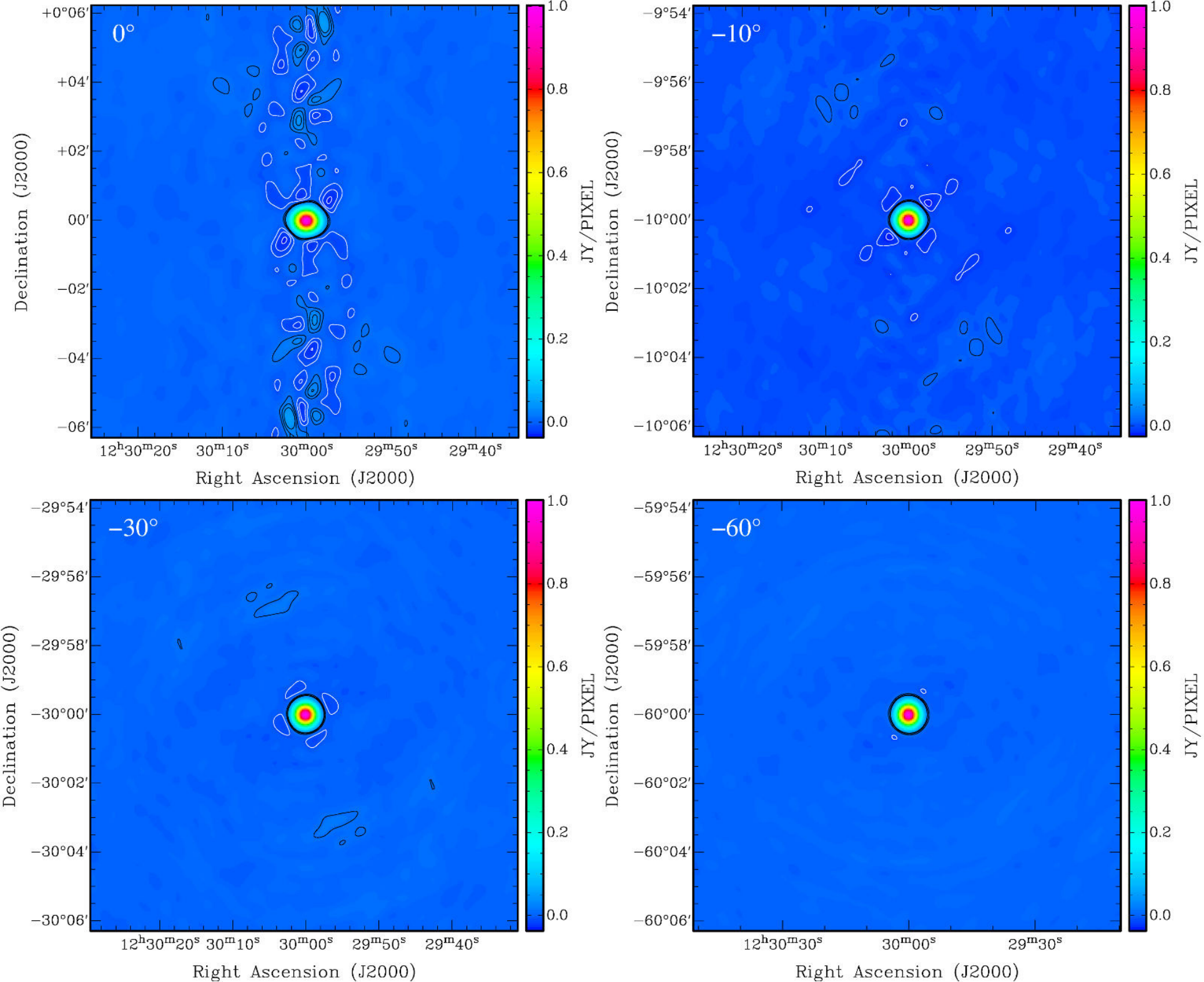}
\caption{Simulation of the ASKAP $uv$-coverage and corresponding synthesized beam for \wal, made using the on-line ASKAP simulator for a set of four declinations ($\delta$ = $0\degr, -10\degr, -30\degr$, and $-60\degr$) with an hour angle coverage of $\pm$4~h. The $uv$-coverage is shown for all 36 antennas, with projected baselines longer than 2~km coloured in red. We used Wiener filtering with `robust' = 0 weighting and $30''$ tapering.}
\label{fig:askap-uvcoverage}
\end{center}
\end{figure*}

ASKAP and its low-frequency counterpart, the Murchison Widefield Array (MWA, Tingay et al. 2013), are part of the Murchison Radio-astronomy Observatory (MRO), which is a designated radio-quiet zone and, moreover, the future Australian SKA site. Nevertheless, like at all other observatories around the world, satellite-based and other RFI increasingly affects the available observing bandwidth. The full 36-antenna ASKAP configuration provides 630 baselines, ranging from 22~m to 6~km. The inner 30 ASKAP dishes are located within a circle of $\sim$2~km diameter, providing a compact core designed for high-sensitivity \HI\ observations of the Local Universe at an angular resolution of $\sim$30~arcsec. ASKAP-36 $uv$-coverages and point spread functions for different declinations are shown in Fig.~\ref{fig:askap-uvcoverage}. 

Each ASKAP dish has been equipped with a Mk\,II Chequerboard PAF, providing a $\sim$30 sq deg field of view and covering a frequency range from 700 to 1800~MHz. While the PAF system temperature, \Tsys, is close to 50~K at 1.4 GHz, the antenna efficiency, $\eta$, is $\sim$0.7, resulting in an effective system temperature of \Tsys/$\eta$ $\sim$ 70~K. Tests are under way on increasing the antenna efficiency by, e.g., installing radio-transparent (fibreglass) feed legs on one antenna. The ASKAP hardware correlator has a nominal bandwidth of 300~MHz\footnote{Currently ASKAP is operating with a bandwidth of 288~MHz, consisting of six 48~MHz blocks, but this may be expanded to 336~MHz. Each coarse 1~MHz channel is divided into 54 narrow channels. One possibility (as yet untested) would be to double the number of beams from 36 to 72 while halfing the bandwidth to focus on the nearby Universe.}, divided into 16\,200 channels. The large $5.5\degr \times 5.5\degr$ field is achieved by forming 36 overlapping beams (FWHM $\sim 1\degr$) from the 188 PAF dipole elements (Chippendale et al. 2014). 

Engineering and system commissioning as well as a variety of early science projects led to a step-by-step deployment of the full ASKAP telescope, the custom-made, real-time data processing pipeline (ASKAPsoft; Whiting et al. 2017) and the dedicated CSIRO ASKAP Science Data Archive (CASDA\footnote{CASDA: \url{https://casda.csiro.au}}). ASKAPsoft is deployed on the Pawsey Supercomputing Centre in Perth and used to (a) ingest the large volumes of ASKAP data, (b) produce image cubes after data flagging, calibration and line-continuum separation, and (c) run the Selavy source finder (Whiting \& Humphreys 2012) to produce the initial bright-source catalogues. The expected data volume for \wal\ is $\sim$10~TB per hour: (a) cross-correlations (630 baselines $\times$ 36 beams $\times$ 4 polarizations $\times$ 16\,200 channels every 5~seconds) and (b) auto-correlations (36 antennas $\times$ 36 beams $\times$ 4 polarizations $\times$ 16\,200 channels every 5~seconds). The \wal\ team has been extensively testing, debugging and improving the ASKAPsoft pipeline using ASKAP early science and more recently pilot survey data, with the aim to perform reliable, robust and real-time data processing once our survey gets under way. Currently we are comparing `robust' and `natural' weighting of the ASKAP-36 $uv$-data,  Gaussian tapers, followed by multi-scale clean (BasisFunction or BasisFunctionMFS) and deep clean of the identified components; for details see Kleiner et al. (2019).

\begin{table*} % Table 1
\begin{center}
\caption{Survey parameters for \hipass\ and \wal; updated table from Koribalski (2012a). --- Note: An angular resolution of 10~arcsec can be achieved for \wal\ once computing resources allow the processing of all ASKAP baselines (up to 6~km), either for a sample of pre-determined galaxies (`postage stamps') or for the full survey.}
\begin{tabular}{lccc}
\tableline
                     & \\
                     & {\bf HIPASS} & {\bf WALLABY}    \\
                     & \\
\hline
                     & \\
telescope            &   Parkes     & ASKAP   \\
                     &  64-m dish   & $36 \times 12$-m dishes \\
                     & \\
baselines            & --- & 22 m to $\sim$2 km \\
% area                & & 4072 m$^2$  \\

                     & \\
receiver             & 21-cm multibeam & Phased Array Feed \\
                     & & (Chequerboard) \\
beams                & 13           & 36 \\
field-of-view        & $\sim$1 sq deg & $\sim$30 sq deg \\
\Tsys/$\eta$         & $\sim$25 K     & $\sim$70 K \\
observing mode       & scanning 
                     & dithering/mosaicking \\
                     & \\
angular resolution   & $\sim$15.5 arcmin 
                     & $\sim$30 arcsec \\
                     & \\
sky coverage         & $\delta < +25\degr$ 
                     & $\delta < +30\degr$ \\
                     & 29 343 sq deg 
                     & 30 940 sq deg \\
cubes/fields         & 538 ($8\degr \times 8\degr$)
       & 1120 ($5.5\degr \times 5.5\degr$) \\
                     & \\
frequency coverage   & 1362.5 -- 1426.5 MHz
                     & 1130 -- 1430 MHz \\
velocity range ($cz$)& --1280 to +12 700\kms
                     & --2000 to +77\,000\kms \\
                     & $z \lesssim 0.04$ 
                     & $z \lesssim 0.26$ \\
bandwidth            & 64 MHz     & 300 MHz \\
no. of channels      & 1\,024     & 16\,200   \\
channel width        & 13.2\kms   & $\sim$4\kms \\
velocity resolution  & 18.0\kms   & $\sim$4\kms \\
                     & \\
rms per channel      & $\sim$13 mJy\,beam$^{-1}$
                     & $\sim$1.6 mJy\,beam$^{-1}$ \\
rms per 0.1~MHz      & $\sim$10 mJy\,beam$^{-1}$
                     & $\sim$0.7 mJy\,beam$^{-1}$ \\
                     & \\
\HI\ detections      & $\sim$5\,500
                     & $\sim$500\,000 \\
mean redshift ($z$)  & $\sim$0.01 
                     & $\sim$0.05 \\
                     & \\
\tableline
\hspace{0.5cm}
\end{tabular}
\end{center}
\end{table*}

\section{WALLABY Parameters} % Section 4

The proposed \wal\ parameters, as compared to \hipass, are listed in Table~1, updated from a similar table in a previous \wal\ paper (Koribalski 2012b). Our aim is to reach a rms noise of 1.6 mJy\,beam$^{-1}$ per 4\kms, allowing us to detect and characterize around 500\,000 galaxies in the proposed survey volume. Table~1 illustrates that while \wal\ will have only slightly larger sky coverage than \hipass, it explores a substantially larger frequency range ($\times$5) at much higher angular resolution ($\times$30), much improved velocity resolution ($\times$4.5) and much higher point source sensitivity ($\times$20). \\

In the following subsections we describe 1) the survey volume \& coverage, 2) survey sensitivity, 3) survey resolution, 4) expected survey detections, 5) source finding \& parametrization, 6) catalogues \& data products, and 7) 3D data visualisation.

\subsection{Survey volume \& coverage} % Section 4.1

\wal\ will cover 75\% of the sky ($-90\degr < \delta < +30\degr$; see Figs.~3 \& 6) to a redshift of $z \lesssim 0.26$, equivalent to a look-back time of $\sim$3~Gyr. The corresponding cosmic volume of the survey is 3.26~Gpc$^3$. Based on the nominal ASKAP bandwidth of 300~MHz, we chose a frequency range of 1.13 to 1.43~GHz, corresponding to a velocity range of $-2000 < cz < 77,000$\kms. Dividing the survey area by the PAF 30 sq deg footprint suggests we need at least 1000 pointings. To maximise survey uniformity and tiling efficieny a number of strategies were explored; one of the solutions, shown in Fig.~\ref{fig:wallaby-coverage}, requires 1120 pointings. We aim to minimize solar sidelobes and satellite interference, as far as possible, and will target complex fields (e.g., containing very bright and extended continuum sources) at night. To achieve uniform sky coverage with the ASKAP PAFs, strategies such as dithering (e.g., Reynolds et al. 2019; Lee-Waddell et al. 2019; Kleiner et al. 2019) and mosaicking of neighbouring fields are also explored. This is necessary because the 36 beams (each $\sim$1$\degr$ FWMH) formed to give a $\sim$30 sq deg field, do not provide Nyquist sampling of the sky. The final strategy will depend strongly on the ASKAP PAF performance (including shape and beam to beam variations; see Hotan et al. 2020, and McConnell et al. 2020), RFI occupancy and computing considerations. The total on-source integration time for \wal\ is projected to be just over two years.

\subsection{Survey sensitivity} % Section 4.2

\begin{figure*}[tb] % Figure 6
\begin{center}
  \includegraphics[width=0.9\linewidth]{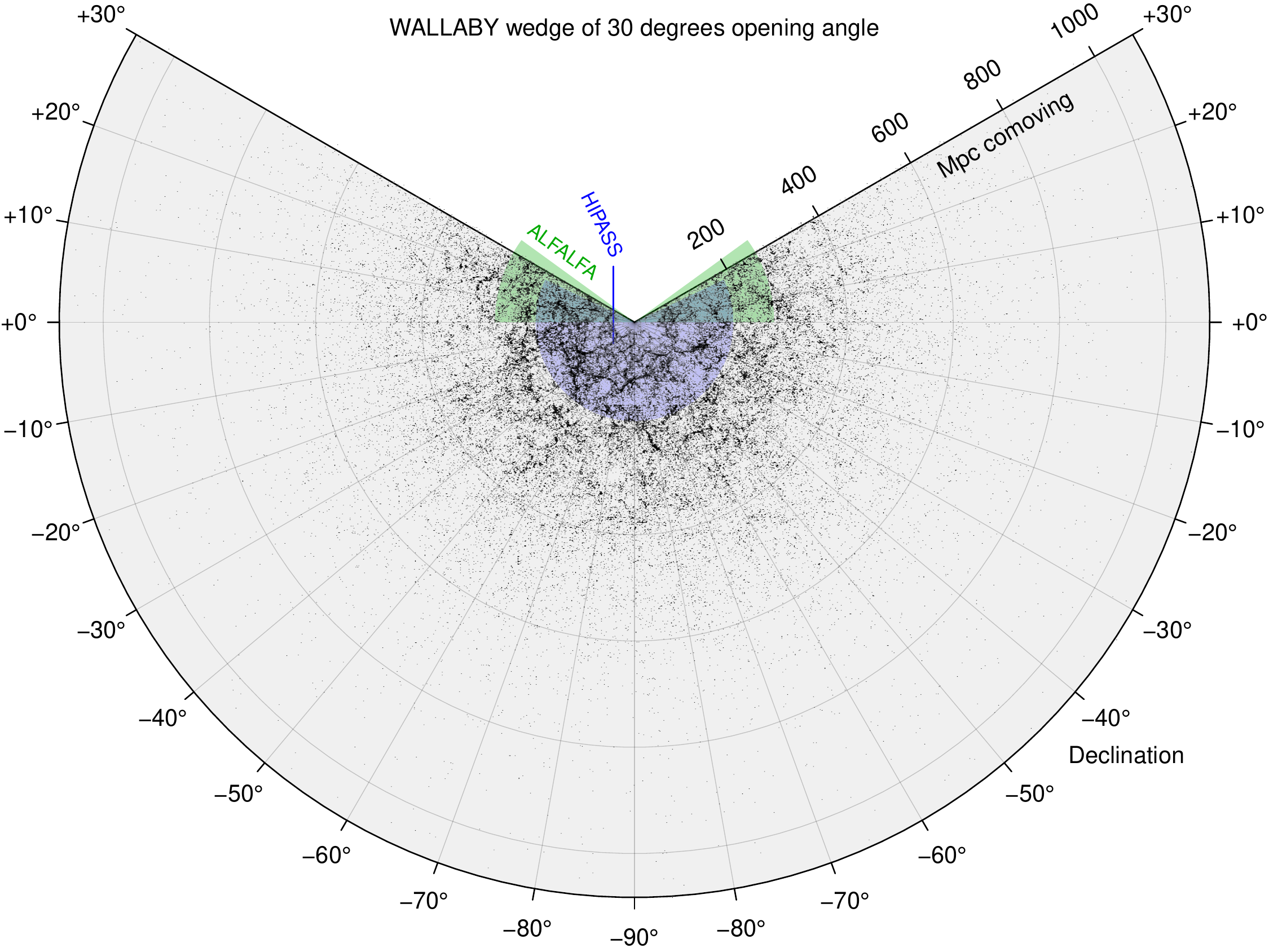}
\caption{Spatial distribution of the mock \wal\ galaxies from the simulation that lie within 15~degrees of the plane defined by $\mathrm{RA} \in \{0~\mathrm{h}, 12~\mathrm{h}\}$. The distance limit of $D_{\rm c} = 1082.6~\mathrm{Mpc}$ corresponds to a cosmological redshift of $z = 0.26$ in the Planck-type cosmology adopted in SURFS (see Elahi et al. 2018). The corresponding maximum volumes of the two largest existing blind \HI\ surveys, \hipass\ and \alfalfa\, are shown in coloured shading. Note that \wal\ and \hipass\ cover all RAs, whereas \alfalfa\ was limited to two RA windows, collectively covering 210$^\circ$ in RA.}
\label{fig:simsky}
\end{center}
\end{figure*}

\begin{figure*}[tb] % Figure 7
\begin{center}
  \includegraphics[width=8cm]{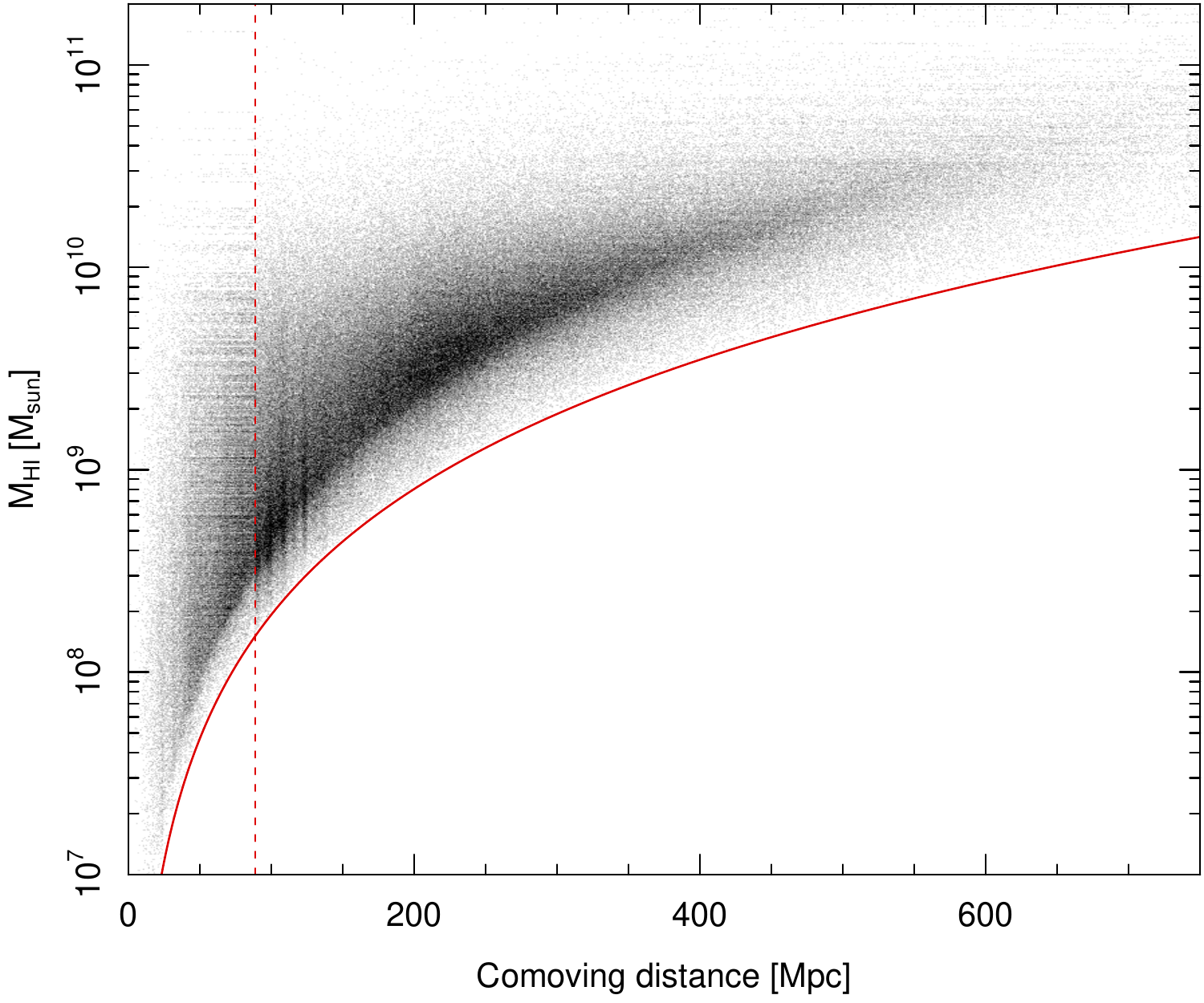}
  \hfill
  \includegraphics[width=8cm]{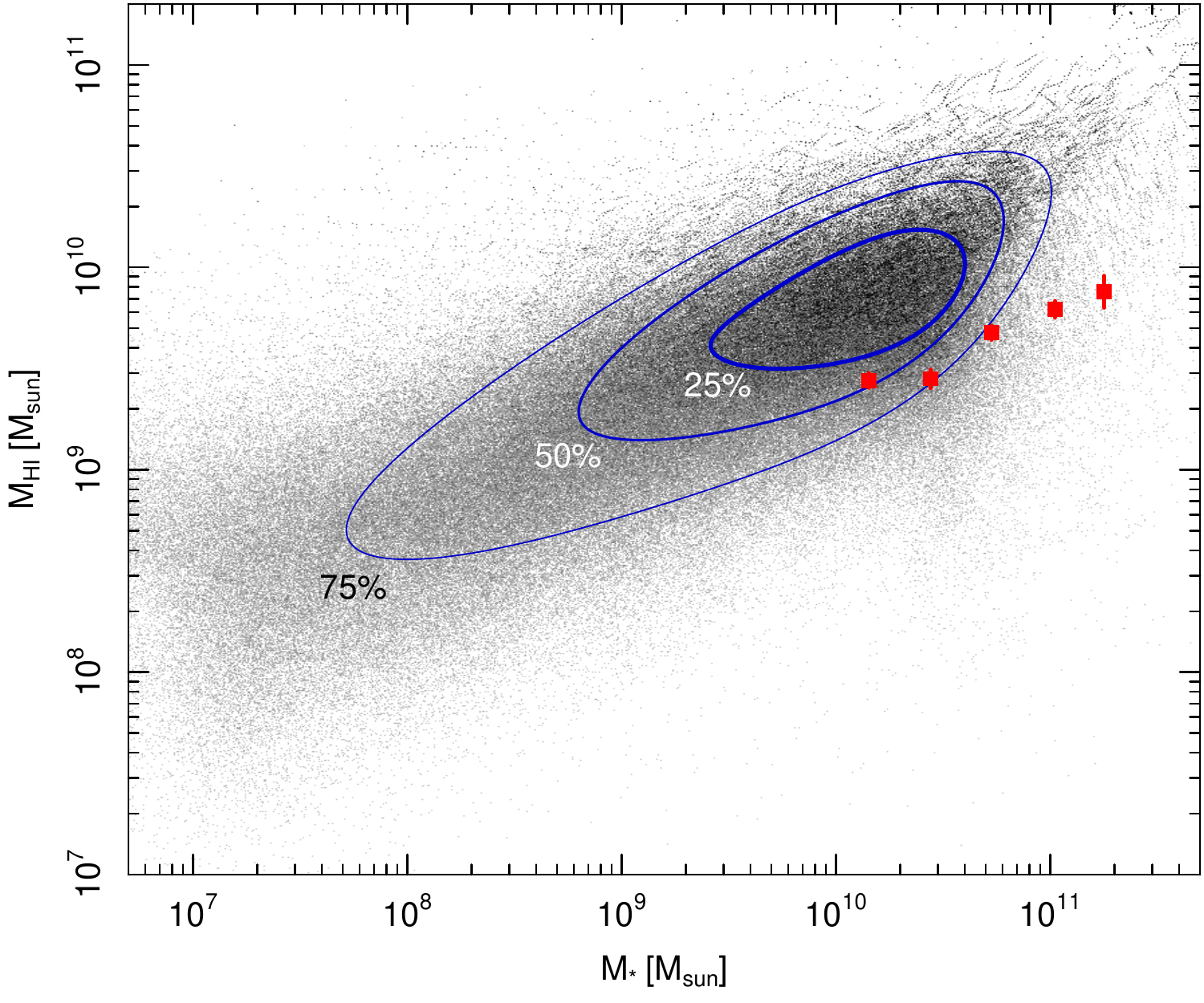}
\caption{Further results from the \wal\ reference simulation. --- {\bf Left:} \HI\ mass plotted against comoving distance, truncated at 750~Mpc. The red solid line is the theoretical detection limit for unresolved (i.e., one beam) and face-on (i.e., minimum number of channels) galaxies. The dashed vertical line shows the transition from the small simulation box to the larger one. Vertical structure in the galaxy distribution comes from real cosmic large-scale structure, whereas horizontal structure comes from (non-real) galaxy repetitions in building the mock sky. --- {\bf Right:} \HI\ mass versus stellar mass. Contours contain the indicated fraction of all galaxies. Red points are the means of the binned \textit{detected} galaxies in the `GALEX Arecibo SDSS Survey' (GASS; Catinella et al. 2013) with error bars showing the statistical uncertainty of the mean. The line-like features at the highest \HI\ and stellar masses come from repetitions of the same galaxies at different cosmic times, thus showing a slight correlated evolution in the masses.}
\label{fig:mhi_distance}
\end{center}
\end{figure*}

To achieve the \wal\ science goals, we aim to reach an rms noise of 1.6~mJy per 30-arcsec beam and 18.52-kHz channel ($\sim$4\kms\ at $z = 0$). Based on the currently available measurement of $T_{\rm sys} / \eta$ = 70~K, we estimate an integration time of 16~h per \wal\ field. This is typically split into $2 \times 8$~h with interleaved footprints of the 36 PAF beams. For the northern most fields, $4 \times 4$~h integrations will be required. \wal\ will detect dwarf galaxies (\MHI\ = 10$^8$\Msun) out to a distance of $\sim$60~Mpc, massive galaxies (\MHI\ = $6 \times 10^9$\Msun) to $\sim$500~Mpc, and super-massive galaxies such as Malin\,1 (\MHI\ = $5 \times 10^{10}$\Msun) to the survey `edge' of 1~Gpc (see Figs.~\ref{fig:simsky}--\ref{fig:wallaby-predictions}). See Table~2 for \MHI\ estimates for different galaxy velocity widths and two \HI\ sizes. At large distances, $<$1.3~GHz ($>$400~Mpc), increasing satellite RFI will (much) reduce the detectability of \HI\ sources. See Table~3 for the impact on the numbers of predicted \wal\ detections.

The 5$\sigma$ \HI\ column density sensitivity of \wal\ is 20, 5 and $2.5 \times 10^{19}$ cm$^{-2}$, assuming a velocity width of 20\kms\ for beams of 30, 60 and 90 arcsec, respectively.

The mean sample redshift is expected to be $z \sim 0.05$ (200~Mpc). The predicted \wal\ source count as a function of signal-to-noise and redshift/velocity is shown in Fig.~\ref{fig:wallaby-predictions} (see Section~4.4 and Table~3). \wal\ will have a flux sensitivity some 20 times better than \hipass\ and will detect $\sim$100 times more galaxies out to distances up to six times further. It will also detect all galaxies in the \alfalfa\ \HI\ catalogue (within the overlap region) with 8$\times$ higher angular resolution, i.e. $\sim$30~arcsec instead of 4~arcmin. The rms noise achieved by \alfalfa\ is $\sim$2 mJy\,beam$^{-1}$ per 10\kms\ channel, i.e. a factor two higher than the predicted \wal\ rms at the same spectral resolution.

\subsection{Survey resolution} % Section 4.3

\wal\ data processing will be restricted to ASKAP-36 baselines shorter than 2~km, delivering excellent $uv$-coverage and an {\bf angular resolution} of $\sim$30~arcsec. Additionally, it may be possible, pending computing resources, to collect ASKAP-36 high-resolution ($\sim$10~arcsec) \HI\ data cubelets (here called `postage stamps') for predetermined galaxies in each \wal\ pointing. 

The ASKAP correlator has a nominal bandwidth of 300~MHz, divided into $300 \times 1$~MHz coarse channels. For \wal\ spectral line mapping the band is divided into 16\,200 narrow channels. Each channel has a fixed frequency width of 18.518~kHz, corresponding to a {\bf velocity resolution} of 3.91\kms\ at $z = 0$ and 4.69\kms\ at $z = 0.2$.

\subsection{Survey predictions} % Section 4.4

Significant progress in galaxy modelling has been made over the past decade. Specifically, semi-analytic models of galaxy formation now routinely use two ISM phases (atomic and molecular hydrogen; e.g., Lagos et al. 2011, Popping et al. 2014, Stevens et al. 2018), while cosmological hydrodynamic simulations are now mature enough to produce realistic galaxy populations in terms of their mass, star formation rates, and colors among other properties (e.g., Schaye et al. 2015, Pillepich et al. 2018, Dav\'e et al. 2017). The latter are now being used to study \HI\ in galaxies across cosmic time and in a wide range of environments (e.g. Bah\'e et al. 2016, Stevens et al. 2019).

The first simulations of source counts for extragalactic \HI\ surveys with ASKAP (Johnston et al. 2008) were empirically-based and used the \HI\ mass function derived from \hipass\ by Zwaan et al. (2003, 2005). Using an angular resolution of 60~arcsec, they estimated that $\sim$600\,000 galaxies would be detected in \HI\ over one hemisphere. In the \wal\ proposal (Koribalski et al. 2009) we estimated that around half a million galaxies will be detected in the full 3$\pi$ sr survey area (assuming an angular resolution of 30~arcsec).

Obreschkow et al. (2009b) and Duffy et al. (2012) presented predictions for the expected galaxy counts from more sophisticated models of galaxy formation and also investigated the spatial, velocity and mass distributions of galaxies and their \HI\ diameters. Duffy et al. showed that the majority of \wal\ galaxies (86\%) will be resolved by more than one beam. About 5\,000 galaxies will have major-axis \HI\ diameters greater than 2.5 arcmin ($> 5$~beams), enabling kinematic studies of their gaseous discs (see Section~5.3). This number would rise to $1.6 \times 10^5$ galaxies if all baselines up to 6~km could be used, resulting in an improved angular resolution of $\sim$10~arcsec. Although these models provided a clear improvement over previous, empirical estimates, they were still based on the post-processing of existing galaxy formation models that treated the ISM of galaxies as a single phase. Hence, it is fit to update these predictions using state-of-the-art galaxy formation simulations that model the multi-phase nature of the ISM in a self-consistent way.

Here, we present a new \wal\ reference simulation that provides a mock catalog of a \textit{perfect} \wal\ survey able to detect all galaxies whose integrated \HI\ signal-to-noise lies 5 times above the \wal\ target noise level (ie., S/N $> 5$). This simulation relies on the `Synthetic UniveRses For Surveys' (SURFS) simulations of cosmic structure (Elahi et al. 2018), the SHARK semi-analytic model of galaxy evolution (Lagos et al. 2018) and the STINGRAY mock sky builder (Obreschkow et al., in prep.). This three-step approach is conceptually identical to that of Chauhan et al. (2019), used for the \alfalfa\ survey. To cover the full dynamic range of \HI\ masses, the reference simulation concatenates two cubic simulation boxes with volumes of $40^3\,h^{-1}~\mathrm{Mpc}$ (micro-SURFS) and $210^3\,h^{-1}~\mathrm{Mpc}$ (medi-SURFS), respectively (where $h = 0.6751$). The catalogue transitions seamlessly from the smaller to the larger box at a comoving distance of $60\,h^{-1}~\mathrm{Mpc}$, chosen as a compromise between the requirements for high resolution in the nearby Universe and the desire to minimise replications of identical cosmic structure.

For each galaxy in the simulation, multi-component rotation curves were produced and convolved with the pressure-based surface density distribution of \HI\ (following Obreschkow et al., 2009a) in order to produce realistic \HI\ line profiles. In doing so, we assumed a constant line-of-sight velocity dispersion of $\sigma$ = 10\kms. The projected surface density profiles and \HI\ line profiles were then used to compute the (integer) number of beams and channels each source subtends in \wal\, assuming a circular beam of 30~arcsec diameter and 4\kms\ channels (see Table~1). Using those parameters and an assumed channel noise of 1.6~mJy per beam, all galaxies with an \textit{integrated} (over all beams and channels) signal-to-noise ratio greater than five were selected.

Five random realisations of the reference simulation are available, which allow a rough estimate of cosmic variance and shot noise statistics. The total number of galaxies in these simulations predicted to be detectable by \wal\ is $490\,000 \pm 6\,000$ (1$\sigma$ interval). This number -- in the absence of any RFI -- is consistent with the $\sim$500\,000 detections estimated in the original \wal\ proposal (Koribalski et al. 2009).

\begin{figure*}[h] % Figure 8 - created by Danail (full page please)
\begin{center}
  \includegraphics[width=10cm]{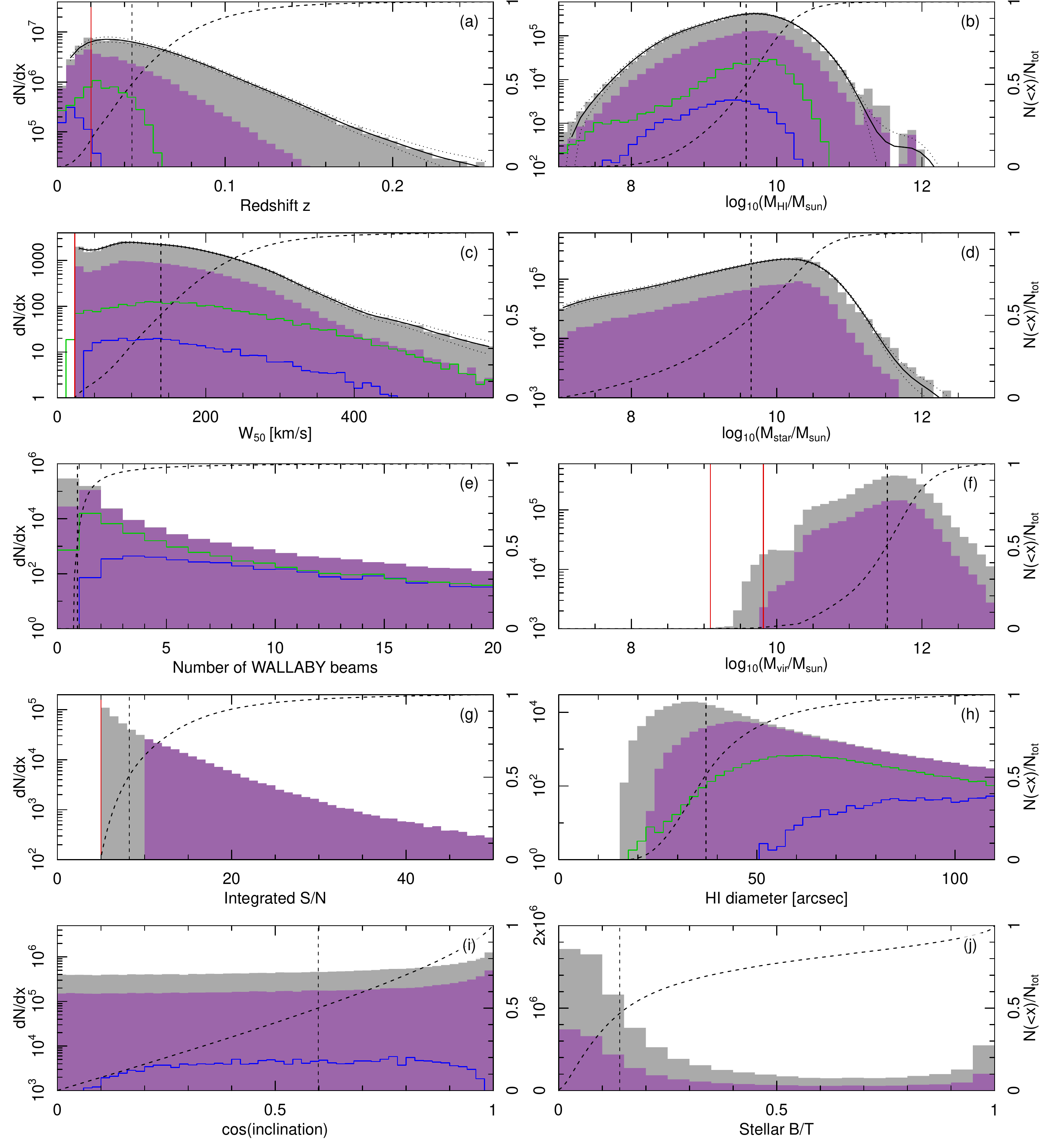}
\caption{Histograms show the differential $dN/dx$ counts from the \wal\ reference simulation, where $x$ is the quantity shown on the horizontal axes. Grey histograms show all galaxies with an integrated $S/N$ above 5. Solid black curves and dotted lines in (a)--(d) show a smoothed mean and standard deviation (cosmic variance + shot noise), determined form spatially distinct sub-samples. Dashed curves show the corresponding cumulative fractions $N(<x)/N_{\rm tot}$ (right axes). Dashed vertical lines are the median values. Purple histograms show the sub-sample of \wal\ detections with $S/N > 10$. Line histograms show the observations from \hipass\ (blue) and \alfalfa{}-100 (green); the \hipass\ inclinations in (i) are from HOPCAT (Doyle et al. 2005); the \HI\ diameters in (h) were inferred from the \HI\ size-mass relation by Wang et al. (2016). Red lines have different meanings in different panels: (a) transition between small and large simulation box; (c) smallest possible line width ($W_{50}$ = 23.5\kms) implied by the intrinsic velocity dispersion of 10\kms; (f) halo mass resolution limit of the two simulation boxes; (g) chosen minimum integrated $S/N$ for blind \HI\ detections.}
\label{fig:wallaby-predictions}
\end{center}
\end{figure*}

The spatial distribution of galaxies for a narrow right ascension slice is presented in Fig.~\ref{fig:simsky}. The redshift limits of \hipass\ ($z \sim 0.042$) and \alfalfa\ ($z \sim 0.06$) are shown for comparison. \wal\ will be able to detect the \HI\ emission from galaxies out to a redshift of 0.26. Thanks to its larger redshift and sky coverage, \wal\ will be detecting significantly more galaxies than \hipass\ ($\sim$5\,500) and \alfalfa\ ($\sim$31\,500), thereby increasing the total number of galaxies detected in \HI\ emission by more than an order of magnitude.

Fig.~\ref{fig:mhi_distance} depicts some basic statistics. In the left-hand panel we present \HI\ mass versus distance of the galaxies expected to be detected by \wal. The simulation predicts that \wal\ will detect several thousand galaxies with \HI\ masses below 10$^{8}$\Msun. Very few galaxies with \HI\ masses $>10^{11}$\Msun\ will be detected by \wal. The right-hand panel of Fig.~\ref{fig:mhi_distance} shows that the survey will be biased towards \HI-rich galaxies, relative to stellar mass, as expected for any blind 21cm survey.

In Fig.~\ref{fig:wallaby-predictions} we present the predicted distributions of \wal\ detections for several important observational parameters: redshift $z$, \HI\ mass (\MHI), velocity width ($W_{50}$), stellar mass ($M_{\rm star}$), number of \wal\ beams over the galaxy \HI\ disc, halo mass ($M_{\rm vir}$), integrated signal-to-noise (S/N), galaxy \HI\ diameter, inclination and bulge-to-total stellar mass ratio ($B/T$). While covering the redshift range of $z < 0.26$, detections are expected to peak at $z \sim 0.03$, with a median redshift across the entire sample of $z \sim 0.045$ (see also Table~3). The different sensitivities of \wal, \hipass\ and \alfalfa\ naturally imply different redshift distributions. In particular, due to its large redshift coverage and moderate sensitivity, \wal\ will be sensitivity-limited up to the highest \HI\ masses. In turn, \hipass\ and \alfalfa\ were only sensitivity-limited for \HI\ masses of $< 10^{10}$\Msun\ (roughly $M_{\rm HI}^{\ast}$), but redshift-limited (i.e., volume-limited) for higher masses. This explains the deficit of \HI\ masses above this limit in \hipass\ and \alfalfa.

The predicted halo mass distribution (Fig.~\ref{fig:wallaby-predictions}f) has a complex shape due to the mixing of central and satellites galaxies, as well as due to the simulation resolution limits at low halo masses (red vertical lines). Interestingly, \wal\ will detect a large number ($\sim$350\,000) of galaxies in low-mass haloes ($<10^{11}$\Msun), where the halo mass function is poorly constrained. For instance, the halo masses inferred from the {\sc gama} group catalogue only extend down to masses of $\approx 5 \times 10^{12}$\Msun\ (e.g., Yang et al. 2007; Robotham et al. 2011; Eckert et al. 2017). Upon assuming that \HI\ line profiles and/or line widths provide sufficient information to estimate even low halo masses, \wal\ will thus be transformational in its ability to constrain the low-mass end of the halo mass function.

\subsection{Source finding \& parametrization} % Section 4.5

Our Source Finding Application (SoFiA; Serra et al. 2015b) combines a range of different filtering, source-finding and parametrization algorithms that will be fine-tuned to produce highly complete and reliable \wal\ source catalogues. Most importantly, SoFiA comes with a built-in reliability estimator based on comparing the number of negative and positive detections in different regions of parameter space (Serra et al. 2012b, 2015b). By setting a suitable reliability threshold, SoFiA can automatically remove unreliable detections such as noise peaks to produce a clean source catalog. SoFiA produces a range of useful outputs, including \HI\ moment maps, source masks, integrated spectra and position-velocity diagrams. Several 2D and 3D source-finding algorithms are discussed in the {\em PASA Special Issue} on ``Source finding and visualisation'' (Koribalski 2012a,b).

Furthermore, SoFiA also provides sophisticated algorithms for the accurate extraction of source parameters. This includes mask optimization algorithms that will automatically grow source masks to ensure that they cover the full spatial and spectral extent of the source. This will help to minimize biases in the calculation of the integrated flux density of a source that would naturally result from applying a source-finding threshold to noisy data. SoFiA can also fit the Busy Function (Westmeier et al. 2014) to the integrated spectrum of a source to extract accurate parameters. The Busy Function is capable of fitting a wide range of \HI\ spectral profiles commonly found in galaxies, including symmetric and asymmetric double-horn profiles as well as simple Gaussian line profiles. For a more detailed overview and testing of the performance of different source finding packages and algorithms, see Westmeier et al. (2012), Popping et al. (2012), Koribalski (2012a), and Westmeier et al. (2018).

\subsubsection{Extended galaxies} % Section 4.5.1
For all extended \wal\ galaxies we aim to measure basic orientation parameters such as diameters and position angles, while more detailed rotating disk models will be applied to well-resolved systems (i.e., $\gtrsim$5 beams along the major axis). We have developed both 2D (Oh et al. 2018) and 3D (Kamphuis et al. 2015) algorithms that automatically estimate the geometries and rotation curves of large \HI\ discs. These algorithms as well as other approaches in the literature (e.g., Di Teodoro \& Fraternali 2015; Sellwood \& Spekkens 2015; Bekiaris et al. 2016) and new algorithms (Bekiaris et al. 2020) will be tested with the aim to deliver the best-fitting kinematic models. For an outline of our science goals see Section~5.4.

\subsubsection{Postage stamps} % Section 4.5.2
\wal\ also intends to obtain high-resolution ($\sim$10 arcsec) \HI\ `postage stamps' of particularly interesting, predefined galaxies, allowing a more detailed analysis and comparison with data at other wavelengths. The number of galaxies for which such high-resolution \HI\ cubes can be processed, in addition to the basic \wal\ data products at $\sim$30~arcsec resolution, will depend on the available computing and data storage capacity. An example from the \wal\ pilot survey is shown in Fig.~\ref{fig:askap-ngc3137}, comparing the \HI\ morphology of the nearby spiral galaxy NGC~3137 at 30, 20 and 10 arcsec.

\subsubsection{HI stacking}
We will also obtain the average \HI\ emission of carefully selected galaxy samples (without individual detections) by co-adding (`stacking') \HI\ spectra at their known positions and redshifts (see, e.g., Delhaize et al. 2013; Brown et al. 2015; Rhee et al. 2013, 2016, 2018; Kleiner et al. 2017). The latter will be obtained from the \taipan\ southern sky galaxy survey (Da Cunha et al. 2017) and other targeted galaxy redshift surveys.

\subsection{Catalogues \& data products} % Section 4.6

Each \wal\ cube will be associated with one or more source catalogues, which provide extensive lists of source properties and reliability when estimated by SoFiA. For unresolved \HI\ sources, we will provide the properties obtained from the integrated \HI\ spectrum and the likely source identification in optical or infrared surveys. For extended \HI\ sources, further properties will be derived, such as their size, position angle, and rotation curve (see Section~4.5.1). We aim to provide separate catalogues for Galactic HVCs as well as detections of \HI\ absorption and hydrogen recombination lines. 

Each catalogued \wal\ source 
% (WALLABY JHHMMSS$\pm$DDMMSS) 
will be connected to a set of pre-defined data products, as listed below. Both the catalogues and the data products will be made publicly available after successful quality assessment by the \wal\ team, once the ASKAP observations and real-time data processing are completed. For each \wal\ detection we aim to provide: (1) a 3D data cube, (2) an integrated \HI\ spectrum, (3) an \HI\ column density map (moment-0), and (4) a radio continuum map. For extended \wal\ sources we aim to also provide: (5) a mean \HI\ velocity field (moment-1), (6) a mean \HI\ velocity dispersion map (moment-2), and (7) a major-axis position-velocity diagram. 
CASDA is designed to store all ASKAP data products and some of the calibrated visibilities. \wal\ team members as well as other users interact with the archive to search, display and download source properties, catalogues and data products. CASDA now also stores all \hipass\ cubes, allowing users to combine ASKAP and Parkes \HI\ data for enhanced sensitivity to extended low-surface brightness structures. 

\wal\ will provide resolved \HI\ image products for several thousand galaxies, similar or better in quality, sensitivity and resolution than the $\sim$500 nearby galaxies recently imaged with mature radio interferometers (e.g., Walter et al. 2008; Ott et al. 2012; Koribalski et al. 2018); for an overview see also Wang et al. (2016) and references therein. 

\subsubsection{Optical/Infrared source identification}

Due to the relatively high angular resolution of \wal\ ($\sim$30~arcsec), optical and/or infrared source identifications will be easy for the majority of catalogued \HI\ detections. Primarily we will use large optical sky surveys such as the Digitised Sky Survey (DSS), the SkyMapper Southern Survey (Wolf et al. 2018), the Panoramic Survey Telescope and Rapid Response System (Pan-Starrs), and SDSS, as well as infrared sky surveys such as the Widefield Survey Infrared Explorer (WISE). Deep optical images are also available for some areas, e.g., covered by DeCam / MegaCam and the DESI Legacy Imaging Surveys (Dey et al. 2019). 

The SkyMapper and LSST surveys cover the whole southern hemisphere in multiple bands, while Pan-Starrs covers $\delta > -30\degr$ and SDSS most of the extragalactic northern sky. Substantial data from LSST is still a couple of years away, while the SDSS, Pan-Starrs\,1 and SkyMapper are already accessible online. The SkyMapper Southern Survey is currently ongoing; its newest data release is DR2 (Onken et al. 2019), which covers $\sim$90\% of the southern hemisphere in $i$ and $z$ bands, and $\sim$40\% of the hemisphere in all six survey bands\footnote{SkyMapper homepage: \url{http://skymapper.anu.edu.au}}. The point-source sensitivity reaches $g \approx 21$~mag and example images are shown in Section~6. Future releases from SkyMapper will improve sky coverage, depth and noise properties as well as provide images of co-added sky tiles. 

The \HI\ position uncertainty for single sources is typically the angular resolution divided by the signal-to-noise. Galaxy identifications will be more difficult in the Zone of Avoidance and for nearby galaxies due to their large size and potential confusion with HVCs and \HI\ clouds.

% \subsubsection{Searching for HI absorption}

% Using the known positions of bright 21-cm radio continuum sources in the \wal\ sky area, the {\sc flash} team will use a separate data pipeline to search the spectral line data cubes for signatures of intervening and associated \HI\ absorption, as described by Allison et al. (2015) and Reeves et al. (2015, 2016).

\subsubsection{OH megamasers}

\wal\ will also allow us to detect OH megamasers in the redshift range $z \approx\ 0.13 - 0.3$. Such masers are known to reside in ultra-luminous infrared galaxies and are associated with merger-induced starbursts (Staveley-Smith et al. 1992; Darling \& Giovanelli 2002; McKean \& Roy 2009). We estimate that at least one hundred OH megamasers will be detected in the 1665/7 MHz doublet with \wal. In addition to detecting OH emission, we can also search for OH lines in absorption against very bright background radio continuum sources. Given the high accuracy to which we can determine the positions of \wal\ spectral line detections, we will be able to get optical/infrared identifications for the vast majority of megamaser candidates. Even more accurate positions will be available for any associated radio continuum sources. The latter is helpful in distinguishing OH from \HI\ lines, especially if no optical redshifts are available for the stellar host galaxy. The shape and flux of the integrated spectrum also provides information about the orientation, size and mass of the associated galaxy (Briggs 1998).

\subsection{3D data visualisation}

Visualisation of \wal\ data products is a critical step in the process of data validation and knowledge discovery. For \wal\ we employ a number of validation steps, including rms noise statistics, flux comparison with existing single-dish and interferometric data, before the data products are publicly released in CASDA. The final quality checks are done by \wal\ experts who will inspect the data products.

Dykes  et al. (2018) demonstrate a practical client-server approach, whereby the remote service performs a distributed rendering task,  and the frames are streamed and displayed in a web browser. The specific case demonstrated in the paper was using the {\sc splotch} software for remote rendering of particle simulations, but this approach can be extended to other visualisation and analysis processes.

The large size of the \wal\ data cubes poses challenges for existing visualisation tools (see, e.g., Koribalski 2012a,b; Punzo et al. 2015), as few were designed for data volumes that exceed desktop memory limits. Advanced techniques, such as scalable algorithms regularly employed in the computer graphics industry, are required to handle and render data sets larger than the available system memory (e.g., Beyer et al. 2015; Wald et al. 2017). \wal\ presents a valuable case study for visualisation research (Fluke et al. 2010), having already motivated new approaches and open source software aimed at moving interactive discovery processes away from standard desktop environments.

Hassan et al. (2013) demonstrated a distributed computing solution harnessing graphics processing units (GPUs) as the computational engine.  Achieving interactive volume rendering (7--10 frames/second) for a 540~GB test data cube, the resulting {\tt GraphTIVA} framework is scalable to a full resolution \wal\ data cube. Recent work has moved {\tt GraphTIVA} to a fully remote-service mode, using the Strudel application\footnote{\url{https://www.massive.org.au/userguide/cluster-instructions/strudel}} to allow the graphical user interface to co-exist on the high performance computing infrastructure. 

Fast, efficient, and highly-optimised codes are an essential step towards tackling the 3D visualisation challenges of \wal\ together with tools and technologies that facilitate data exploration and scientific discovery. This includes novel qualitative and quantitative visualisation software designed for standard desktop environments (Punzo et al. 2017; Vohl et al. 2017), or in the form of more immersive hardware such as tiled display walls, CAVEs, and virtual reality equipment (Vohl et al. 2016, Fluke \& Barnes 2018). \wal\ could test the effectiveness of visualisation systems to aid scientific discovery.

\section{WALLABY Key Science} % Section 5

Using \wal\ we will produce the largest sample of galaxies that is possible to detect in \HI\ emission in a given observing time with ASKAP, allowing us to address fundamental science questions about galaxy formation, galaxy evolution and cosmology. We will examine the \HI\ properties and large-scale distribution of galaxies in the Local Universe in order to study: (1) the composition of the Local Volume, (2) star formation in galaxies, (3) environment-driven galaxy evolution, (4) the structure of \HI\ discs, (5) \HI\ scaling relations, (6) the \HI\ mass function and its variation with galaxy density and redshift, (7) intergalactic gas and the cosmic web, and (8) cosmological parameters.  \\

\wal\ data will be suitable for a large range of Galactic and extragalactic studies. While the extragalactic studies form the main \wal\ science, our proposed velocity coverage includes \HI\ mapping of the Milky Way and Galactic high velocity clouds (e.g., Westmeier 2018). Beyond our Galaxy, we expect to discover a large number of new Local Volume galaxies, in particular dwarf irregular and transitional galaxies, ultra-diffuse and low surface brightness galaxies (e.g., Irwin et al. 2007; Ryan-Weber et al. 2008; Westmeier et al. 2015; Adams \& Oosterloo 2018; For et al. 2019), as well as the occasional \HI\ cloud complex (e.g., Kilborn et al. 2000; Ryder et al. 2001; Koribalski et al. 2004). Examples of such new galaxy discoveries in the vicinity of known spirals are presented in the `Local Volume \HI\ Survey' (\lvhis) galaxy atlas (Koribalski et al. 2018). \wal\ will measure the \HI\ content of galaxy groups, quantify the amount of intra-group gas, highlight any on-going tidal interactions, and characterize their evolutionary state (e.g., Serra et al. 2015a; Lee-Waddell et al. 2019). Similarly, mapping \HI\ in galaxy clusters will reveal filaments and streams resulting from violent galaxy interactions and stripping processes (e.g., Oosterloo \& van Gorkom 2005; Chung et al. 2009; Wolfinger et al. 2013; Scott et al. 2018).   \\

\wal\ will also measure the velocity structure of galaxies and enable inference of their dark matter distribution. The widefield coverage of \wal\ will allow for the discovery of \HI\ in many previously uncatalogued galaxies, as well as provide redshifts for a large number of catalogued galaxies (see, e.g., Kleiner et al. 2019; For et al. 2019). It will trace the \HI\ structure and kinematics of interacting galaxies (Hibbard et al. 2001) and pinpoint candidate tidal dwarf galaxies (e.g., Lee-Waddell et al. 2018, 2019). We note that all \wal\ galaxies will be spectrally resolved. \\

The volume covered by \wal\ ($\sim$0.4~Gpc$^3$ for $M_{\rm HI}^*$ galaxies) is large enough for the measurement of cosmological parameters without substantial cosmic variance. We expect to accurately measure the density-dependence of the \HI\ mass function, the clustering and bias parameter for gas-rich galaxies and the local flow field (e.g., Beutler et al. 2011, Papastergis et al. 2013). Further, we expect a possible detection of Baryonic Acoustic Oscillations (BAOs) at the lowest redshift. Independently of forthcoming optical surveys, \wal\ could reduce the errors on the dark energy parameter $w$ and the Hubble constant \Ho\ by up to a factor of two when compared to {\sc Planck}-only data. 

\wal\ will serve as an accurate low-redshift anchor for later SKA \HI\ surveys of the distant Universe and allow optimization of SKA \HI\ survey design. It will provide an important pathfinder for key SKA \HI\ science projects. Combined with existing and new surveys from optical, infrared, X-ray and millimeter/sub-millimeter facilities, \wal\ will be part of an exciting new generation of panchromatic studies of the Local Universe. In the following we provide more details on the broad \wal\ science goals outlined above.

\begin{table} % Table 2
\caption{\wal\ \MHI\ detection limits for four different galaxy velocity widths and two source sizes, where $D$ is the galaxy distance in Mpc.}
\begin{tabular}{ccc}
\hline
line width  & \multicolumn{2}{c}{5$\sigma$ \MHI\ limit [\MMsun]} \\
            & single beam & over 10 beams \\
\hline
25   & $1.9 \times 10^4 D^2$ 
     & $6.0 \times 10^4 D^2$ \\
50   & $2.7 \times 10^4 D^2$ 
     & $8.4 \times 10^4 D^2$ \\
     & \\
100  & $3.8 \times 10^6 (D/10)^2$ 
     & $1.2 \times 10^7 (D/10)^2$ \\
200  & $5.3 \times 10^6 (D/10)^2$ 
     & $1.7 \times 10^7 (D/10)^2$ \\
\hline
\end{tabular}
\end{table}
% Notes - for single beam:
% M(HI) = 2.356e+5 * 5sigma/sqrt(w/dv) * w * D^2
% Notes - for 10 beams, multiply by sqrt(10)

\begin{table*} % Table 3 
\begin{center}
\caption{Expected galaxy detections from the \wal\ reference simulation down to \MHI\ = 10$^6$\Msun\ in four bins of comoving distance and for four different cuts in signal-to-noise (with Poisson uncertainties); see Section~4.4 and Fig.~\ref{fig:wallaby-predictions}.}
\begin{tabular}{ccccc}
\hline
 distance range &  S/N $>$ 5  & S/N $>$ 10 & S/N $>$ 15 & S/N $>$ 20 \\
\hline
   0 --  10 Mpc &      $300 \pm    200$ &      $300 \pm    200$ 
                &      $200 \pm    100$ &      $200 \pm    100$ \\
  10 -- 100 Mpc &  $95\,000 \pm 3\,000$ &  $60\,000 \pm 2\,000$ 
                &  $43\,800 \pm 1\,000$ &  $30\,700 \pm    600$ \\ 
 100 -- 400 Mpc & $330\,000 \pm    400$ & $117\,000 \pm 2\,000$
                &  $45\,100 \pm    900$ &  $16\,400 \pm    400$ \\
  $>$400 Mpc    &  $61\,000 \pm    300$ &   $6\,870 \pm     80$
                &      $920 \pm     30$ &      $180 \pm     10$ \\
\hline
  Total         & $490\,000 \pm 6\,000$ & $184\,000 \pm 3\,000$
                &  $90\,000 \pm 2\,000$ &  $48\,000 \pm 1\,000$ \\
\hline
\end{tabular}
\end{center}
\end{table*}

\subsection{The composition of the Local Group} % Section 5.1 

\begin{figure}[tb] % Figure 9 - created by Tobias
\begin{center}
  \includegraphics[width=8cm]{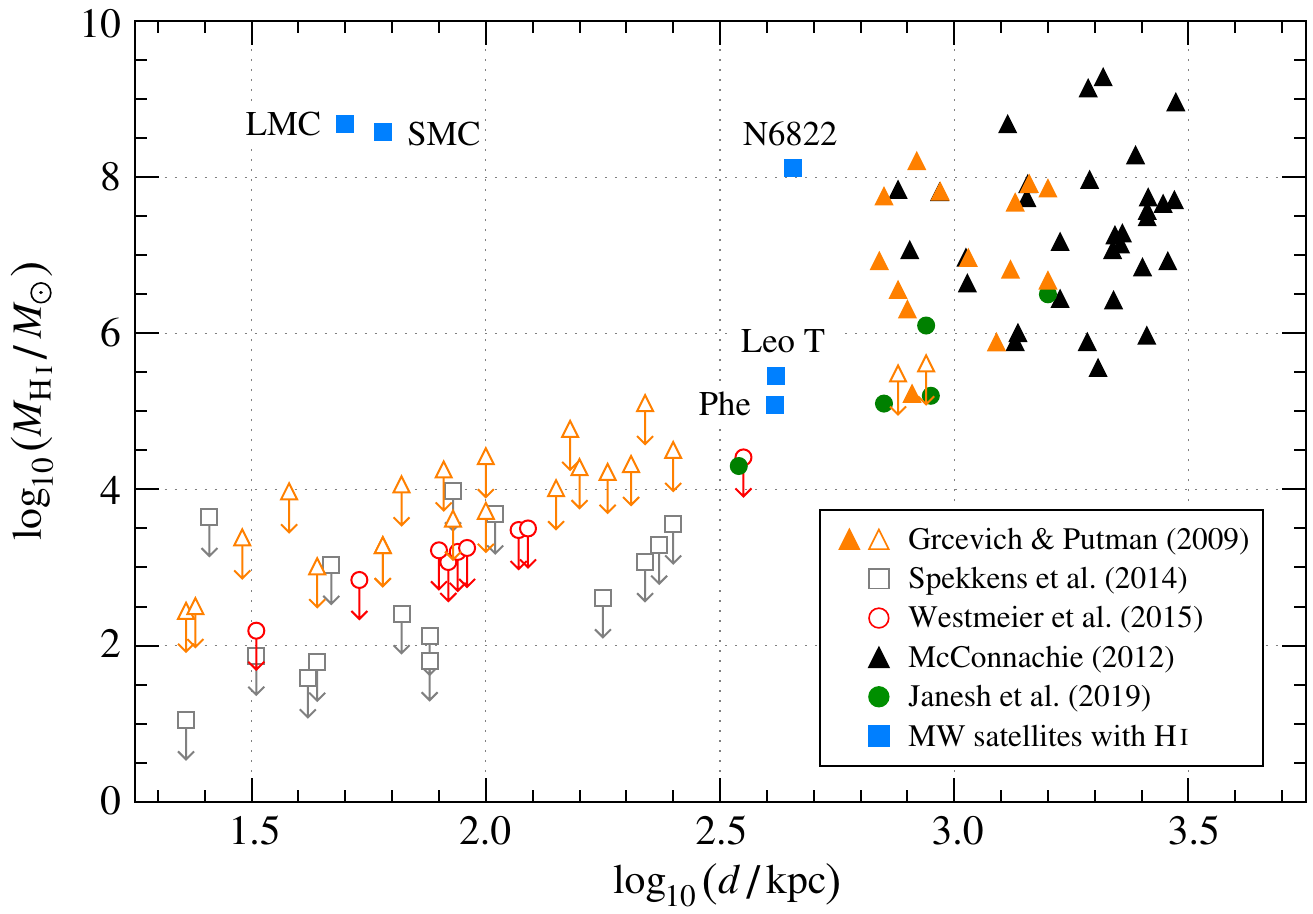}
\caption{\HI\ mass of Local Group dwarf galaxies as a function of distance. Detections are shown as filled symbols, while open symbols with arrows mark upper limits from non-detections. The values are taken from Grcevich \& Putman (2009), McConnachie (2012), Spekkens et al. (2014), Westmeier et al. (2015) and Janesh et al. (2019). \wal\ will detect all Local Group galaxies with \HI, including transition-type galaxies like Leo\,T. Discovering more dwarf galaxies in and around the Local Group will help us understand the mechanisms for gas retention in the smallest known galaxies.}
\label{fig:wallaby-LG}
\end{center}
\end{figure}

\begin{figure*}[tb] % Figure 10 - created by Karen
\begin{center}
 \includegraphics[width=16cm]{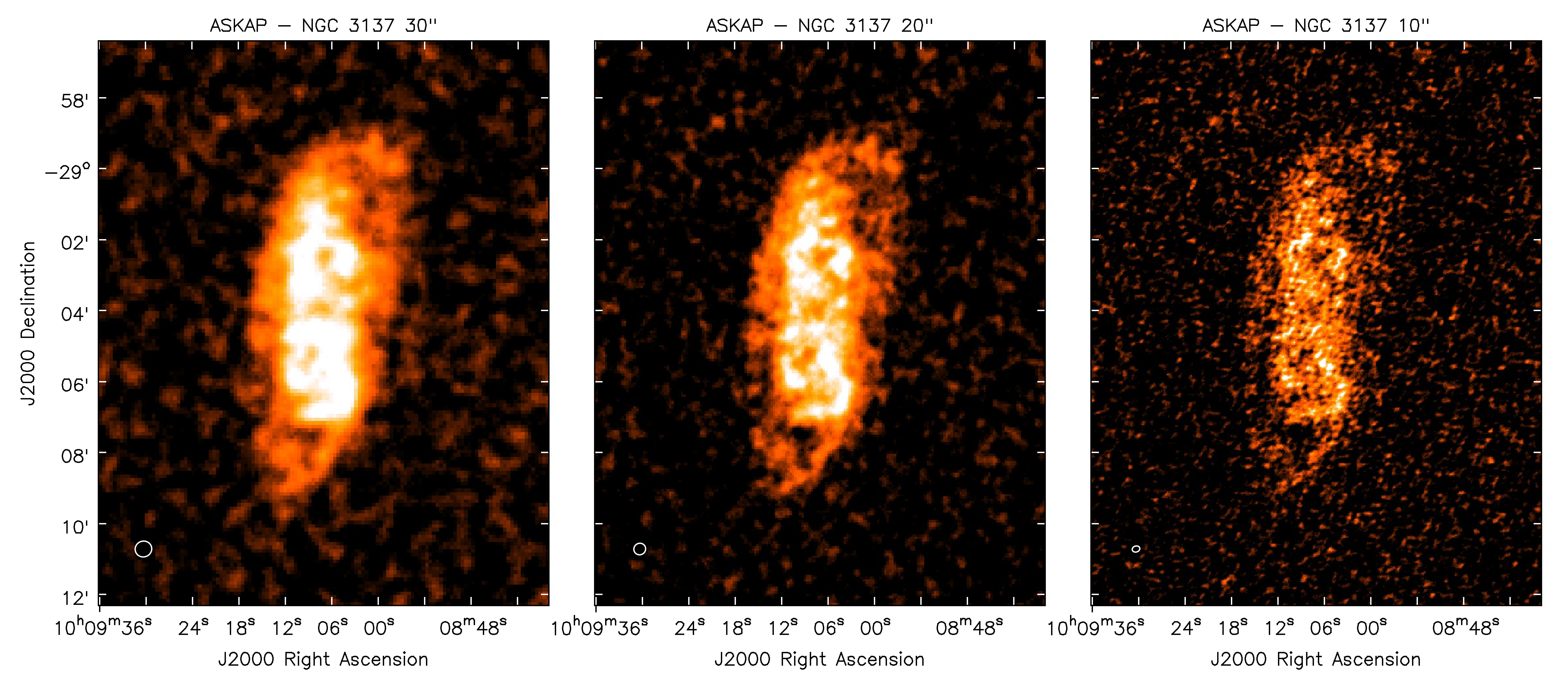}
\caption{ASKAP-36 integrated \HI\ column density maps of the nearby spiral galaxy NGC~3137 from \wal\ pilot observations of the Hydra cluster. The data was processed using ASKAPsoft, achieving angular resolutions of 30~arcsec (left), 20~arcsec (middle) and 10~arcsec (right). This illustrates the desirability of high-resolution \wal\ `postage stamps' for the detailed study of galaxy morphology.}
\label{fig:askap-ngc3137}
\end{center}
\end{figure*}

The Local Group provides a unique test bed for galaxy formation and subsequent evolution. No other family of galaxies can be observed in such detail and fine mass resolution (see Fig.~\ref{fig:wallaby-LG} and references therein). Since 2005, more than 40 new dwarf galaxies have been uncovered in orbit around the Milky Way and M\,31. New widefield optical imaging surveys, such as the Dark Energy Survey (DES), continue to discover faint dwarf galaxy candidates (e.g., Bechtol et al. 2015). While the new discoveries have shrunk the gap between the number of observed galaxies and the hundreds of satellites predicted by cosmological simulations (Klypin et al. 1999), a missing dwarf galaxies problem has emerged (Klypin et al. 2015). Some authors argue that luminosity bias alleviates the problem (Tollerud et al. 2008), while others require significant suppression of star formation in low-mass galaxies (Koposov et al. 2009), perhaps due to cosmic reionization changing the Jeans mass or tidal stripping of gas due to the proximity of a much larger parent (e.g., Ricotti \& Gnedin 2005; Mayer et al. 2001; Bovill \& Ricotti 2009). These comparisons extrapolate the number of known dwarfs by at least a factor of 10, thus the total expected number of galaxies is still highly uncertain. \\

One of the most important recent discoveries in the Local Group was the galaxy Leo\,T (Irwin et al. 2007; Ryan-Weber et al. 2008). Based on new WSRT data, Adams \& Oosterloo (2018) give an \HI\ mass of $4.1 \times 10^5$\Msun\ (for $D$ = 420~kpc), an \HI\ radius of $\sim$3.3~arcmin (400~pc), and a dynamical mass of at least $1.4 \times 10^7$\Msun. Leo\,T is a gas-rich `transitional' dwarf galaxy (see Fig.~\ref{fig:wallaby-LG}) with both an old stellar population as well as new stars. It does not fit the idea that reionization suppressed star formation in low-mass dwarfs, although simulations suggest that some gas may remain (Revaz et al. 2009) or is subsequently gained through late accretion (Ricotti 2009). Evidence suggests that this gas can survive galactic winds and supernovae explosions as highlighted in the case of the Phoenix dwarf, where the gas is adjacent to the stars but regularly falls back to trigger more star formation (Young et al. 2007). Most baryons in Leo\,T are in \HI, highlighting the potential to discover many more gas-rich dwarf galaxies in a widefield and high-resolution \HI\ survey such as \wal\ (DeFelippis, Putman \& Tollerud 2019). We expect most new \HI\ detections at distances between 300 and 1500 kpc from the Milky Way; our \HI\ mass detection limit for dwarf galaxies ($\sim$25\kms\ velocity width) is \MHI\ $\gtrsim 1.9 \times 10^4 D^2$\Msun\ (see Table~2). 

Fig.~\ref{fig:wallaby-LG} summarises our current knowledge of dwarf galaxy \HI\ masses and mass limits as a function of distance from the Milky Way. It highlights the lack of \HI-detected dwarf galaxies closer than 300 kpc, where gas may have been removed by tidal and/or ram pressure forces, and the increasing number of \HI\ detections further out (see also Tolstoy, Hill \& Tosi 2009). \\

Together with parallel data from optical/IR surveys, we will address the following questions:

\begin{itemize}
\item How do galaxies retain/gain gas and form stars\,? We will be able to compile physical evidence of gas stripping and galaxy `harassment' as a function of environment. The discovery of more transitional (dIrr/dSph) galaxies, typically with small amounts of \HI\ gas offset from the stellar body, can lead to a better understanding of the underlying physical processes.

\item By moving further from the `zone of destruction', i.e., beyond 250--350 kpc from the Milky Way / M\,31 (see Fig.~\ref{fig:wallaby-LG}), will we find more gas-rich dwarfs with $M_{\rm vir} \sim 10^7$\Msun\,?

\item Is there a common minimum halo mass for star formation (see, e.g., Warren et al. 2007)\,?  

\item Is a galaxy's star formation history and \HI\ mass-to-light ratio dominated by initial conditions (total mass, baryon fraction), or distance to the nearest spiral\,?

\item Are there any gas-rich galaxies without stars\,? The vicinity of the Local Group would be the best place to find them since low-mass galaxies have the lowest detected star formation efficiencies.

\item Is there a minimum mass for the smallest galaxies\,? Such a limit can give information on whether dark matter is cold or warm (e.g., Macci\`o et al. 2019).
\end{itemize}

\subsubsection{The Milky Way and HVCs} % Section 5.1.1

The Milky Way itself provides an ideal local laboratory to study some of the above questions in great detail, in particular with respect to gas inflow and outflow (e.g., Kalberla \& Dedes 2008; Putman, Peek \& Joung 2012; Lockman \& McClure-Griffiths 2016). Such processes are crucially important in linking gas physics to the physics of star formation and galaxy evolution (e.g., Hopkins et al. 2008). The lowest redshift \wal\ data will therefore be used for an all-sky study of high-velocity clouds (see, e.g., Westmeier 2018), the disc-halo interface (e.g., For et al. 2014, 2016), and the Galactic Plane, by supplementing data collected by {\sc gaskap}. 

Recently, the Gaussian decomposition of HI4PI survey data (Kalberla \& Haud 2018) underlined that we have to distinguish between three \HI\ phases. For the very first time, interferometric absorption line studies (Murray et al. 2018) were successful at evaluating the \HI\ spin-temperature, \Tspin, of all three phases: the cold neutral medium (CNM; \Tspin\ $< 250$~K), the thermally unstable neutral medium (UNM; 250 $<$ \Tspin\ $<$ 1000~K), and the warm neutral medium (WNM; \Tspin\ $> 1000$~K), leading to estimates of the corresponding filling factors. However, significant discrepancies between single-dish and interferometric studies remain. 

By combining \wal\ with single dish \HI\ data to form high resolution maps of all \HI\ phases, new light will be shed on these potential discrepancies. The phase transitions of gas are inherently related to local density, dust content and magnetic field strength (Krumholz 2013). In fact, recent work has shown that density variations in the ISM are accompanied by variations in magnetic field strength (Kalberla et al. 2017; Kalberla \& Kerp 2016; Zaroubi et al. 2015). Only a combined analysis of the depolarized radio continuum emission with \HI\ observations will allow us to identify the origin of highly polarized, but radio dark filaments.

\subsection{Star formation in galaxies} % Section 5.2

Star formation is a key process in galaxy evolution that is still not fully understood. Great strides have been made with targeted, high-resolution \HI\ surveys such as {\sc things} (Walter et al. 2008), {\sc vla-angst} (Ott et al. 2012), {\sc little things} (Hunter et al. 2012) and \lvhis\ (Koribalski et al. 2018), in combination with multi-wavelength surveys (e.g., Kennicutt et al. 2003; Meurer et al. 2006; Dalcanton et al. 2009; Leroy et al. 2009; Saintonge et al. 2011; Wang et al. 2017), towards linking the distribution of cold gas in the inner discs of nearby galaxies with regions of active star formation (see Fig.~11). \wal\ will  increase the sample of well-resolved galaxy \HI\ discs $\sim$10-fold, much enhancing our understanding of the formation and evolution of nearby galaxies.

The star formation law or the Kennicutt-Schmidt relationship between gas surface densities and star formation rate (SFR) densities is largely understood in M$^*$-type galaxies (e.g., Bigiel et al. 2008; Leroy et al. 2008). The roles of angular momentum and disk stability in the regulation of \HI\ content and star formation have also been an active area of recent research (e.g., Obreschkow et al. 2016; Wong et al. 2016; Stevens et al. 2018). On the other hand, there are still large gaps in our understanding of star formation at low surface brightness (LSB) and in low-mass dwarf galaxies (Lee et al. 2009; Kova\v{c} et al. 2009; Bigiel et al. 2010; Roychowdhury et al. 2017; L\'opez-S\'anchez et al. 2018). It is still unclear whether the initial mass function (IMF) for star formation varies globally between galaxies (e.g., Hoversten \& Glazebrook 2008; Meurer et al. 2009).  

The star formation theory from Krumholz (2013) suggests that stars can form directly from low temperature gas whether or not the gas is in its molecular state (i.e., when the free-fall timescale is less than the chemical timescales, it is not necessary for molecular gas to form first). Can such direct formation of stars occur in the outer regions of disk galaxies (and LSB dwarf galaxies) where \HI\ appears to be the dominant phase of cold gas\,? In addition, there exist extreme conditions where the ISM appears to be \HI-rich but CO-deficient (e.g., Bicalho et al. 2019). For example, the ISM in collisional ring galaxies, which experience compression-triggered star formation, is one such star-forming ISM that is \HI-rich but CO-poor (e.g., Wong et al. 2017; Higdon et al. 2011). \wal\ will much increase the \HI\ samples of low-mass and LSB galaxies as well as ring galaxies (see Bait et al. 2020).

Furthermore, we note that the majority of \wal\ catalogued spiral galaxies will have 20-cm radio continuum emission detectable by the ASKAP radio continuum survey, {\sc emu} (Norris et al. 2011). This also means that \wal\ may contribute close to half a million redshifts and optical/infrared identifications to {\sc emu}. The \wal\ team aims to measure local and global star formation (SF) rates for all gas-rich spirals and compare their SF and \HI\ distributions.

\subsection{Environment-driven galaxy evolution}

\wal\ will enable detailed studies of the \HI\ content of galaxies through all environments, from the densest clusters, to compact and loose groups, to voids. We will combine the \wal\ data with widefield optical and infrared surveys such as SkyMapper and WISE (see Section~4.6.1), to investigate the relation between \HI\ mass, environment, SFR, and morphological type, and how these dependencies change with redshift. The resolution of \wal\ will enable investigation of the shapes and frequency of asymmetries in the \HI\ distribution, e.g., lopsidedness and warping, to determine the impact of gravitational interactions (e.g., mergers, harassment) and hydrodynamical interactions (e.g., ram pressure stripping, evaporation). We will compare our observations with theoretical predictions of galaxy formation, including semi-analytic models to describe the large-scale \HI\ distribution and changes in \HI\ mass function (e.g., Power et al. 2010), and hydrodynamic simulations (e.g., Marasco et al. 2016; Crain et al. 2017; Stevens et al. 2019) to investigate the physical processes occurring in the galaxies, both internal (e.g., feedback, outflows) and external (e.g., ram pressure stripping, gravitational interactions). By characterizing the global \HI\ spectra of around half a million galaxies, we will obtain a wealth of properties, including their systemic and rotational velocities, \HI\ and total dynamical mass estimates, and their profile shape. The latter can be an excellent indicator of environmental effects, with highly symmetric \HI\ profiles typically found in isolated spiral galaxies (Espada et al. 2011) and asymmetric \HI\ profiles in galaxies affected by tidal interactions and/or ram pressure stripping (Giese et al. 2016, Scott et al. 2018).

The Local Universe can be used as a test bed for galaxy evolution theory. The standard model for structure formation and evolution states that galaxies arise from early perturbations in a $\Lambda$CDM Universe that grow and merge to become the galaxies we observe today. This model is able to reproduce the large-scale features observed in the Universe such as the galaxy two-point correlation function, the galaxy clustering amplitude and baryon acoustic oscillations (Springel et al. 2005). However, the complex physics of the ISM and its interaction with stellar populations largely remains beyond the ability of these simulations to model in a self-consistent manner. \HI\ is a vital component of the ISM, and \wal\ observations provide stringent tests for any models of galaxy evolution.

\begin{figure*}[tb] % Figure 11 - created by BSK
\begin{center}
 \includegraphics[width=16cm]{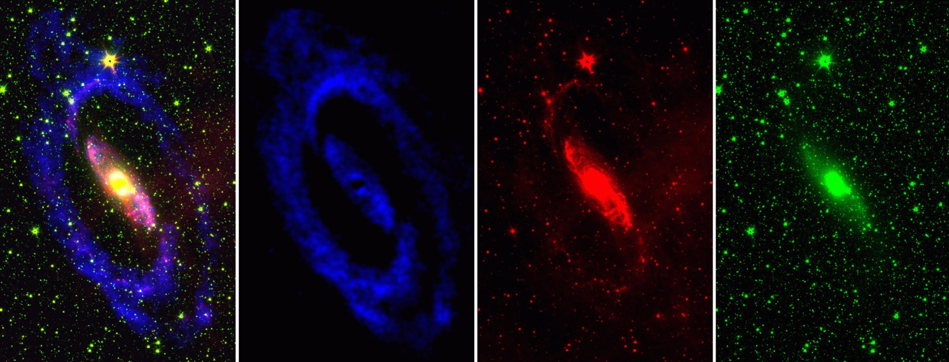}
\caption{The inner disk of the nearby Circinus Galaxy as observed by the ATCA in \HI\ (blue) and the Spitzer Space Telescope at 8$\mu$m (red; warm dust) and 3.4$\mu$m (green; stars). The three-colour composite image is shown on the left, emphasizing the need for multi-wavelength observations in studying galaxies. For details see For et al. (2012) and Koribalski et al. (2018).}
\label{fig:circinus-rgb}
\end{center}
\end{figure*}

\subsubsection{Gas accretion} % Section 5.3.1

Gas accretion from the IGM or through minor and major mergers plays a fundamental role in galaxy evolution and depends strongly on the environment a galaxy lives in. Accretion influences the gas content, SFR, stellar populations, and the galaxy morphology over cosmic time (e.g., Bournaud et al. 2007; Keres et al. 2005, 2009). There is some observational evidence for accretion of neutral gas onto spiral galaxies, including from gas-rich dwarf galaxies, extra-planar gas, and warped discs (e.g., Sancisi et al. 2008), although Di Teodoro \& Fraternali (2014) find minor mergers to play a minor role. For example, gas accretion by the nearby dwarf galaxy NGC~5253, which is located in the M\,83 galaxy group (Koribalski et al. 2018), may have triggered a powerful starburst (L\'opez-S\'anchez et al. 2012). Similarly, the nearby spiral galaxy NGC~1512, which is interacting with the blue compact dwarf galaxy NGC~1510 (Koribalski \& L\'opez-S\'anchez 2009), is accreting intergalactic metal-poor gas (L\'opez-S\'anchez et al. 2015). Analysis of the HVC population surrounding the Milky Way has provided particular insights and limits into potential accretion (Wakker et al. 1999; Putman 2006). 

The {\sc halogas} survey (Heald et al. 2011), a systematic deep and resolved \HI\ survey of 24 nearby galaxies, also failed to detect large numbers of infalling clouds or filaments down to a column density of $\sim$10$^{19}$ cm$^{-2}$. The {\sc halogas} upper limits imply an \HI\ accretion rate that is an order of magnitude less than what is required to sustain star formation. The direct detection of extragalactic cold gas accretion in \HI\ emission is therefore difficult and requires sensitivities to column densities of 10$^{18}$ cm$^{-2}$ or lower (Pingel et al. 2018). Pisano et al. (2004) did not detect large reservoirs of intra-group \HI\ gas in nearby groups, concluding it may be tied up in warmer, ionised gas reserves, confirming hydrodynamic simulations (e.g., Joung et al. 2012; Crain et al. 2017). 

The goal of ongoing and future surveys such as {\sc mhongoose} on MeerKAT (de Blok et al. 2016) and {\sc imagine} (led by A. Popping) on the ATCA is to push the column density sensitivity another order or magnitude deeper to constrain the accretion rates at the 10$^{17}$ cm$^{-2}$ column density level. 

Surveys such as \lvhis\ have used their large sample sizes to identify a handful of candidates that could be accreting (Koribalski et al. 2018). \wal's sample sizes will be many times larger, and we will study the \HI\ morphology of galaxies as a function of environment, e.g., identify asymmetries, close companions and tidal features. By analysing the galaxy outskirts we can estimate gas accretion rates across a range of environments and galaxy densities (see, e.g., Serra et al. 2015a; Lee-Waddell et al. 2019). In conjunction with simulations, the dependence of galaxy asymmetries on different environments can be used to assess the relative importance of ram pressure, tidal interactions and `harassment' in driving galaxy evolution. In the most isolated galaxies, where gas captured from companions and galactic fountains (due to SF and AGN) is minimized, as shown in the {\sc amiga} project (Verdes-Montenegro et al. 2005), cold gas accretion from the cosmic web should stand out prominently. 

\subsubsection{Groups and Clusters} % Section 5.3.2

Comparing the properties of relatively isolated galaxies with those in denser surroundings, such as groups and clusters, is a primary \wal\ goal. We will explore typical signatures and impact of internal versus external processes in galaxy groups (Verdes-Montenegro et al. 2005; D\'enes et al. 2014; Jones et al. 2018a), galaxy clusters (Verheijen 2001; Chung et al. 2009) and voids (Kreckel et al. 2012). We will use the \wal\ galaxy catalogue, supplemented by multi-wavelength data (see Fig.~11), to extract samples of (a) {\sc amiga}-like isolated galaxies and (b) Hickson-like compact groups. The properties of the isolated galaxies can then be used as a reference for studies of environmental impact, scaling relations and \HI\ mass functions. The \wal\ group catalogue will enhance our understanding of gas removal processes, including the extreme \HI\ deficiencies observed in Hickson Compact Groups (HCGs) in their later stages of evolution (Verdes-Montenegro et al. 2001). 

Gas-rich galaxies residing in clusters are affected by their massive potential, hot intra-cluster medium, and galaxy-galaxy interactions. Late-type galaxies in clusters typically are \HI\ deficient compared to isolated galaxies of similar optical morphology (e.g., Solanes et al. 2001; Chung et al. 2009; Brown et al. 2017). Studies of the Virgo cluster revealed \HI\ filaments stripped from galaxies in this extreme environment (e.g., Koopmann et al. 2008; Oosterloo \& van Gorkom 2005; Boselli et al. 2018a,b). The conditions for stripping, in particular ram pressure stripping, are becoming better understood through the increasing efforts in observations and modelling. Galaxy clusters also influence the morphologies of galaxies at large distances from their cores: observations in the outskirts of the Virgo cluster found galaxies affected by the cluster potential before they reach the central regions (Chung et al. 2007; Yoon et al. 2017), so-called pre-processing. Furthermore, the SFRs of cluster members are affected well outside the virial radius (Lewis et al. 2002), indicating an early effect on the hydrogen component, which is the fuel for star formation.

With the majority of galaxies in the Local Universe residing in groups (e.g., Eke et al. 2004; Yang et al. 2007), this medium-dense environment also plays an important role in driving galaxy evolution. Galaxy-galaxy interactions are thought to be more effective at stripping gas from galaxies in groups than their counterparts in clusters, owing to the smaller velocity dispersions in the former. However, the effectiveness of other mechanisms such as ram pressure stripping and strangulation in the group environment is still under investigation (e.g., Rasmussen et al. 2008; Westmeier et al. 2011; Stevens \& Brown 2017). Systematic investigations of both compact (e.g., Verdes-Montenegro et al. 2001; Borthakur et al. 2010) and loose (e.g., Kilborn et al. 2009; Freeland et al. 2009; Hess \& Wilcots 2013; Westmeier et al. 2017; Hess et al. 2019) groups can act as test cases for ram pressure stripping and galaxy merger models. \HI\ studies of galaxies in underdense regions, like the carefully selected sample of void galaxies by Kreckel et al. (2012), are also important. An \alfalfa\ study of the Leo Region revealed a significant population of low-mass, optically faint, gas-rich dwarf galaxies that have been missed in previous surveys (Stierwalt et al. 2009), placing tight new constraints on the low-mass end of the galaxy mass function in the group environment.

\subsubsection{AGN feedback and Radio Galaxies}

% removed on request by the FLASH PIs  
% The spectrum of every radio-bright AGN in the survey volume will be searched for associated \HI\ absorption, i.e. absorption due to the ISM in the AGN host galaxy, by the {\sc flash} team. About 30\% of radio-bright AGN have such associated \HI\ absorption (Ger\'eb et al. 2015). Studying the statistical properties of such \HI\ absorption gives valuable information on the role of gas in AGN activity and, in particular, on how AGN affect the ISM and star formation activity in galaxies through feedback effects. In $\sim$5\% of radio AGN, the \HI\ absorption profiles show direct evidence that the energy released by the AGN is strongly affecting the ISM, for example in the form of fast ($>$500\kms) outflows of cold gas. Many of these fast \HI\ outflows are also detected as outflows of molecular gas. So far, searches for associated \HI\ absorption have been done through targeted surveys. The ASKAP \HI\ surveys allow blind searches for associated \HI\ absorption and will increase the number of known absorption systems by several orders of magnitude. This much larger sample will make it possible to obtain a much better understanding of the systematics of the role of AGN feedback in galaxy evolution. It will also provide many excellent targets for follow up observations of the molecular outflows with ALMA. Morganti \& Oosterloo (2018) estimate that \wal\ will make $\sim$2300 detections of associated \HI\ absorption. \\

Radio galaxies are special cases (e.g., Emonts et al. 2007) where it is possible to study one of the most active periods in the evolution of galaxies when gas accretes onto the nuclear region, giving rise to important heating mechanisms which may lead, in different circumstances, to both suppression and triggering of star formation. \wal\ will provide, for the first time, \HI\ information on a large statistical sample ($\sim$150 detections with \MHI\ $>$ few times 10$^8$\Msun) of well-defined types of nearby radio galaxies as well as a very large radio-quiet comparison sample.

\subsection{The structure of \HI\ discs} % Section 5.4

\wal\ will provide detailed, high-sensitivity, high-fidelity \HI\ images for $\sim$5\,000 galaxies with \HI\ diameters in excess of 150~arcsec (five beams). The majority of these are known and correspond to the catalogued \hipass\ galaxies (Koribalski et al. 2004, Meyer et al. 2004, Wong et al. 2006), for which we calculated expected \HI\ diameters based on the \HI\ size-mass relation (Wang et al. 2016).  

\subsubsection{HI rotation curves and velocity width}

We aim to derive reliable rotation curves for over a thousand \wal\ galaxies using currently available techniques (e.g., Kamphuis et al. 2015; Bekiaris et al. 2016, 2020; Di Teodoro \& Fraternali 2015; Oh et al. 2018). These studies suggest that $\sim$5 beams across the major axis will -- in most cases -- be sufficient to derive a rotation curve at intermediate inclinations. \wal\ rotation curves will allow detailed mass dissections when supplemented with other readily available data (as described above). The main questions are: 

1. We will address the controversies about the mass distributions of disk galaxies, which constitute a challenge to the $\Lambda$CDM model (e.g., Dutton et al. 2018). One of the open questions concerns the amount of self-gravity of the disc, i.e. the question on whether the disk is maximal or not. This can be addressed by combining \HI\ observations with data at other wavelengths. A significant number of dwarf galaxies will be resolved by \wal\ and will contribute to much improved number statistics re-assessing the core-cusp problem (e.g., de Blok 2010). High resolution (10 arcsec) postage stamps would be highly beneficial for this study.

2. On average, rotation curves in the outer disk decline for very massive spirals and continue to rise for low-luminosity spirals. The optimum \HI\ line width needed for the Tully-Fisher (TF) relation should therefore be based on v$_{\rm flat}$, derived from the outer parts of the rotation curves (e.g., Verheijen 2001; Ponomareva et al. 2016). We will use this estimate for galaxy population studies involving the TF relation itself and the baryonic version of it (e.g., Oh et al. 2011). 

3. Galaxy rotation curves will provide a direct constraint on the halo-to-\HI\ velocity relation that is critical to the cosmological interpretation of the velocity function (e.g., Papastergis \& Shankar 2016). Even though rotation curve mass models appear to be consistent with bursty star formation predictions (e.g., Oh et al. 2011; Katz et al. 2016; but see Pace 2016), the rotation curve shapes are inconsistent with the relationship between halo and \HI\ velocity required by the velocity function. Precise predictions require larger and more homogeneous samples. Moreover, we need simulations of the resolved galaxy population probed by \wal, and detailed performance tests of available algorithms in the marginally resolved regime. 

\wal\ data will have sufficient spectral resolution to measure the \HI\ velocity dispersion in disk galaxies. Comparison of the turbulent energy in the ISM with the underlying stellar populations will provide important insights into the effects of stellar wind and supernovae feedback. 

\subsubsection{Spin, warps and kinematics}

With the data on the \HI\ kinematics and radial distribution of \wal\ galaxies, we can analyse the spin orientation and warp structure of \HI\ discs. By comparing this with optical data, we will identify cases of kinematical misalignments between gas and stars (Serra et al. 2014), which serve as a tracer for the degree to which accretion by the IGM or satellite galaxies has been coherent (see, e.g., Stevens et al. 2016; Bryant et al. 2019). Recently, Welker et al. (2020) found mass-dependent spin alignments of stellar galaxy discs with respect to cosmic filaments as predicted by simulations (Dubois et al. 2014; Lagos et al. 2018). 

With \wal, this study can be extended for the larger galaxies. At large radii (typically beyond the 25th $B$-band magnitude), the spin vectors of \HI\ discs of spiral discs often appear to be tilted with respect to the inner discs, i.e., galactic discs become warped where the stellar component is fading. The warp, the transition from one orientation to the other, often appears in a confined radial range, indicating a two-regime structure of an inner bright disk and a faint, inclined outer disk (e.g., Briggs 1990; van der Kruit 2007; J\'ozsa 2007; Kam et al. 2017). The transition is sometimes marked by a break in surface brightness as well as in their kinematics (J\'ozsa 2007). Warps have long been suggested to trace a large-scale alignment (Battaner et al. 1990), can potentially trace accretion from the IGM with misaligned angular momentum (van der Kruit 2007), and have been suggested to trace IGM properties in different environments (Haan \& Braun 2014). While attempts to use warps as tracers of large-scale accretion in optical wavelength (Lopez-Corredoira et al. 2008) are hampered by the fact that the optical disk tapers off before a warp is detectable, the same is not true for well-resolved \HI\ discs. The environment also influences the intrinsic warp structure and the warp kinematics, providing clues on warp formation.

We will parametrize the spin orientation, warp orientation, warp amplitude, and warp symmetry of the \wal\ galaxies in \HI. These quantities will be related to the galaxies' environment and their other optical properties. In addition we aim to investigate the kinematical and surface-brightness structure in the context of the warp structure (warp radius). This will become a benchmarking test for cosmological simulations, such as the recent analysis of {\sc eagle} simulations (Ganeshaiah Veena et al. 2019). 

\subsubsection{The Local Volume}

The Local Volume (LV; $D < 10$ Mpc) contains more than 700 known galaxies (Karachentsev et al. 2013) about 200 of which are detected in \hipass\ ($\delta < +25\degr$; Koribalski et al. 2004, Meyer et al. 2004, Wong et al. 2006). New LV discoveries continue to be made in deep optical, ultraviolet, infrared imaging and high-resolution \HI\ spectral line surveys (e.g., M\"uller et al. 2018; Adams \& Oosterloo 2018; Koribalski et al. 2018; Lee-Waddell et al. 2018). For example, five new dwarf galaxies were identified in the \lvhis\ project\footnote{\lvhis\ homepage: \url{www.atnf.csiro.au/research/lvhis/}} (Koribalski et al. 2018), which provides ATCA \HI\ observations of $\sim$100 southern LV galaxies. \lvhis\ galaxy maps are similar --- in terms of sensitivity and resolution --- to those expected from \wal. During ASKAP commissioning \lvhis\ results and data products together with other \HI\ surveys are used for \wal\ data validation.

\subsection{\HI\ scaling relations} % Section 5.5 

Gas scaling relations link the total gas content of galaxies to their structural and star formation properties (e.g., Wang et al. 2016; Catinella et al. 2018), thus are key to our understanding of galaxies and have a unique constraining power for galaxy formation models (e.g., Marasco et al. 2016; Crain et al. 2017; Lagos et al. 2018; Diemer et al. 2019; Stevens et al. 2019). \wal\ will have a major impact in this field, providing large galaxy samples for studies of their \HI\ content across the Hubble sequence and the color-magnitude diagram. 

While the main trends of decreasing \HI\ content (normalised by a scale parameter such as luminosity) for more luminous, redder and earlier-type galaxies have been known for over two decades (e.g., review by Roberts \& Haynes 1994), this remains an active area of research. Practical applications of gas scaling relations include: quantifying environmental effects (\HI-deficiency, e.g., Giovanelli \& Haynes 1985; Boselli \& Gavazzi 2006; Cortese et al. 2011; D\'enes et al. 2014; Brown et al. 2017); predicting gas content when \HI\ observations are not available (photometric gas fractions, e.g., Zhang et al. 2009; Eckert et al. 2015), identifying interesting populations of galaxies that are outliers in some of those relations (e.g., \HI-excess galaxies, Lutz et al. 2017; Ger\'eb et al. 2016, 2018).

Gas scaling relations have been obtained from both \hipass\ (D\'enes et al. 2014) and \alfalfa\ (Huang et al. 2012). However, \HI-blind surveys detect only the most gas-rich systems over their full volumes, hence care must be taken to recover gas scaling relations that are representative of the local galaxy population (see, e.g., Huang et al. 2012).

Significant effort has been devoted in the past decade to measure \HI\ properties of galaxies that are typically missed by \HI-blind surveys, such as early-type galaxies (ATLAS-3D, Serra et al. 2012) and stellar mass selected samples (GASS, Catinella et al. 2010, 2013; xGASS, Catinella et al. 2018; see also RESOLVE, Stark et al. 2016). In particular, the gas fraction limited nature of the GASS and xGASS surveys probed the gas-poor regime for a statistical sample of $\sim$1200 galaxies, providing benchmark gas scaling relations in the Local Universe ($z \leq 0.05$) that have been widely used to test semi-analytic models and hydrodynamic simulations (e.g., Lagos et al. 2014; 2015; Popping et al. 2015; Marasco et al. 2016; Crain et al. 2017; Dav\'e et al. 2017; Zoldan et al. 2017; Stevens et al. 2019). 

From a theoretical viewpoint it has long been expected that the \HI\ content of galaxies correlates with the angular momentum in the disc. Several empirical studies supported this view (e.g., Huang et al. 2012). Based on analytical stability considerations, Obreschkow et al. (2016) proposed a parameter-free model to quantitatively predict the \HI\ fraction as a function of the mass and angular momentum of galactic discs in dynamic equilibrium. They showed that local \HI\ selected galaxies (from \hipass, {\sc things} and {\sc little things}) follow the model prediction remarkably well. Moreover, isolated \HI-excess galaxies (see Lutz et al. 2017, 2018) and \HI-deficient late-type galaxies (see Murugeshan et al. 2019), with similar stellar masses and optical morphology are observed to follow this relation consistently. Such a disparity will naturally bring about scatter in the \MHI\ -- M$_{\rm star}$ scaling relations. Therefore, understanding this relation is crucial to understanding a potential source of scatter in \HI\ scaling relations. \wal\ will resolve the \HI\ discs of $\sim$5\,000 galaxies with five or more beams (corresponding roughly to the catalogued \hipass\ detections), enabling us to derive accurate outer rotation curves and thus to calculate the angular momentum for large samples of galaxies from different environments (low to high densities). Such a study will for the first time allow us to understand the origin of the scatter in \HI\ scaling relations from an angular momentum perspective (see Kurapati et al. 2018). 

One of the most interesting aspects of gas scaling relation studies is to determine what physical processes drive their scatter. This requires the ability to bin galaxies simultaneously by multiple properties, providing larger statistics than provided by the xGASS survey, described above, and is best tackled with spectral stacking. As mentioned in Section~1, \HI\ spectral stacking is a powerful technique to exploit \HI-blind surveys well below their nominal sensitivity limits, making use of spectra of non-detected galaxies that are present within their volumes. In addition to constraining the \HI\ cosmic density up to redshifts as high as $z \sim 1.265$ (Kanekar et al. 2016), spectral stacking is an excellent tool to investigate gas scaling relations (Fabello et al. 2011, 2012; Brown et al. 2015, 2017). By applying this technique, for instance, to a stellar mass-selected sample of $\sim$25\,000 galaxies within the \alfalfa\ volume (mostly undetected in \HI), Brown et al. (2015) showed that the scatter of the gas fraction-stellar mass relation is driven by a specific star-formation rate, and its slope is determined by the bimodality of local galaxies (with gas-rich, star-forming systems dominating in numbers at the low stellar mass end, and gas-poor, quiescent systems found preferentially at the high stellar mass end). Even more interestingly, Brown et al. (2017) presented for the first time convincing evidence for systematic environmental suppression of gas content at both fixed stellar mass and fixed specific SFR in satellite galaxies, and showed that gas suppression begins in halo masses typical of the group regime, well before galaxies reach the cluster environment. \wal\ will advance these studies not only by providing \HI\ spectra (see Fig.~\ref{fig:hipass-spectra}) and their properties for around half a million galaxies, but also and especially by covering an unprecedented volume, which will be a goldmine for spectral stacking analyses of optically-selected samples. 

In addition to the scaling relations between the total \HI\ mass and stellar properties, other scaling relations also provide useful tools. The \HI\ size-mass relation (Broeils \& Rhee 1997, Wang et al. 2016) provides an estimate of the \HI\ disk size for unresolved \HI\ masses and also puts strong constraints on the star formation and stellar feedback recipes adopted in cosmological simulations of galaxy formation (e.g., Bah\'e et al. 2016). The TF relation between stellar or baryonic mass and \HI\ velocity width implies the significant role that the dark matter halo may have played in the assembly of the galaxies and has also been studied with spectral stacking (Meyer et al. 2016).

\subsection{\HI\ Mass Functions} % Section 5.6

The \HI\ mass function (HIMF) describes the volume density of galaxies of a given \HI\ mass and is therefore the neutral hydrogen equivalent of the optical luminosity function. The HIMF is one of the most important statistical tools that astronomers use to describe and characterize gas-rich galaxy populations. One important parameter that can be derived from the \HI\ mass function is the cosmic \HI\ mass density, $\Omega_{\rm HI}$: the number of neutral hydrogen atoms per unit volume. This $\Omega_{\rm HI}$ is a key input into the understanding of the evolution of the cosmic SFR density. The evolution of the \HI\ mass density can be mapped out and compared to models of gas and galaxy evolution when the local 21-cm-derived $\Omega_{\rm HI}$ is combined with damped \Lya\ measurements at higher redshifts (e.g., Noterdaeme et al. 2012, and references therein). In particular, the comparison with the predictions of semi-analytic or hydrodynamic models can provide insights into the processes governing the distribution and evolution of cool gas (e.g., Obreschkow et al. 2009a; Lagos et al. 2011; Diemer et al. 2019). \\

The shape of the \HI\ mass function at $z = 0$ is an equally important constraint in cosmological models of galaxy evolution. For example, semi-analytic models (Obreschkow et al. 2009a; Lagos et al. 2011; Neistein \& Weinmann 2010; Kim et al. 2011; Stevens et al. 2016) use the \HI\ mass function shape as an essential input. Compared to the stellar mass function, the \HI\ mass function provides additional limits because the correlation between halo mass and cold gas mass is very different from that between halo mass and stellar mass. In the models, the most gas-rich galaxies often have low halo masses. Kim et al. (2012) show that the \HI\ mass function is a more sensitive probe of the effects of cosmological reionization on galaxy formation models than the luminosity function. Furthermore, these authors show how different star formation laws affect the shape of the \HI\ mass function. 

The first direct measurements of the \HI\ mass function were the result of Arecibo surveys in the late 1990s (Zwaan et al. 1997; Spitzak \& Schneider 1998; Rosenberg \& Schneider 2002). These surveys detected up to a few hundred galaxies over small areas of the sky, but were able to determine the mass function for \HI\ masses more than a few times $10^7$\Msun. Like the optical luminosity function, the \HI\ mass function was demonstrated to follow a Schechter function with a power-law faint-end slope and an exponential fall-off at the high mass end. However, different surveys disagreed on the slope of the low-mass end and hence the significance of the low \HI\ mass population remained uncertain (see e.g., Zwaan et al. 1997; Schneider et. al. 1998; Kova\v{c} et al. 2009). 

The accuracy of measured \HI\ mass functions improved considerably with the new generation of blind 21-cm surveys: \hipass\ and \alfalfa. \hipass\ produced a catalogue of several thousand galaxies and put particular effort into understanding the completeness and reliability of the catalogue (Zwaan et al. 2004). With \hipass, it became possible to divide the sample into different sub-samples to study, for instance, the effect of environment on the shape of the mass function. However, controversy still surrounds the magnitude and sign of the density-dependence of the \HI\ mass function, possibly due to depth and cosmic variance issues with existing shallow surveys (Springob et al. 2005; Zwaan et al. 2003, 2005). \alfalfa\ covers a much smaller region of the sky than \hipass, but is considerably deeper. Jones et al. (2018b) derive the \HI\ mass function from the final \alfalfa\ catalogue as well as separately for the two survey regions (the Virgo cluster and the local void). They find a large discrepancy in the low-mass slope between the two regions ($-1.29$ vs. $-1.15$), likely due to the difference in galaxy density. Their global slope of $\alpha = -1.25 \pm 0.02 \pm 0.03$ is slightly below the global \hipass\ slope of $\alpha = -1.37 \pm 0.06$ from Zwaan et al. (2005). The observed differences are statistically significant and could indicate systematic variations in the slope of the HIMF with environmental density as traced by the different survey regions (e.g., Said et al. 2019). 

\begin{figure}[tb] % Figure 12 - new figure created by Martin Zwaan
\begin{center}
  \includegraphics[width=8cm]{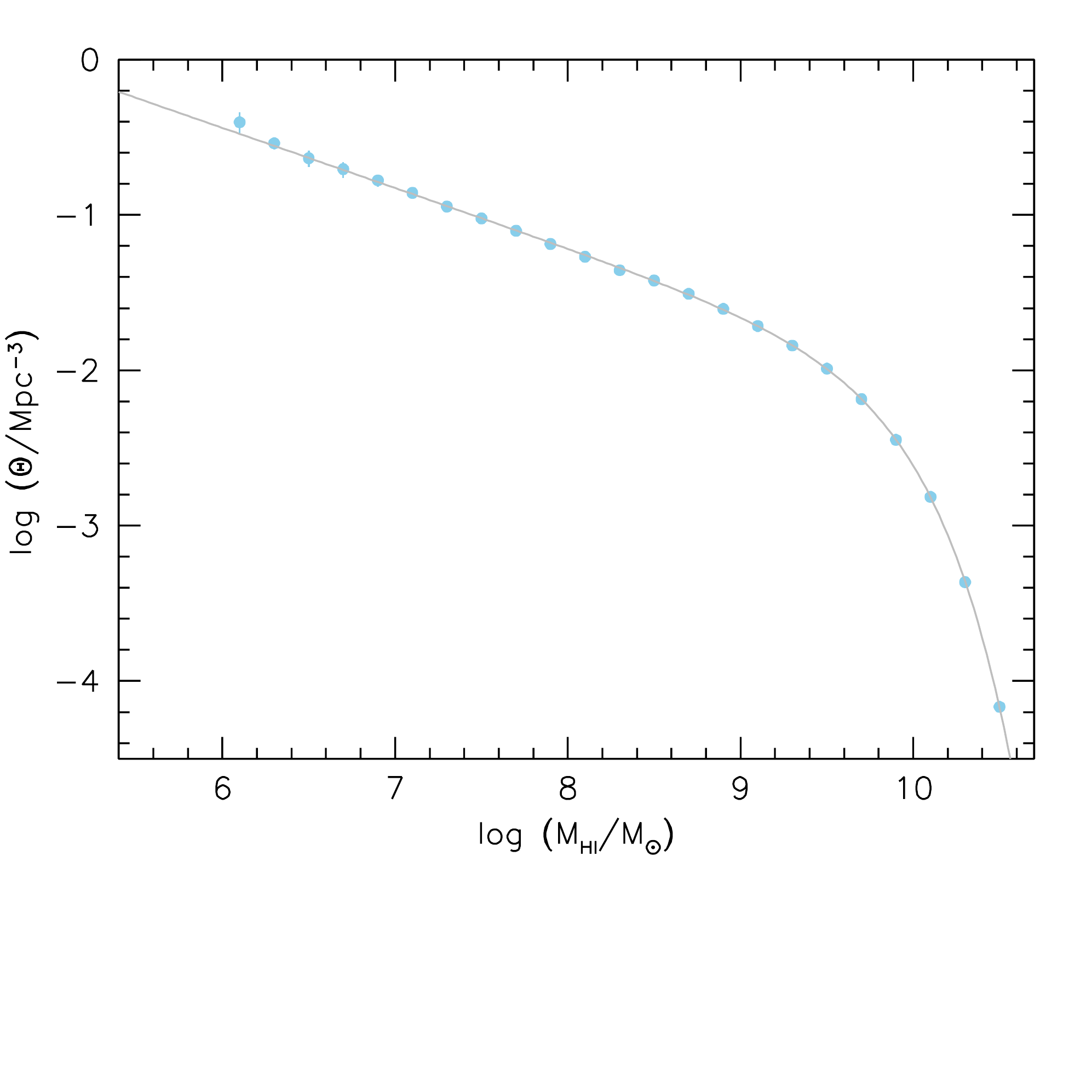}
  \vspace{-2cm}
\caption{The reconstructed \HI\ mass function for the simulated \wal\ volume. A 2D stepwise maximum likelihood method (SWML; Zwaan et al. 2003) method has been used to measure the space density of galaxies, which can be measured accurately down to \HI\ masses of \MHI\ = 10$^{6}$\Msun. The grey line is a Schechter fit to the data points.} 
\label{fig:himf}
\end{center}
\end{figure}

The number of galaxies detected by \wal\ will be $>$10 times larger than that found by \hipass\ and \alfalfa\ combined. Compared to \hipass, \wal\ can detect low-mass galaxies over a $\sim$100 times larger volume and is therefore much less sensitive to cosmic variance. Fig.~\ref{fig:himf} shows that with \wal\ the local \HI\ mass function can be reliably measured down to masses below \MHI\ = $10^6$\Msun. In addition, the large number of \wal\ detections will allow for a detailed look at the connection between \HI\ and halo mass through measurements of the conditional HIMF (similar to the conditional stellar mass function; Yang et al. 2005, 2008; Reddick et al. 2013). Projection of the Conditional HIMF along its two dimensions results in measurements of the full HIMF and the occupancy of \HI\ galaxies within halos of a given mass, and so provides insight into the formation of \HI\ galaxies. Thanks to the much larger number of detections and the larger survey volume, the question of how environment, evolution and halo mass affect the \HI\ mass function in the Local Universe will be put to rest by \wal.

\subsection{Intergalactic gas and the cosmic web} % Section 5.7

There are two key aspects for \wal\ studies of the intergalactic medium (IGM): at low redshift \wal\ will detect the IGM in emission, in the form of intra-group and intra-cluster \HI\ gas, typically present in nearby groups/clusters, where tidal interactions and harassment/stripping are on-going (see Section~5.3). Examples are: the M\,96 group (Schneider et al. 1983), the M\,81 group (de Blok et al. 2018; Sorgho et al. 2019), the NGC~2434 group (Ryder et al. 2001), the NGC~3263 group (English et al. 2010), the IC\,1459 group (Serra et al. 2015a; Saponara et al. 2018), compact groups (Verdes-Montenegro et al. 2001; Serra et al. 2013), the Virgo cluster (Kent et al. 2009; Oosterloo \& van Gorkom 2005; Kent et al. 2009; Chung et al. 2009) and the Abell~1367 cluster (Scott et al. 2018). At the \wal\ resolution of $\sim$30~arcsec, an \HI\ cloud with a diameter of 7~kpc can be spatially resolved to a redshift of 0.011 ($cz$ = 3400\kms). Assuming a velocity width of 25\kms, the 5$\sigma$ \HI\ detection limit is $4 \times 10^{19}$~cm$^{-2}$. Therefore, \wal\ will be able to locate and map all the emission from \HI\ gas at $cz < 3400$\kms\ that is capable of producing a DLA. At the high redshift end of our band, the IGM can be characterized via intervening \HI\ absorption line systems in spectra towards bright continuum background sources. Morganti \& Oosterloo (2018) estimate that \wal\ will make $\sim$2300 detections of associated \HI\ absorption (see their Table~2).

\wal\ will provide a first-ever large-scale, unbiased look at the denser parts of the IGM, which so far has mainly been seen in \Lya\ absorption lines against background quasars. About one-third of the $z \sim 0$ baryons are in this photoionized phase with $T \approx 10^4$~K (Danforth \& Shull 2008), with the remaining two-thirds in galactic concentrations and the warm-hot IGM. Wakker \& Savage (2009) show that 50\% of the photoionized \Lya\ forest lies within 400~kpc of galaxies and that the highest column density forest lines mostly occur close to galaxies. \Lya\ absorption lines are very rare in the Local Universe because their detection requires space-based ultraviolet observations. Of the highest column density absorbers, the DLA and sub-DLA systems, only $\sim$10 instances are known at $z<0.2$. Of the nearest four, two originate in the disk of a spiral galaxy, one in a dwarf galaxy, and one (seen toward the quasar PG\,1216+069) appears to be an intergalactic cloud behind the Virgo Cluster (see Chengalur et al. 2015, and references therein). This will allow us to answer questions about the exact relationship between these intergalactic clouds and galaxies, such as what fraction originates in galaxy discs and how far from galaxies they occur. It will also give us a zero-redshift marker from which we can track evolution through higher-redshift absorption line studies.

\subsection{Large-scale Structure and Cosmological parameters} % Section 5.8

A series of landmark observations have confirmed many of the predictions of the standard $\Lambda$CDM cosmology. However, uncertainties remain over the validity of $\Lambda$CDM because there is no physical understanding of either dark matter or dark energy, and there remains the possibility that there may exist other, as yet undetected, components such as warm dark matter and primordial black holes. The unique widefield capabilities of ASKAP allow \wal\ to detect a large number of galaxies over a large volume of the local Universe and allow several cosmological experiments to be conducted. These experiments will provide tests of General Relativity, the $\Lambda$CDM paradigm, and refine cosmological parameters.

\wal\ will determine the 3D locations of large numbers of local galaxies. Firstly, this will allow the measurement of the baryonic acoustic oscillations (BAO) scale at the present epoch. When combined with other surveys such as 2dFGRS, SDSS and WiggleZ (which have measured the BAO scale 2--3 billion and seven billion years ago, respectively), \wal\ will provide complete information on cosmic expansion to the present day. In the Local Universe, \wal\ will improve on the 6dFGS measurement of the Hubble constant (Beutler et al. 2011), and provide a valuable complement to the \taipan\ optical spectroscopic survey (Da Cunha et al. 2017). Together \wal\ and \taipan\ will yield the present-day expansion rate of the universe, \Ho, to better than 1\% precision. 

Secondly, $\Lambda$CDM also makes specific predictions as to how the gravitational attraction of dark matter produces variations in the otherwise smooth expansion of galaxies. This gives rise to structure in the distribution of galaxies and also gives rise to non-Hubble velocities and redshift-space distortions. These can be measured via the density and velocity/momentum power spectra. Combined with cosmic microwave background (CMB) derived priors, the full-shape of the clustering measured with \wal\ can be used to constrain the $\Lambda$CDM cosmological model and its extensions, providing constraints on: $\Omega_m$ and $\Omega_b$, the total matter and baryon densities defined relative to the critical density; $h$, the Hubble constant in units of 100\kms\,Mpc$^{-1}$, $f$, the growth rate of structurer; $\sigma_8$, the present amplitude of the matter perturbations; $n_{\rm s}$, the spectral index of the density fluctuations; and finally the equation of state parameter for dark energy, $w$, where $w = -1$ is the cosmological constant. The main impact of \wal\ is to reduce the error on $h$ and $w$ by a factor two (Beutler et al. 2011; Duffy, Moss \& Staveley-Smith 2012) relative to the value achieved with the Planck satellite alone; and to improve the constraints on $f \sigma_8$ at $z \approx 0$ by a factor of three (Koda et al. 2014; Howlett et al. 2017a) compared to 6dFGS (Beutler et al. 2012; Adams \& Blake 2017). This latter improvement results from the fact that \wal\ will provide both a redshift sample and redshift-independent distances.

\subsubsection{Redshift-independent distances} % Section 5.8.1

About 10\% of the \wal\ \HI\ detections will be of suitably inclined galaxies with profiles of sufficient S/N ratio to yield reliable distances via the TF relation. Such distances rely on quality infrared or near-infrared photometry from surveys such as SkyMapper (Wolf et al. 2018), 2MASS (Skrutskie et al. 2006), WISE (Wright et al. 2010; Jarrett et al. 2012, 2019) and the VISTA Hemispheric Survey (Sutherland et al. 2015). Such a sample will provide the most comprehensive TF dataset to date, exceeding the recent 2MTF survey (Hong et al. 2019) by an order of magnitude. It will also complement the existing 6dFGS peculiar velocity survey of 10\,000 early-type galaxies (Campbell et al. 2014) and the future \taipan\ survey of up to 100\,000 early-type galaxies (Da Cunha et al. 2017), potentially similar numbers of supernova from the Zwicky Transient Facility (ZTF) and LSST (Howlett et al. 2017c; Graham et al. 2019), and will significantly add to the totality of redshift-independent distances available in the cosmicflows compilation (Courtois et al. 2017).

Crucially, the differences in sample type and distance indicator between \wal\ and other surveys will provide two completely independent probes of the galaxy motions and density field over a volume with considerable overlap. Furthermore, while the early-type galaxies in \taipan\ tend to be found in clusters and other over-densities, the late-type galaxies of \wal\ will better trace the low-density filaments and the large-scale structures, including across the Zone of Avoidance which otherwise masks some of the most massive regions in the local Universe (Staveley-Smith et al. 2016). The combined power of \wal\ TF data, the \taipan\ Fundamental Plane data and future supernova samples will provide the best route to estimate the distribution of mass in the Local Universe and the total matter density (e.g. Erdogdu et al. 2006; Graziani et al. 2019). Combined measurements of \HI\ line width and central stellar velocity dispersion can improve the TF relation and extend it to a generalised scaling relation (Cortese et al. 2014; Tiley et al. 2019; Barat et al. 2019) and reduce the uncertainties in the velocities of individual galaxies, currently sitting at $\sim$20\%, and expand the range of application to galaxies of all morphological types.

Such velocity fields can be used to provide better values for the bulk and shear motion of galaxies over greater depths in the local Universe (Scrimgeour et al. 2016; Hong et al. 2019; Qin et al. 2019a). $\Lambda$CDM provides testable predictions for the expected range for these measurements. The velocity fields can also be compared with the density distributions inferred from redshift surveys, including \wal\ itself. These can also be used to test $\Lambda$CDM and General Relativity by constraining to growth factor $f \sigma_8$ at low redshifts (Springob et al. 2016). Furthermore, the velocity or momentum auto power and their cross-power spectrum with the density field can be used to accurately measure $f \sigma_8$ (Koda et al. 2014; Howlett 2019; Qin et al. 2019b).

Studies of the TF relation require high signal-to-noise, double-horn \HI\ spectra of undisturbed galaxies, ideally with steep wings. Using the Busy Function (Westmeier et al. 2014) to characterize the \wal\ galaxy \HI\ spectra may allow us to use a lower signal-to-noise threshold than previously employed (e.g., Hong et al. 2013, 2019).

\subsubsection{Large-scale structure}

The observed large-scale structure of the Universe, as traced by the spatial distribution of galaxies, is a complex web of sheets, filaments and knots (e.g., Einasto et al. 1997). Numerical simulations have been successful at reproducing the basic properties of the formation and evolution of the cosmic web (e.g., Colberg et al. 2005), although the processes regulating the baryonic component are still not fully understood.

We now know that filaments of galaxies stretch between clusters, and that these provide the environment in which the overwhelming majority of galaxies form and spend a significant fraction of their lives. The 2dFGRS and SDSS surveys have helped to establish the monotonic decline of star formation in galaxies from the low-density field to the cores of clusters (e.g., Lewis et al. 2001; Kauffmann et al. 2004). What comes as a surprise is the high incidence of star formation in the outskirts of clusters (at 1--2 times the virial radii), particularly for clusters that are part of a filamentary network (e.g., Porter et al. 2008). This seems to indicate that the web of filaments plays a very important role in processing a large fraction of galaxies before they are assimilated in modern-day clusters.

Within $z < 0.1$, there are several remarkable superclusters that consist of $>$20 rich clusters and associated filaments, namely Shapley (Raychaudhury 1989), Horologium-Reticulum (Fleenor et al. 2005), Pisces-Cetus (Porter \& Raychaudhury 2005) and possibly the Vela Supercluster (Kraan-Korteweg et al. 2017). Such systems provide a wide range of environments, in local density and local velocity dispersion, within a small volume. The measurement of \HI\ deficiency and the morphology of neutral gas will allow the study of effects like the interaction between the galaxies and the intra-cluster medium (stripping and strangulation), galaxy-galaxy interactions (harassment, mergers), and star formation in clusters and on the inter-cluster filaments (e.g., D\'enes et al. 2014).

\begin{figure}[tb] % Figure 13 - created by Chris Wolf and Karen LW
\begin{center}
 \includegraphics[width=8cm]{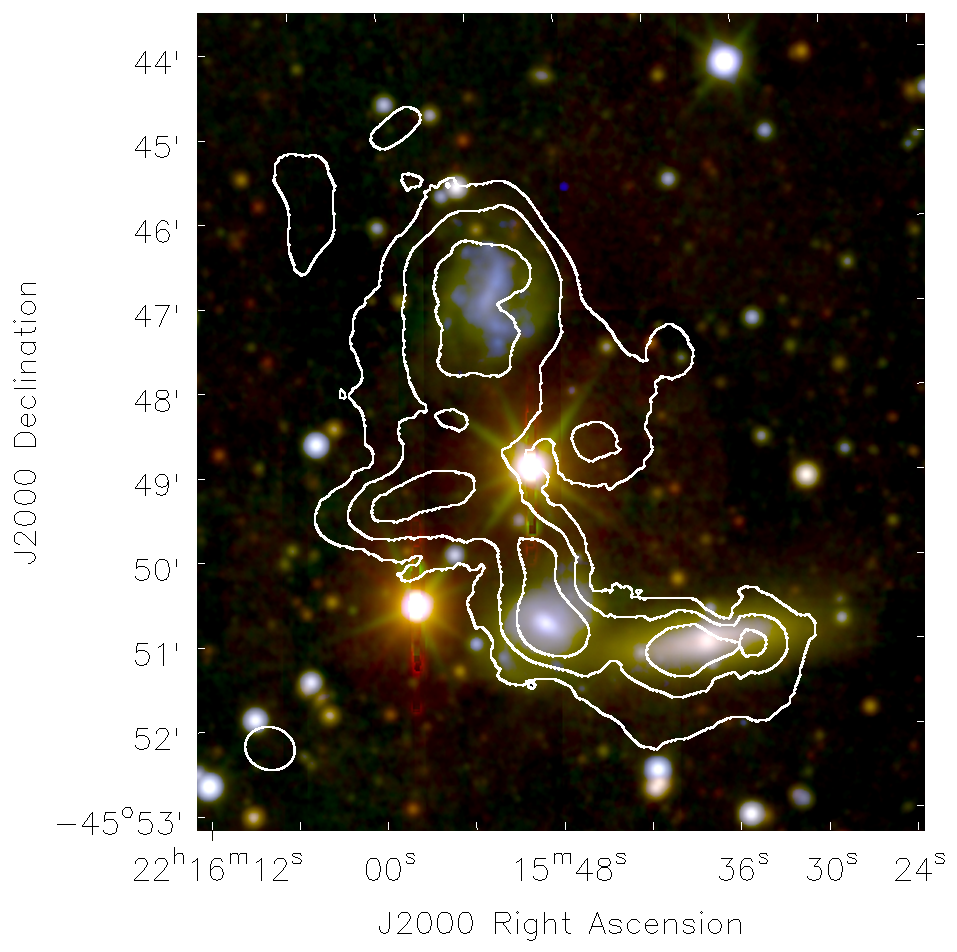}
\caption{The NGC~7232 galaxy triplet: ASKAP \HI\ contours from \wal\ Early Science data (Lee-Waddell et al. 2019) overlaid onto a SkyMapper image. The \HI\ contour levels are at 1, 3, and $6 \times 10^{20}$ atoms cm$^{-2}$. }
\label{fig:n7232}
\end{center}
\end{figure}

\begin{figure*}[htb] % Figure 14 - created by Tobias
\begin{center}
 \includegraphics[width=13cm]{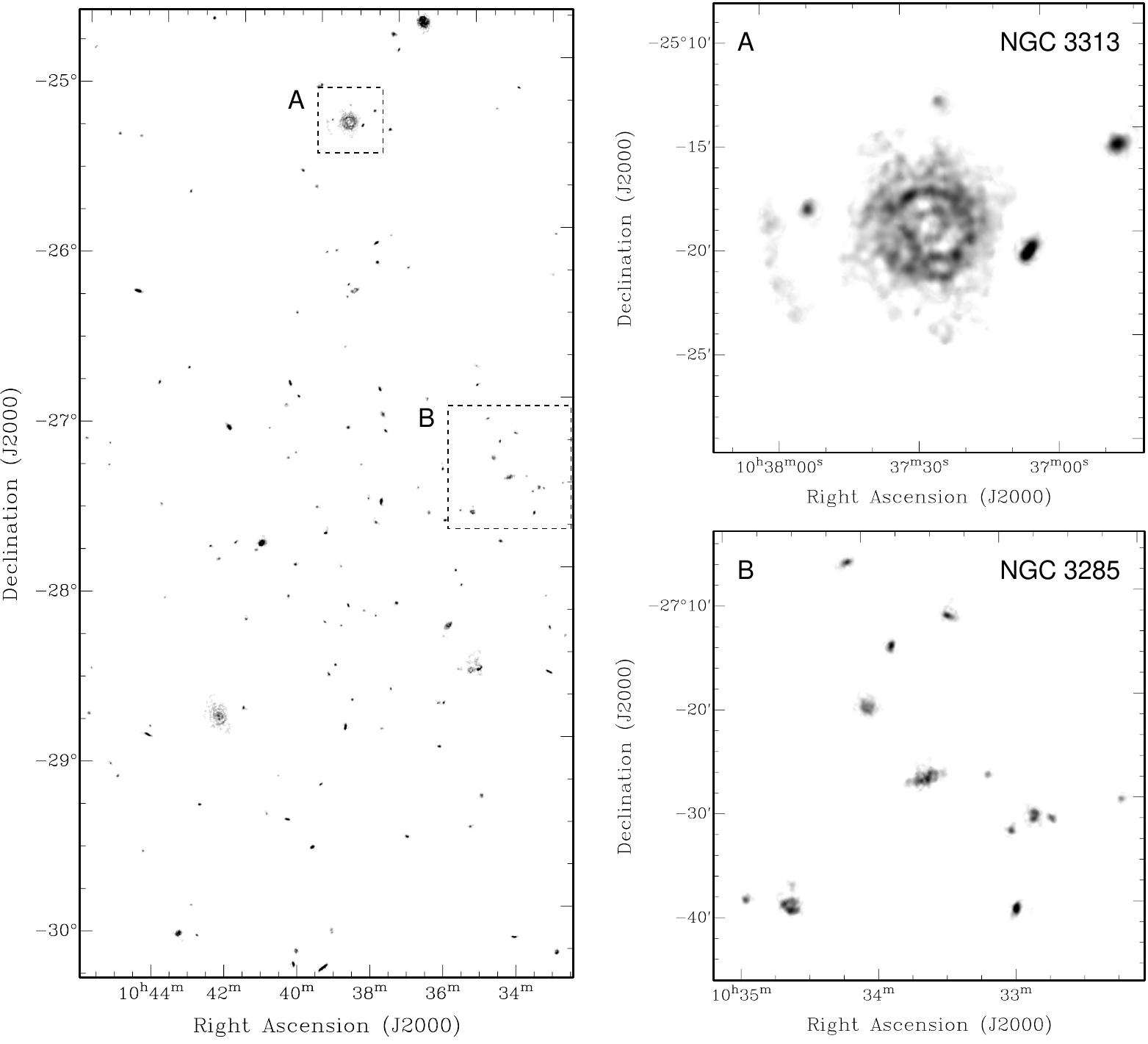}
 \includegraphics[width=13cm]{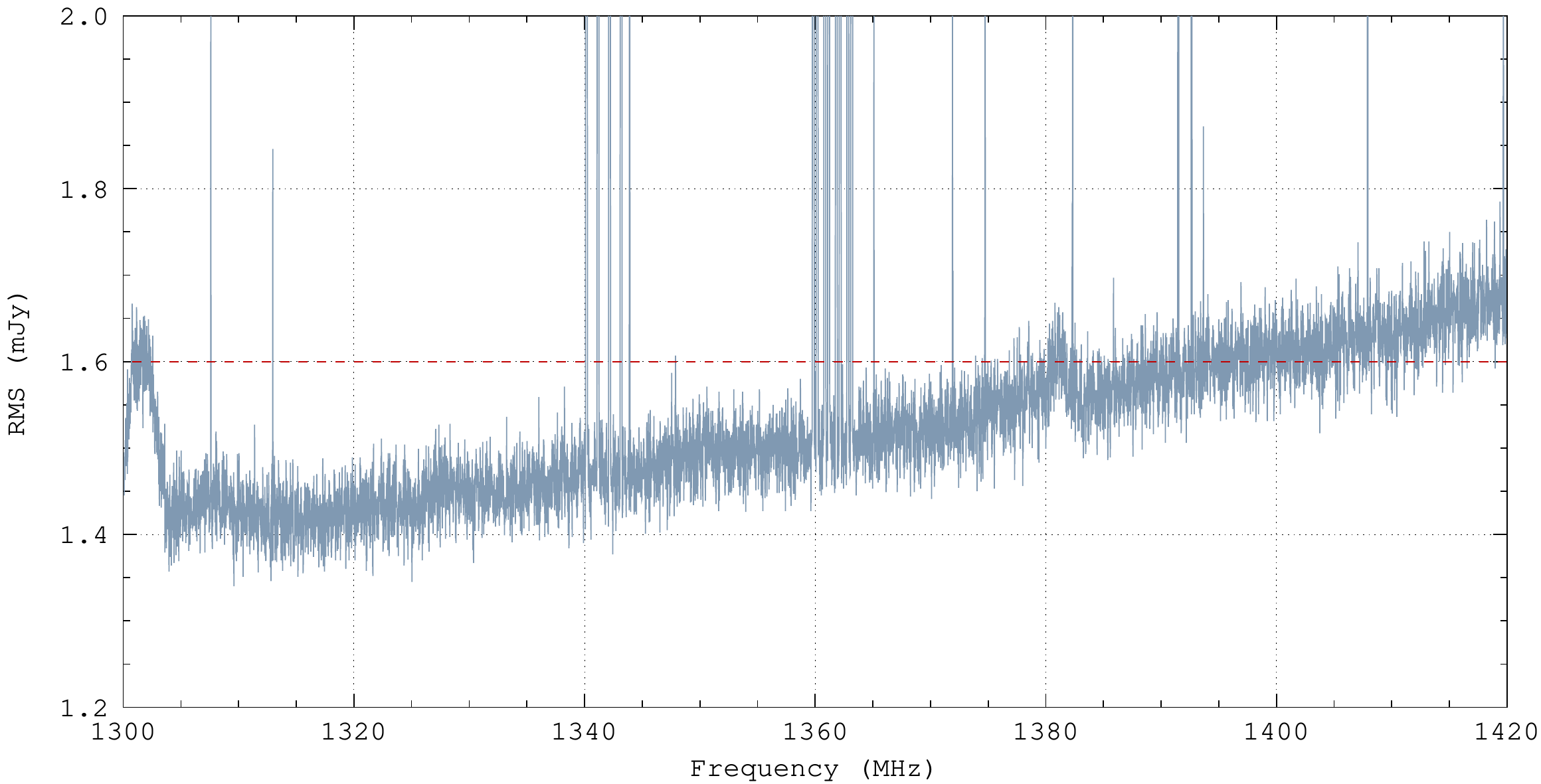}
\caption{\wal\ pilot survey results for the Hydra cluster, obtained with the full ASKAP-36 array. At the top left we show the SoFiA-derived H\,{\sc i} intensity map of $\sim$150 reliably detected galaxies over approximately a third of the observed field and a velocity range from 500 to 15\,000\kms\ ($z < 0.05$). The full Hydra field consists of two sets of interleaved 36-beam observations for two adjacent 30 sq degr fields. The four data cubes were combined and we achieve a median rms of 1.5 mJy\,beam$^{-1}$, as demonstrated in the bottom plot. At the top right we show zoom-ins for regions A and B, centred on the galaxies NGC~3313 (face-on) and NGC~3285 (edge-on). The synthesized beam is $\sim$30~arcsec. }
\label{fig:hydra}
\end{center}
\end{figure*}

\section{WALLABY Early Science and Pilot Surveys} % Section 6

During ASKAP Early Science (Oct 2016 -- Mar 2018) the \wal\ team obtained $\sim$800~h of \HI\ spectral line observations for four target fields with an array of up to 16 PAF-equipped antennas. Using all 36 beams, typically arranged in a square $6 \times 6$ pattern, we achieved the desired field of view of 30 sq deg. Our aim was to reach full \wal\ depth (see Table~1), which required spending $\sim$150~h per field spread over multiple nights. By targeting nearby, gas-rich galaxy groups and clusters, we set out to explore the \HI\ content, kinematics, and star formation of galaxies and their surroundings as a function of local environment. The \wal\ Early Science fields were ideal commissioning targets, delivering not only \HI\ emission and absorption data, and studies of radio transients, but also radio continuum and polarization maps. All \wal\ Early Science fields were selected to contain numerous \hipass-catalogued spiral galaxies for data validation purposes.

For the first field, centered on the NGC~7232 galaxy group, we were able to use only a limited ASKAP bandwidth of 48 MHz, divided into $\sim$2500 channels. Most observations were conducted from Oct to Dec 2016 with an array of up to 12 PAF-equipped ASKAP antennas. \wal\ Early Science results obtained from target areas within this field are published by Lee-Waddell et al. (2019) on the interacting NGC~7232/3 galaxy triplet and surroundings (see Fig.~\ref{fig:n7232}), Kleiner et al. (2019) on the spiral galaxy IC\,5201 and companions, and Reynolds et al. (2019) on the NGC~7162 galaxy group. The second field targeted the Fornax Cluster in Dec 2016; the available observing bandwidth was 192 MHz. Our third and fourth fields were centered on the Dorado and M\,83 galaxy groups, respectively, covering a full 24 hour LST range suitable for continuous observing in 2016/7. We took particular care to minimise solar sidelobes entering the ASKAP dishes and the ASKAP PAFs for our day-time field (towards the M\,83 group). This set of data allowed us to re-assess the day-time observing strategy as well as to create a metric to detect artefacts associated with solar interference. \wal\ Early Science results from the two fields are published by Elagali et al. (2019), focused on the spiral galaxy NGC~1566, and For et al. (2019) on galaxy groupings behind the M\,83 group. Our Source Finding Application (SoFiA; Serra et al. 2015b) has been essential to the success of \wal\ Early Science, used in each of the above publications, and continues to evolve in readiness for creating \wal\ catalogues.

In Mar 2019, the first observations were conducted with the full array of 36 PAF-equipped ASKAP antennas (ASKAP-36) and a bandwidth of 288~MHz targeting the Eridanus galaxy group. This was followed by the first round of pilot science surveys between Aug 2019 and May 2020. \wal\ pilot fields targeted the Hydra and Norma galaxy clusters as well as an equatorial field around the NGC~4636 galaxy group, allowing us to finalise the ASKAPsoft pipeline parameters and sky tiling strategy. 

In Figs.~10 \& 14 we show our first \HI\ results from ASKAP-36 observations of the Hydra galaxy cluster. An area of $\sim$60 sq degr was observed using four interleaved pointings (8h each). Each data set was then calibrated and imaged using the ASKAPsoft pipeline (Whiting et al. 2017, Kleiner et al. 2019) and a 30 arcsec restoring beam. The resulting cubes were combined before running SoFiA to find and extract all reliable \HI\ detections. Due to significant RFI in the lower part of the band, we focus on the 1.3 -- 1.42 GHz range, obtaining a median rms of 1.5 mJy\,beam$^{-1}$ for the central beams. The rms noise does increase by a factor $\sim$1.4 towards the outer beams. The $\sim$150 \wal\ detections in the displayed part of the Hydra cluster (see Fig.~14, top left) compare to $\sim$50 known \HI\ galaxies, including eight HIPASS detections. 

A full space sampling for nearby, large galaxies can be achieved by combining \wal\ data with data from single-dish telescopes. We have started a pilot campaign of synergetic observations with FAST on a 20 sq deg field around the center of the NGC~4636 group at a distance of 15.9 Mpc, with the aim of detecting the faint, diffuse \HI\ stripped by environmental effects. The FAST image is under reduction and expected to achieve a column density limit of a few times 10$^{18}$ cm$^{-2}$. We expect a depth of $\sim$10$^{19}$ when combining FAST and \wal\ data with roughly equal weights.

\section{Summary} % Section 7

With the full ASKAP-36 telescope now in operation, this overview paper provides a timely description of the \wal\ project, its observing parameters, software tools and science goals. The first round of ASKAP pilot surveys were recently completed, and we highlight some initial \wal\ results, including high-resolution postage stamps, from the Hydra cluster field. While data processing is likely continuing until the second round of pilot surveys commences later this year, \wal\ results and discoveries are being prepared for publication.

\wal\ is an extragalactic \HI\ survey which builds on recent large-area surveys conducted with multibeam feed arrays on single-dish telescopes (e.g., \hipass, \alfalfa), and complements planned northern surveys with new instruments (Apertif, FAST). \wal\ will become a low-redshift reference survey which will anchor our knowledge of the \HI\ properties of galaxies and the effect of environment on the formation and evolution of galaxies. This sets the scene for several deep \HI\ surveys with ASKAP and MeerKAT ({\sc dingo, mightee-hi, laduma}) and, later, the SKA and better informs us on the evolution of the gaseous properties of galaxies, and will expand our knowledge of gas accretion, outflow and galaxy interactions. 

A major advantage of \wal\ over previous surveys, an overview of which is provided here,  will be its ability to clearly separate individual galaxies in group and cluster environments, and to well-resolve thousands of nearby galaxies, thus allowing accurate studies of their kinematics and interstellar medium. We confirm previous predictions that \wal\ is expected to detect around 500\,000 galaxies, the vast majority with redshifts $z<0.1$. Furthermore, as \wal\ will have complete three-dimensional sampling of the sky south of $\delta = +30\degr$, stacking studies will be possible in situations where redshifts are available from optical spectroscopic studies. 

Our \wal\ Early Science results (based on a limited number of ASKAP antennas) and recent ASKAP-36 pilot surveys, both achieving full \wal\ depth, have already demonstrated the scientific potential of large \HI\ surveys with widefield Phased-Array Feeds. They have also shown the capability of the ASKAPsoft data reduction pipeline and our post-processing software tools. Further refinement of analysis pipelines using the full 36-antenna ASKAP telescope is presently occuring, and it is expected that full \wal\ observations will commence in late 2021 after a second round of pilot surveys. Fully-reduced \wal\ data are made available to the astronomy community following quality control, with advanced data products (verified source catalogues, galaxy parameters, velocity fields etc.) being released in CASDA and updated at regular intervals.

\section*{Acknowledgements}
% \acknowledgements
We thank both referees for carefully reading the full manuscript and providing excellent feedback. \\

The Australian SKA Pathfinder (ASKAP) is part of the Australia Telescope National Facility (ATNF) which is managed by CSIRO. Operation of ASKAP is funded by the Australian Government with support from the National Collaborative Research Infrastructure Strategy (NCRIS). ASKAP uses the resources of the Pawsey Supercomputing Centre. Establishment of ASKAP, the Murchison Radio-astronomy Observatory (MRO) and the Pawsey Supercomputing Centre are initiatives of the Australian Government, with support from the Government of Western Australia and the Science and Industry Endowment Fund. We acknowledge the Wajarri Yamatji as the traditional owners of the Observatory site. \\

% Frank Bigiel and Dane Kleiner
FB and DK acknowledge funding from the European Union's Horizon 2020 research and innovation programme; grant agreement No 726384 (FB) and No 679627 (DK).

% Michelle E. Cluver
MEC is a recipient of an Australian Research Council Future Fellowship (project number FT170100273) funded by the Australian Government.

% Rob Crain
RAC is a Royal Society University Research Fellow.

% Renee Kraan-Korteweg, G.I.G. Jozsa
RKK acknowledges the South African Research Chairs Initiative of the Department of Science and Innovation. RKK and GIGJ acknowledge the South African National Research Foundation for their support. 

% Michael Jones and Lourdes Verdes-Montenegro
MGJ and LVM acknowledge support from the grants AYA2015-65973-C3-1-R (MINECO/FEDER, UE) and RTI2018-096228-B-C31 (MICIU/FEDER, EU), as well as from the State Agency for Research of the Spanish MCIU through the "Center of Excellence Severo Ochoa" award for the Instituto de Astrof\'isica de Andaluc\'ia (SEV-2017-0709). MGJ is supported by a Juan de la Cierva formaci\'{o}n fellowship (FJCI-2016-29685).

% Peter Kamphuis
PK is partially supported by the BMBF project 05A17PC2 for D-MeerKAT. 

% Danail Obreschkow
DO is a recipient of an Australian Research Council Future Fellowship (FT190100083) funded by the Australian Government.

% Kristine Spekkens
KS acknowledges support from the Natural Sciences and Engineering Research Council of Canada (NSERC).

% Thijs van der Hulst
JMvdH acknowledges support from the European Research Council under the European Union's Seventh Framework Programme (FP/2007-2013) / ERC Grant Agreement nr. 291531. \\

Parts of this research were supported by the Australian Research Council Centre of Excellence for All Sky Astrophysics in 3 Dimensions (ASTRO~3D), through project number CE170100013.

%%%%%%%%%%%%%%%%%%%% REFERENCES %%%%%%%%%%%%%%%%%%

% Don't change these lines
% \bsp	% typesetting comment
% \label{lastpage}


\begin{thebibliography}{99}
\bibitem[]{ao18}Adams E.A.K., Oosterloo T.A. 2018, A\&A 612, 26 
\bibitem[]{ab17}Adams C. \& Blake C. 2017, MNRAS 471, 839 
\bibitem[]{allen86}Allen R.J., Atherton P.D., Tilanus R.P.J. 1986, Nature 319, 296
\bibitem[]{allen97}Allen R.J., Knapen, J.H., Bohlin R., Stecher T.P. 1997, ApJ 487, 171
\bibitem[]{allison15}Allison J., et al. 2015, MNRAS 453, 1249 
\bibitem[]{anderson18}Anderson C.J., et al. 2018, MNRAS 476, 3382 
\bibitem[]{arjun19}Dey A., et al. 2019, AJ 157, 168 % >8 authors
\bibitem[]{athanassoula16}Athanassoula E., Rodionov S.A., Peschken N., Lambert J.C. 2016, ApJ 821, 90
\bibitem[]{auld06}Auld R., et al. 2006, MNRAS 371, 1617  % >8 authors
\bibitem[]{bahe16}Bah\'e Y.M., et al. 2016, MNRAS 456, 1115
\bibitem[]{bait20}Bait O., Kurapati S., Duc P.A., Cuillandre J.-C., Wadadekar Y., Kamphuis P., Barway S. 2020, MNRAS 492, 1    
\bibitem[]{barnes01}Barnes D.G., et al. 2001, MNRAS, 322, 486 % >8 authors
\bibitem[]{battaglia06}Battaglia G., Fraternali F., Oosterloo T., Sancisi R. 2006, A\&A 447, 49 
\bibitem[]{bechtol15}Bechtol K., et al. 2015, ApJ 807, 50  % >8 authors
\bibitem[]{begum05}Begum A., Chengalur J.N., Karachentsev I.D. 2005, A\&A 433, L1
\bibitem[]{begum08}Begum A., Chengalur J.N., Karachentsev I.D., Sharina M.E., Kaisin S.S. 2008, MNRAS 386, 1667 
\bibitem[]{bekiaris16}Bekiaris G., Glazebrook K., Fluke C.J., Abraham R. 2016, MNRAS 455, 754 
\bibitem[]{bekiaris20}Bekiaris G., et al. 2020, in prep. % new GBKFIT
\bibitem[]{bekki05a}Bekki K., Koribalski B.S., Ryder S.D., Couch W. 2005a, MNRAS 357, L21
\bibitem[]{bekki05b}Bekki K., Koribalski B.S., Kilborn V.A. 2005b, MNRAS 363, L21
\bibitem[]{beutler11}Beutler F., et al. 2011, MNRAS 416, 3017 
\bibitem[]{beyer15}Beyer J., Hadwiger M., Pfister H. 2015, Computer Graphics Forum 34, 13
\bibitem[]{bicalho19}Bicalho I.C., Combes F., Rubia M., Verdugo C., Salome P. 2019, A\&A 623, 66 
\bibitem[]{bb12}Bigiel F., Blitz L. 2012, ApJ 756, 183 
\bibitem[]{bigiel08}Bigiel F., Leroy A., Walter F., Brinks E., de Blok W.J.G., Madore B., Thornley M.D. 2008, AJ 136, 2846
\bibitem[]{bird03}Bird T.S., Cort\'es-Medell\'in G. 2003, IEEE 1, 116
\bibitem[]{blecher19}Blecher T., Deane R., Heywood I., Obreschkow D. 2019, MNRAS 484, 3681 
\bibitem[]{blyth16}Blyth S., et al. 2016, Proc. of MeerKat Science: On the Pathway to the SKA, 25--27 May, 2016 Stellenbosch, South Africa, p.\,4 % >8 authors
\bibitem[]{bg06}Boselli A., Gavazzi G. 2006, PASP 118, 517
\bibitem[]{bg14}Boselli A., Gavazzi G. 2014, A\&AR 22, 74 
\bibitem[]{boselli18a}Boselli A., et al. 2018a, A\&A 614, 56 % >8 authors
\bibitem[]{boselli18b}Boselli A., et al. 2018b, A\&A 615, 114 % >8 authors
\bibitem[]{bovill09}Bovill M.S., \& Ricotti M. 2009, ApJ 693, 1859
\bibitem[]{bosma81a}Bosma A. 1981a, AJ 86, 1791
\bibitem[]{bosma81b}Bosma A. 1981b, AJ 86, 1825
\bibitem[]{bosma17}Bosma A. 2017, ASSL 434, 209
\bibitem[]{boyce01}Boyce et al. 2001, ApJ 560, L127
\bibitem[]{briggs90}Briggs F.H. 1990, ApJ 352, 15 
\bibitem[]{briggs98}Briggs F.H. 1998, A\&A 336, 815 
\bibitem[]{broeils97}Broeils A.H., Rhee M.-H. 1997, A\&A 324, 877
\bibitem[]{brown15}Brown T., et al. 2015, MNRAS 452, 2479
\bibitem[]{brown17}Brown T., et al. 2017, MNRAS 466, 1275
\bibitem[]{bryant19}Bryant J.J., et al. 2019, MNRAS 483, 458 
\bibitem[]{campbell14}Campbell L.A., et al. 2014, MNRAS 443, 1231 % >8 authors
\bibitem[]{carignan-freeman}Carignan C., Freeman K. C. 1985, ApJ 294, 494
\bibitem[]{carignan16}Carignan C., Libert Y., Lucero D.M., Randriamampandry T.H., Jarrett T.H., Oosterloo T.A., Tollerud E.J. 2016, A\&A 587, L3
\bibitem[]{catinella08}Catinella B., Haynes M., Giovanelli R., Gardner J.P., Connolly A.J. 2008, ApJL 685, L13
\bibitem[]{catinella10}Catinella B., et al. 2010, MNRAS 403, 683
\bibitem[]{catinella13}Catinella B., et al. 2013, MNRAS 436, 34
\bibitem[]{cc15}Catinella B., Cortese L. 2015, MNRAS 446, 3526 
\bibitem[]{catinella18}Catinella B., et al. 2018, MNRAS 476, 875 
\bibitem[]{chang10}Chang T.-C., Pen U.-L., Bandura K., Peterson J.B. 2010, Nature 466, 463
\bibitem[]{chauhan19} Chauhan G., Lagos, C.D.P., Obreschkow, D., Power, C., Oman, K., Elahi, P. 2019, MNRAS 488, 5898
\bibitem[]{chengalur01}Chengalur J.N., Braun R., Wieringa M. 2001, A\&A 372, 768
\bibitem[]{chengalur15}Chengalur J.N., Ghosh T., Slater C.J., Kanekar N., Momjian E., Keeney B.A., Stocke J.T. 2015, MNRAS 453, 3135
\bibitem[]{chippendale14}Chippendale A.P., Hayman D., Hay S.G 2014, PASA 31, 19
\bibitem[]{chung07}Chung A., van Gorkom J.H., Kenney J.D.P., Vollmer B. 2007, ApJ 659, L11
\bibitem[]{chung09}Chung A., van Gorkom J.H., Kenney J.D.P., Crowl H., Vollmer B. 2009, AJ 138, 1741
\bibitem[]{cluver10}Cluver M., et al. 2010, ApJ 710, 248 % >8 authors
\bibitem[]{colberg05}Colberg J.M., Krughoff, K.S., Connolly, A.J. 2005, MNRAS 359, 272 
\bibitem[]{cormier18}Cormier D., et al. 2018, MNRAS 475, 3909 % 8 authors
\bibitem[]{cortese08}Cortese L., et al. 2008, MNRAS 383, 1519 % 8 authors
\bibitem[]{cortese11}Cortese L., et al. 2011, MNRAS 415, 1797
\bibitem[]{cortese14}Cortese L., et al. 2014, ApJ 795, L37 % >8 authors
% \bibitem[]{courtois12}Courtois H., Hoffman Y., Tully R.B., Gottl\"ober S. 2012, ApJ 744, 43
% \bibitem[]{courtois13}Courtois H., Pomar\`ede D., Tully R.B., Hoffman Y., Courtois D. 2013, AJ 146, 69 
\bibitem[]{courtois17}Courtois H., Tully R.B., Hoffman Y., Pomar\`ede D., Graziana R., Dupuy A. 2017, ApJ 847, L6 
\bibitem[]{crain17}Crain R., et al. 2017, MNRAS 464, 4204 % >8 authors
\bibitem[]{dacunha17}Da Cunha E., et al. 2017, PASA 34, 47
\bibitem[]{danforth08}Danforth C.W., Shull J.M., 2008, ApJ 679, 643 
\bibitem[]{dg02}Darling J., Giovanelli R. 2002, AJ 124, 100 
\bibitem[]{dave99}Dav\'e R., Hernquist L., Katz N., Weinberg D.J. 1999, ApJ 511, 521
\bibitem[]{dave17}Dav\'e R., Rafieferantsao M.H., Thompson R.J., Hopkins P.F. 2017, MNRAS 467, 115
\bibitem[]{deblok02}de Blok W.J.G., Bosma A. 2002, A\&A 385, 816 
\bibitem[]{deblok08}de Blok W.J.G., Walter F., Brinks E., Trachternach C., Oh S.-H., Kennicutt R.C. Jr. 2008, AJ 136, 2648 
\bibitem[]{deblok10}de Blok W.J.G. 2010, Advances in Astronomy, id.\,789293
\bibitem[]{deblok16}de Blok W.J.G., et al. 2016, Proc. of MeerKat Science: On the Pathway to the SKA, 25--27 May, 2016 Stellenbosch, South Africa, p.\,7 % >8 authors
\bibitem[]{deblok18}de Blok W.J.G., et al. 2018, ApJ 865, 26 % >8 authors
\bibitem[]{delhaize13}Delhaize J., Meyer M.J., Staveley-Smith L., Boyle B.J. 2013, MNRAS 433, 1398 
\bibitem[]{denes14}D\'enes H., Kilborn V.A., Koribalski B.S. 2014, MNRAS 444, 667 
\bibitem[]{dfelippis19}DeFelippis D., Putman M., Tollerud E. 2019, ApJ 879, 22
\bibitem[]{dickey13}Dickey J.D., et al. 2013, PASA 30, 3 % >8 authors
\bibitem[]{diemer19}Diemer B., et al. 2019, MNRAS 487, 1529 % >8 authors
\bibitem[]{df14}Di Teodoro E.M., Fraternali F. 2014, A\&A 567, 68
\bibitem[]{3d-barolo}Di Teodoro E.M., Fraternali F. 2015, MNRAS 451, 3021
\bibitem[]{doyle05}Doyle M., et al. 2005, MNRAS 361, 34 
\bibitem[]{driver11}Driver S.P., et al. 2011, MNRAS 413, 971
\bibitem[]{driver16}Driver S.P., et al. 2016, MNRAS 455, 3911 
\bibitem[]{dubois14}Dubois Y., et al. 2014, MNRAS 444, 1453 % >8 authors
\bibitem[]{duffy12a}Duffy A.R., et al. 2012, MNRAS 426, 3385 
\bibitem[]{duffy12b}Duffy A.R., Moss A., Staveley-Smith L. 2012, PASA 29, 202 
% \bibitem[]{dupuy19}Dupuy A., Courtois H., Kubik B. 2019, MNRAS 486, 440 
\bibitem[]{dykes18}Dykes T., Hassan A., Gheller C., Croton D., Krokos M. 2018 MNRAS 477, 1495 
\bibitem[]{einasto97}Einasto M., Tago E., Jaaniste J., Einasto J., Andernach H. 1997, A\&AS 123, 119 
\bibitem[]{eke04}Eke V.R., et al. 2004, MNRAS 355, 769 % >8 authors
\bibitem[]{elagali19}Elagali A., et al. 2019, MNRAS 487, 2797 % >8 authors
\bibitem[]{elahi18}Elahi P., et al. 2018, MNRAS 475, 5338
\bibitem[]{english10}English J., Koribalski B.S., Bland-Hawthorn J., Freeman K.C., McCain C. 2010, AJ 139, 102
\bibitem[]{erdogdu06}Erdogdu P., et al. 2006, MNRAS 373, 45 % >8 authors
\bibitem[]{espada11}Espada D., Verdes-Montenegro L., Huchtmeier W.K., Sulentic J., Verley S., Leon S., Sabater J. 2011 A\&A 532, 117
\bibitem[]{ewen51}Ewen H.I., Purcell E.M. 1951, Nature 168, 356
\bibitem[]{fernandez16}Fernandez X., et al. 2016, ApJL 824, 1 % >8 authors
\bibitem[]{fleenor05}Fleenor M.C., Rose J.A., Christiansen W.A., Hunstead R.W., Johnston-Hollitt M., Drinkwater M.J., Saunders W. 2005, AJ 130, 957
\bibitem[]{floeer12}Fl\"oer L., Winkel B. 2012, PASA 29, 244 
\bibitem[]{floeer14}Fl\"oer L., Winkel B., Kerp J. 2014, A\&A 569, 101 
\bibitem[]{fluke10}Fluke C.J., Barnes D.G., Hassan A.H., 2010, Proceedings of 6th IEEE International Conference on e-Science Workshops, 15
\bibitem[]{fluke18}Fluke C.J., Barnes D.G. 2018, PASA 35, 26 
\bibitem[]{for12}For B.-Q., Koribalski B.S., Jarrett T.H. 2012, MNRAS 425, 1934
\bibitem[]{for14}For B.-Q., Staveley-Smith L., Matthews D., McClure-Griffiths N.M. 2014, ApJ 792, 43
\bibitem[]{for16}For B.-Q., Staveley-Smith L., McClure-Griffiths N.M., Westmeier T., Bekki K. 2016, MNRAS 461, 892
\bibitem[]{for19}For B.-Q., et al. 2019, MNRAS 489, 5723 % >8 authors
\bibitem[]{fossati18}Fossati A., et al. 2018, A\&A 614, 57 % >8 authors
\bibitem[]{freudling11}Freudling W., et al. 2011, ApJ 727, 40 % 9 authors
\bibitem[]{veena19}Ganeshaiah Veena P., Cautun M., Tempel E., van de Weygaert R., Frenk C.S. 2019, MNRAS 487, 1607
\bibitem[]{gereb16}Ger\'eb K., Catinella B., Cortese L., Bekki K., Moran S.M., Schiminovich D. 2016, MNRAS 462, 382
\bibitem[]{gereb18}Ger\'eb K., Janowiecky S., Catinella B., Cortese L., Kilborn V. 2018, MNRAS 476, 896
\bibitem[]{giese16}Giese N., van der Hulst T., Serra P., Oosterloo T. 2016, MNRAS 461, 1656 
\bibitem[]{giovanelli05} Giovanelli R., et al. 2005, AJ 130, 2598 % >8 authors
\bibitem[]{graham19} Graham M.J., et al. 2019, PASP 131, 078001 % >8 authors
\bibitem[]{graziani19}Graziani R., Courtois H.M., Lavaux G., Hoffman Y., Tully R.B., Copin Y., Pomar\`ede D. 2019, MNRAS 488, 5438 
\bibitem[]{grcevich09}Grcevich, J., Putman, M.E. 2009, ApJ, 696, 385
\bibitem[]{gupta16}Gupta N., et al. 2016, Proc. of MeerKat Science: On the Pathway to the SKA, 25--27 May, 2016 Stellenbosch, South Africa, p.\,14 % >8 authors
\bibitem[]{hassan13}Hassan A.H., Fluke C.J., Barnes D.G., Kilborn V.A. 2013, MNRAS 429, 2442 
\bibitem[]{haynes18}Haynes M.P., et al. 2018, ApJ 861, 49 % >8 authors
\bibitem[]{heald11}Heald G., et al. 2011, A\&A 526, 118 % >8 authors 
\bibitem[]{henning10}Henning P.A., et al. 2010, AJ 139, 2130 % >8 authors 
\bibitem[]{hess17}Hess K., Cluver M.E., Yahya S., Leisman L., Serra P., Lucero D.M., Passmoor S.S., Carignan C. 2017, MNRAS 464, 957  % 8 authors
\bibitem[]{hess19}Hess K., et al. 2019, MNRAS 484, 2234 % >8 authors
\bibitem[]{heywood16}Heywood I., et al. 2016, MNRAS 457, 4160 % >8 authors 
\bibitem[]{hi4pi16}HI4PI Collaboration 2016, A\&A, 594, A116 % >8 authors
\bibitem[]{hibbard01}Hibbard J.E., van Gorkom J.H., Rupen M.P., Schiminovic D. 2001, ASPC 240, 657
\bibitem[]{hong13}Hong T., et al. 2013, MNRAS 432, 1178
\bibitem[]{hong19}Hong T., et al. 2019, MNRAS 487, 2061 % >8 authors  
\bibitem[]{hoppmann15}Hoppmann L., Staveley-Smith L., Freudling W., Zwaan M.A., Minchin R.F., Calabretta M.R. 2015, MNRAS 452, 3726 
\bibitem[]{hotan14}Hotan A.W., et al. 2014, PASA 31, 41 % >8 authors
\bibitem[]{hotan20}Hotan A.W., et al. 2020, PASA, in prep. % >8 authors
\bibitem[]{hoversten08}Hoversten E.A., Glazebrook K. 2008, ApJ 675, 163 
\bibitem[]{howlett19}Howlett C. 2019, MNRAS 487, 5209
\bibitem[]{howlett17a}Howlett C., Staveley-Smith L., Blake C. 2017a, MNRAS 464, 2517
\bibitem[]{howlett17b}Howlett C., et al. 2017b, MNRAS 471, 3135 % >8 authors
\bibitem[]{howlett17c}Howlett C., Robotham A.S.G., Lagos C.D.P., Kim A.G. 2017c, ApJ 847, 128
\bibitem[]{hunter12}Hunter D.A., et al. 2012, AJ 144, 134 % >8 authors
\bibitem[]{irwin07}Irwin M., et al. 2007, ApJL 656, L13
\bibitem[]{jachym14}J\'achym P., Combes F., Cortese L., Sun M., Kenney J.D.P. 2014, ApJ 792, 11
\bibitem[]{james04}James P.A., et al. 2004, A\&A 414, 23 % >8 authors
\bibitem[]{janesh19}Janesh, W., Rhode, K.L., Salzer, J.J., Janowiecki, S., Adams, E.A.K., Haynes, M.P., Giovanelli, R., Cannon, J.M., 2019, AJ, 157, 183
\bibitem[]{jarrett12}Jarrett T.H., et al. 2012, AJ 144, 68 % >8 authors
\bibitem[]{jarrett19}Jarrett T.H., Cluver M.E., Brown M.J.I., Dale D.A., Tsai C.W., Masci F. 2019, ApJS 245, 25
\bibitem[]{jarvis16}Jarvis M., et al. 2016, Proc. of MeerKat Science: On the Pathway to the SKA, 25--27 May, 2016 Stellenbosch, South Africa, p.\,6 % >8 authors
\bibitem[]{johnston07}Johnston S., et al. 2007, PASA 24, 174 % >8 authors
\bibitem[]{johnston08}Johnston S., et al. 2008, ExA 22, 151 % >8 authors
\bibitem[]{jonas16}Jonas J., MeerKAT Team 2016, Proc. of MeerKat Science: On the Pathway to the SKA, 25--27 May, 2016 Stellenbosch, South Africa, p.\,1 
\bibitem[]{jones18a}Jones M.G., Espada D., Verdes-Montenegro L., et al. 2018a, A\&A 609, 17
\bibitem[]{jones18b}Jones M.G., Haynes M.P., Giovanelli R., Moorman C. 2018b, MNRAS 477, 2 
\bibitem[]{jozsa07}J\'ozsa G.I.G 2007, A\&A 468, 903
\bibitem[]{jkko07}J\'ozsa G.I.G, Kenn F., Klein U., Oosterloo T.A. 2007, A\&A 468, 731
\bibitem[]{kalberla05}Kalberla P.M.W., et al. 2005, A\&A 440, 775 % LAB survey
\bibitem[]{kd08}Kalberla P.M.W., \& Dedes L. 2008, A\&A, 487, 951
\bibitem[]{kk09}Kalberla P.M.W., Kerp J. 2009, AR\&A 47, 27 
\bibitem[]{kalberla17}Kalberla P.M.W., Kerp J., Haud U., Haverkorn M. 2017, A\&A 607, 15
\bibitem[]{kh15}Kalberla P.M.W., Haud U. 2015, A\&A 578, 78
\bibitem[]{kk16}Kalberla P.M.W., Kerp J. 2016, A\&A 595, 37
\bibitem[]{kh18}Kalberla P.M.W., Haud U. 2018, A\&A 619, 58
\bibitem[]{kam17}Kam S.Z., Carignan C., Chemin L., Foster T., Elson E., Jarrett T.H. 2017, AJ 154, 41
\bibitem[]{kamphuis15}Kamphuis P., J\'ozsa G.I.G., Oh S.-H., Spekkens K., Urbancic N., Serra P., Koribalski B.S., Dettmar R.-J. 2015, MNRAS 452, 3139 
\bibitem[]{kanekar16}Kanekar N., Sethi S., Dwarakanath K.S. 2016, ApJL 818, L28
\bibitem[]{kara13}Karachentsev I.D., Makarov D.I., Kaisina E.I. 2013, AJ 145, 101 
\bibitem[]{Kennicutt03}Kennicutt R.C.Jr., et al. 2003, PASP 115, 928 
\bibitem[]{Kennicutt08}Kennicutt R.C.Jr., Lee J.C., Funes, J.G., Sakai S., Akiyama S. 2008, ApJS 178, 247 
\bibitem[]{khandai12}Khandai N., et al. 2012, MNRAS 415, 2580 % 7 authors !
\bibitem[]{2004MNRAS.353..713K}  Kauffmann G., White S.D.M., Heckman T.M., M\'enard B., Brinchmann J., Charlot S., Tremonti C., Brinkmann J. 2004, MNRAS 353, 713
\bibitem[]{kenney08}Kenney J.D.P., Tal T., Crowl H.H., Feldmeier J., Jacoby G.H. 2008, ApJ 687, L69
\bibitem[]{kent09}Kent B.R., Spekkens K., Giovanelli R., Haynes M.P., Momjian E., Cort\'es J.R., Hardy E., West A.A. 2009, ApJ 691, 1595
\bibitem[]{kerp11}Kerp J., Winkel B., Ben Bekhti N., Fl\"oer L., Kalberla P.  2011, AN 332, 637
\bibitem[]{kerp16}Kerp J., Kalberla P. M. W., Ben Bekhti N., Fl\"{o}er L., Lenz D., Winkel B. 2016, A\&A, 589, A120
\bibitem[]{kerr54}Kerr F.J., Hindman J.F., Robinson B.J., 1954, Aust. J. Phys, 7, 297
\bibitem[]{kilborn00}Kilborn V.A. et al. 2000, AJ 120, 1342 
\bibitem[]{kilborn09}Kilborn V.A., Forbes D., Barnes D.G., Koribalski B.S., Brough S., Kern K. 2009, MNRAS 400, 1962
\bibitem[]{kleiner17}Kleiner D., Pimblett K.A., Jones D.H., Koribalski B.S., Serra P. 2017, MNRAS 466, 4692
\bibitem[]{kleiner19}Kleiner D., Koribalski B.S., et al. 2019, MNRAS 488, 5352 
\bibitem[]{klypin99}Klypin A., et al. 1999, ApJ, 522, 82
\bibitem[]{klypin15}Klypin A., Karachentsev I., Makarov D, Nasonova O. 2015, MNRAS 454, 1798
% \bibitem[]{kolatt95}Kolatt T., Dekel A., Lahav O. 1995, MNRAS 275, 797
\bibitem[]{koposov09}Koposov S.E., et al. 2009, ApJ 696, 2179
\bibitem[]{koribalski03}Koribalski B.S., Gordon S., Jones K. 2003, MNRAS 339, 1203
\bibitem[]{koribalski04}Koribalski B.S., et al. 2004, AJ 128, 16 % HIPASS BGC
\bibitem[]{kdickey04}Koribalski B.S., Dickey J. 2004, MNRAS 348, 1255 
\bibitem[]{koribalski-etal09}Koribalski B.S., et al. 2009, \wal\ proposal 
  % \url{https://www.atnf.csiro.au/research/WALLABY/proposal.html} 
\bibitem[]{koribalski09}Koribalski B.S., L\'opez-S\'anchez A.R. 2009, MNRAS 400, 174 
\bibitem[]{koribalski12a}Koribalski B.S., 2012a, PASA 29, 359 
\bibitem[]{koribalski12b}Koribalski B.S., 2012b, PASA 29, 213 % Editor
\bibitem[]{koribalski15}Koribalski B.S., 2015, IAU S309, p.\,39
\bibitem[]{koribalski17a}Koribalski B.S. 2017, IAU S321, p.\,232
\bibitem[]{koribalski18}Koribalski B.S., et al. 2018, MNRAS 478, 1611 % LVHIS 
\bibitem[]{kovac09}Kova\v{c} K., Oosterloo T.A., van der Hulst J.M. 2009, MNRAS 400, 743 
\bibitem[]{kraan17}Kraan-Korteweg R.C., et al. 2017, MNRAS 466, L29 % 8 authors!
\bibitem[]{kreckel12}Kreckel K., et al. 2012, AJ 144, 16 
\bibitem[]{krumholz13}Krumholz M.R. 2013, MNRAS 436, 2747
\bibitem[]{kurapati18}Kurapati S, Chengalur J.N., Pustilnik S. Kamphuis P. 2018, MNRAS 479, 228
\bibitem[]{lagos11}Lagos C.D.P., Baugh C.M., Lacey C.G., Benson A.J., Kim H.-S., Power C. 2011, MNRAS 418, 1649
\bibitem[]{lagos14}Lagos C.D.P., Baugh C.M., Zwaan M.A., et al. 2014, MNRAS 440, 920
\bibitem[]{lagos15}Lagos C.D.P., et al. 2015, MNRAS 452, 3815
\bibitem[]{lagos18}Lagos C.D.P., Tobar, R.J., Robotham A.S.G., Obreschkow D., Mitchell P.D., Power C., Elahi P.J. 2018, MNRAS 481, 3573
\bibitem[]{lang03}Lang et al. 2003, MNRAS 342, 738 
\bibitem[]{lee-waddell12}Lee-Waddell K., et al. 2012, MNRAS 427, 2314
\bibitem[]{lee-waddell18}Lee-Waddell K., et al. 2018, MNRAS 474, 1108
\bibitem[]{lee-waddell19}Lee-Waddell K., Koribalski, B.S., et al. 2019, MNRAS 487, 5248 % >8 authors 
\bibitem[]{leroy2008}Leroy A.K., Walter F., Brinks E., Bigiel F., de Blok W.J.G., Madore B., Thornley M.D. 2008, AJ 136, 2782 
\bibitem[]{leroy2009}Leroy A.K., et al. 2009, AJ 137, 4670 
\bibitem[]{li18}Li, D. et al. 2018, IEEE Microwave Magazine 19, 112 
\bibitem[]{lockman16}Lockman F.J., McClure-Griffiths N.M. 2016, ApJ 826, 215 
\bibitem[]{ls12}L\'opez-S\'anchez A.R., Koribalski B.S., van Eyemeren J., Esteban C., Kirby E., Jerjen H., Lonsdale N. 2012, MNRAS 419, 1051
\bibitem[]{ls15}L\'opez-S\'anchez A.R., Westmeier T., Esteban C., Koribalski B.S. 2015, MNRAS  450, 3381
\bibitem[]{ls18}L\'opez-S\'anchez A.R., Lagos C.D.P., Young T., Jerjen H. 2018, MNRAS 480, 210
\bibitem[]{lutz17}Lutz K.A., et al. 2017, MNRAS 467, 1083 % 9 authors
\bibitem[]{lutz18}Lutz K.A., et al. 2018, MNRAS 476, 3744 % 9 authors
\bibitem[]{maccio19}Macci\'o A.V., et al. 2019, MNRAS 484, 5400 
\bibitem[]{marasco16}Marasco A., Crain R.A., Schaye J., Bah\'e Y.M., van der Hulst T., Theuns T., Bower R.G. 2016, MNRAS 461, 2630 
\bibitem[]{mateo98}Mateo M. 1998, ARA\&A, 36, 435
\bibitem[]{mayer01}Mayer L., et al. 2001, ApJ, 547, 123
\bibitem[]{mcclure05}McClure-Griffiths N.M., Dickey J.M., Graensler B., Green A.J., Haverkorn M., Strasser S. 2005, ApJS 158, 178 
\bibitem[]{mcclure09}McClure-Griffiths N.~M., Pisano D.~J., Calabretta M.~R., Ford H.~A., Lockman F.~J. 2009, ApJS, 181, 398 
\bibitem[]{mcconnachie12}McConnachie, A. 2012, AJ 144, 4
\bibitem[]{mcconell16}McConnell, D., et al. 2016, PASA 33, 42 
\bibitem[]{mcconell20}McConnell, D., et al. 2020, PASA, in prep. % >8 authors
\bibitem[]{mr09}McKean J., Roy A.L. 2009, Proc. Panoramic Radio Astronomy, p.\,60 
\bibitem[]{meurer96}Meurer G.R., Carigan C., Beaulieu S.F, Freeman K.C. 1996, AJ 111, 1551
\bibitem[]{meurer06}Meurer G.R. et al. 2006, ApJ 487, 171 % 
\bibitem[]{meurer09}Meurer G.R. et al. 2009, ApJ 695, 765 % >8 authors
\bibitem[]{meyer04}Meyer M., et al. 2004, MNRAS 350, 1195
\bibitem[]{meyer08}Meyer M., Zwaan M.A., Webster R.L., Schneider S., Staveley-Smith L. 2008, MNRAS 391, 1712
\bibitem[]{meyer16}Meyer S., et al. 2016, MNRAS 455, 3136
\bibitem[]{morganti18}Morganti R., Oosterloo T.A. 2018, A\&A Rev. 26, 4 
\bibitem[]{muller51}Muller C.A., Oort J.H., 1951, Nature, 168, 357
\bibitem[]{murray18}Murray C.E., et al. 2018, ApJS 238, 14 
\bibitem[]{murugeshan19}Murugeshan C., Kilborn V.A., Obreschkow D., Glazebrook K., Lutz K., Dzudzar R., D\'enes H. 2019, MNRAS 483, 2398 
\bibitem[]{norris11}Norris R.P., et al. 2011, PASA 28, 215 % >8 authors
\bibitem[]{noterdaeme12}Noterdaeme P., et al. 2012, A\&A 547, L1
\bibitem[]{obreschkow2009a}Obreschkow D., Croton, D., De Lucia, G., Khochfar, S. Rawlings, S. 2009a, ApJ 698, 1467
\bibitem[]{obreschkow2009b}Obreschkow D., Kl\"ockner H.-R., Heywood I., Levrier F., Rawlings S. 2009b, ApJ 703, 1890
\bibitem[]{obreschkow2011}Obreschkow D., Heywood I., Rawlings S. 2011, ApJ 743, 84
\bibitem[]{obreschkow16}Obreschkow D., Glazebrook K., Kilborn V.A., Lutz K. 2016, ApJ 824, 26 
\bibitem[]{oey07}Oey M.S., et al. 2007, ApJ 661, 801 % SINGG; >8 authors
\bibitem[]{oh08}Oh S.-H., de Blok W.J.G., Walter F., Brinks E., Kennicutt R.C. Jr 2008, AJ 136, 2761
\bibitem[]{oh11}Oh S.-H., de Blok W.J.G., Brinks E., Walter F., Kennicutt R.C. Jr 2011, AJ 141, 193 
\bibitem[]{oh18}Oh S.H., Staveley-Smith L., Spekkens K., Kamphuis P., Koribalski B.S. 2018, MNRAS 473, 3256
\bibitem[]{onken19}Onken, C.A., et al. 2019, PASA 36, 33 % >8 authors
\bibitem[]{oosterloo05}Oosterloo T.A., van Gorkom J. 2005, A\&A 437, L19 
\bibitem[]{oosterloo07}Oosterloo T.A., et al. 2007, AJ 134, 1019 
\bibitem[]{oosterloo10}Oosterloo T.A., Verheijen M., van Cappellen W. 2010, Proc. ISKAF Science Meeting 
\bibitem[]{ott12}Ott J., Stilp A.M., Warren S.R., Skillman E.D., Dalcanton J.J., Walter F., de Blok W.J.G., Koribalski B.S., West A.A. 2012, AJ 144, 123 
\bibitem[]{papastergis13}Papastergis E., Giovanelli R., Haynes M.P., Rodrigues-Puebla A., Jines M.G. 2013, ApJ 776, 43 
\bibitem[]{papastergis16}Papastergis E., Shankar F. 2016, A\&A 591, 58
\bibitem[]{parker05}Parker Q., et al. 2005, MNRAS 362, 689 % >8 authors
\bibitem[]{pearson16}Pearson et al. 2016, MNRAS 459, 1827 % >8 authors
\bibitem[]{peek18}Peek, J.E.G., et al. 2018, ApJS 234, 2 % >8 authors
\bibitem[]{pen09}Pen U.-L., Staveley-Smith L., Peterson J.B., Chang T.-C. 2009, MNRAS 394, L6 
\bibitem[]{perley11}Perley R.A, Chandler C.J., Butler B.J., Wrobel J.M. 2011, ApJ 739, L1 
\bibitem[]{pillepich18}Pillepich A., et al. 2018, MNRAS 473, 4077
\bibitem[]{pineda13}Pineda J.L., Langer W.D., Velusamy T., Goldsmith, P.F. 2013, A\&A 554, 103 
% \bibitem[]{planck16}Planck Collaboration et al. 2016, A\&A 594, A13 
\bibitem[]{planck18}Planck Collaboration et al. 2018, submitted 
\bibitem[]{popping12}Popping A., Jurek R., Westmeier T., Serra P., Fl\"{o}er L., Meyer M., Koribalski B.~S., 2012, PASA 29, 318 
\bibitem[]{popping14}Popping G., Sommerville R.S., Trager S.C. 2014, MNRAS 442, 2398
\bibitem[]{popping15}Popping G., Behroozi P.S., Peeples M.S. 2015, MNRAS 449, 477
\bibitem[]{porter05}Porter S.C., Raychaudhury S. 2005, MNRAS 364, 1387 
\bibitem[]{porter08}Porter S.C., Raychaudhury S., Pimbblet K.A., Drinkwater, M.J. 2008, MNRAS 388, 1152 
\bibitem[]{power10}Power C., Baugh C.M., Lacey C.G. 2010, MNRAS 406, 43
\bibitem[]{punzo15}Punzo D., van der Hulst J.M., Roerdink J.B.T.M., Oosterloo T.A., Ramatsoku M., Verheijen M.A.W. 2015, A\&C 12, 86 
\bibitem[]{punzo17}Punzo D., van der Hulst J.M., Roerdink J.B.T.M., Fillion-Robin J.C., Yu L. 2017, A\&C 19, 45 
\bibitem[]{putman02}Putman M.E., et al. 2002, AJ 123, 873 
\bibitem[]{putman03}Putman M.E., et al. 2003, ApJ 586, 170
\bibitem[]{putman12}Putman M.E., Peek, J., Joung M.R. 2012, ARA\&A 50, 491
\bibitem[]{qin18}Qin F., Howlett C., Staveley-Smith L., Hong T. 2018, MNRAS 477, 5150
\bibitem[]{qin19a}Qin F., Howlett C., Staveley-Smith L., Hong T. 2019a, MNRAS 482, 192
\bibitem[]{qin19b}Qin F., Howlett C., Staveley-Smith L. 2019b, MNRAS 487, 5235
\bibitem[]{rao06}Rao S.M., Turnshek D.A., Nestor D.B., 2006, ApJ 636, 610
\bibitem[]{ray89}Raychaudhury S. 1989, Nature 342, 251
\bibitem[]{reddick13}Reddick R.M., Wechsler R.H., Tinker J.L., Behroozi P.S. 2013, ApJ 771, 30
\bibitem[]{reeves15}Reeves S., Sadler E., Allison J.A., Koribalski B.S., Curran S.J., Pracy M.B. 2015, MNRAS 450, 926 
\bibitem[]{reeves16}Reeves S., Sadler E., Allison J.A., Koribalski B.S., Curran S.J., Pracy M.B., Phillips C.J., Bignall H.E., Reynolds C. 2016, MNRAS 457, 2613
\bibitem[]{revaz09}Revaz Y., et al. 2009, A\&A 501, 189
\bibitem[]{reynolds17}Reynolds T.N., et al. 2017, PASA 34, 51 % >8 authors
\bibitem[]{reynolds19}Reynolds  T.N., Westmeier T., Staveley-Smith L., et al. 2019, MNRAS 482, 3591 % >8 authors
\bibitem[]{rhee13}Rhee J., Zwaan M.A., Briggs F.H., Chengalur J.N., Lah P., Oosterloo T., van der Hulst, T. 2013, MNRAS 435, 2693
\bibitem[]{rhee16}Rhee J., Lah P., Chengalur J.N., Briggs F.H., Colless M. 2016, MNRAS 460, 2675
\bibitem[]{rhee18}Rhee J., Lah P., Briggs F.H., Chengalur J.N., Colless M., Wilner S.P., Ashby M.L.N., Le F\`evre O. 2018, MNRAS 473, 1879 
\bibitem[]{ricotti05}Ricotti M., \& Gnedin N.Y. 2005, ApJ, 629, 259
\bibitem[]{ricotti09}Ricotti M. 2009, MNRAS 392, L45
\bibitem[]{rosenberg00}Rosenberg J.L,, Schneider S. 2000, ApJS 130, 177 % ADBS obs
\bibitem[]{rosenberg02}Rosenberg J.L,, Schneider S. 2002, ApJ 567, 247 % ADBS HIMF
\bibitem[]{rd03}Rossa J., Dettmar R.-J. 2003, A\&A 406, 505 
\bibitem[]{roychowdhury17}Roychowdhury S., Chengalur J., Shi Y., 2017 A\&A 608, 24
\bibitem[]{ryanweber02}Ryan-Weber E.V., Koribalski B.S. et al. 2002, AJ 124, 1954 % >8 authors 
\bibitem[]{ryanweber08}Ryan-Weber E.V., et al. 2008, MNRAS 384, 535
\bibitem[]{ryder01}Ryder S., Koribalski B.S. et al. 2001, ApJ 555, 232
\bibitem[]{said19}Said K., Kraan-Korteweg R.C., Staley-Smith L. 2019, MNRAS 486, 1796 
\bibitem[]{saintonge11}Saintonge A., et al. 2011, MNRAS 415, 52 % >8 authors
\bibitem[]{sb19}Sanchez-Barrantes, M., et al. 2019, AJ 158, 234 % >8 authors
\bibitem[]{sancisi08}Sancisi R., Fraternali F., Oosterloo T., van der Hulst T. 2008, A\&AR 15, 189
\bibitem[]{saponara18}Saponara J., Koribalski B.S., Benaglia P., Fern\'adez L\'opez M. 2018, MNRAS 473, 3358
\bibitem[]{saul12}Saul D.R., et al. 2012, ApJ 758, 44 % >8 authors
\bibitem[]{schaye15}Schaye J., Crain R., et al. 2015, MNRAS 446, 521
\bibitem[]{schneider83}Schneider S.E., Helou G., Salpeter E.E., Terzian Y., 1983, ApJ 273, L1
\bibitem[)]{schroeder19}Schr\"{o}der A.C., Fl\"{o}er L., Winkel B., Kerp J. 2019, MNRAS 489, 2907 
\bibitem[]{scott18}Scott T.C., Brinks E., Cortese L., Boselli A., Bravo-Alfaro H. 2018, MNRAS 475, 4648 
\bibitem[]{scrimgeour16}Scrimgeour M.I., et al. 2016, MNRAS 455, 386
\bibitem[]{skrutski06}Skrutskie M.F., et al. 2006, AJ 131, 1163 % >8 authors
\bibitem[]{serra12a}Serra P., et al. 2012a, MNRAS 422, 1835 % ATLAS-3D
\bibitem[]{serra12b}Serra P., Jurek R., Fl\"{o}er L. 2012, PASA, 29, 296 
\bibitem[]{serra13}Serra P., Koribalski B.S., et al. 2013 MNRAS 428, 370
\bibitem[]{serra15a}Serra P., Koribalski B.S., et al. 2015a, MNRAS 452, 2680 
\bibitem[]{serra15b}Serra P., Westmeier T., et al. 2015b, MNRAS 448, 1922 
\bibitem[]{serra16}Serra P., et al. 2016, Proc. of MeerKat Science: On the Pathway to the SKA, 25--27 May, 2016 Stellenbosch, South Africa, p.\,8 % >8 authors
\bibitem[]{solanes01}Solanes J.M., et al. 2001, ApJ 548, 97 
\bibitem[]{sorgho19}Sorgho A., et al. 2019, MNRAS 482, 1248 
\bibitem[]{spekkens07}Spekkens K., Sellwood J.A. 2007, ApJ 664, 204 
\bibitem[]{spekkens14}Spekkens K., Urbancic N., Mason B.S., Willman B., Aguirre J. E. 2014, ApJ 795, L5 
\bibitem[]{springel05}Springel V., et al. 2005, Nature 435, 629 % >8 authors
\bibitem[]{springob05}Springob C., Haynes M.P., Giovanelli R., Kent B.R. 2005, ApJS 160, 149 
\bibitem[]{springob07}Springob C., et al. 2007, ApJS 172, 599 % 5 authors ! 
\bibitem[]{springob16}Springob C., et al. 2016, MNRAS 456, 1886 % >8 authors
\bibitem[]{lss92}Staveley-Smith L., Norris R.P., Chapman J.M., Allen D.A., Whiteoak J.B., Roy A.L. 1992, MNRAS 258, 725 
\bibitem[]{lss16}Staveley-Smith L., Kraan-Korteweg R.C., Schr\"oder A.C., Henning P.A., Koribalski B.S., Stewart I.M., Heald G. 2016, AJ 151, 52 
\bibitem[]{stevens16}Stevens A.R.H., Croton D.J., Mutch S.J. 2016, MNRAS 461, 859
\bibitem[]{sb17}Stevens A.R.H., Brown T. 2017, MNRAS 471, 447
\bibitem[]{stevens18}Stevens A.R.H., Lagos C.P., Obreschkow D., Sinha M. 2018, MNRAS 481, 5543
\bibitem[]{stevens19}Stevens A.R.H., et al. 2019, MNRAS 483, 5334 % >8 authors
\bibitem[]{stil06}Stil J.M., et al. 2006, AJ 132, 1158 
\bibitem[]{sun07}Sun M., Donahue M., Voit G.M. 2007, ApJ 671, 190 
\bibitem[]{sutherland15}Sutherland W., et al. 2015, A\&A 575, 25 
\bibitem[]{tiley19}Tiley A.L., et al. 2019, MNRAS 482, 2166 % >8 authors
\bibitem[]{tingay13}Tingay S.J., et al. 2013, PASA 30, 7 % >8 authors 
\bibitem[]{tolstoy09}Tolstoy E., Hill V., Tosi M. 2009, ARA\&A 47, 371
\bibitem[]{tollerud08}Tollerud E.J., et al. 2008, ApJ 688, 277
\bibitem[]{tollerud15}Tollerud E.J., Geha M.C., Grcevich J., Putman M.E., Stern D. 2015, ApJL 798, L21
\bibitem[]{vanalbada85}van Albada T.S., Bahcall J.N., Begeman K., Sancisi R. 1985, ApJ 295, 305 
\bibitem[]{vandehulst45}van de Hulst H.C. 1945, Nederlands Tijdschrift voor Natuurkunde 11, 201
% XXX - alternately: Van de Hulst 1957, IAU Symp. 4, 3
\bibitem[]{vanderkruit07}van der Kruit P.C. 2007, A\&A 466, 883
\bibitem[]{vanderkruit17}van der Kruit P.C., Freeman K.C. 2017, ARA\&A 49, 301
\bibitem[]{vandriel16}van Driel et al. 2016, A\&A 595, 118 
\bibitem[]{vm01}Verdes-Montenegro, L., Yun, M.S., Williams, B.A., et al. 2001, A\&A 377, 812 
\bibitem[]{vm02}Verdes-Montenegro, L., Bosma A., Athanassoula L.E., A\&A 389, 825
\bibitem[]{vm05}Verdes-Montenegro, L., Sulentic, J., Lisenfeld, U. et al. 2005, A\&A 436, 443 
\bibitem[]{verheijen01}Verheijen M.A.W. 2001, ApJ 563, 694 
\bibitem[]{verheijen07}Verheijen M., van Gorkum J.H., et al. 2007, ApJL 668, L9
\bibitem[]{vohl16}Vohl D., et al. 2016, PeerJCS 2016, 88 % >8 authors
\bibitem[]{vohl17}Vohl D., Fluke C.J., Barnes D.G., Hassan A.H. 2017, MNRAS 471, 3323 
\bibitem[]{wakker99}Wakker B.P., et al. 1999, Nature 402, 388 % >8 authors
\bibitem[]{wakker09}Wakker B.P., Savage B.D. 2009, ApJS 182, 378
\bibitem[]{wald17}Wald I., Johnson G.P., Amstutz J., Brownlee C., Knoll A., Jeffers J., Gunther J., Navratil P., 2017, IEEE Transactions on Visualization and Computer Graphics, 23, 931
\bibitem[]{walter08}Walter F., et al. 2008, AJ 136, 2563 
\bibitem[]{wang15}Wang J., et al. 2015, MNRAS 453, 2399 % >8 authors
\bibitem[]{wang16}Wang J., Koribalski B.S., Serra P., van der Hulst T., Roychowdhury S., Kamphuis P., Chengalur J.N. 2016, MNRAS 460, 2143 
\bibitem[]{wang17}Wang J., Koribalski B.S., Serra P., et al. 2017, MNRAS 472, 3029 % >8 authors
\bibitem[]{warren04}Warren B.E., Jerjen H., Koribalski B.S. 2004, AJ 128, 1152 
\bibitem[]{warren07}Warren B.E., Jerjen H., Koribalski B.S. 2007, AJ 134, 1849
\bibitem[]{welker20}Welker C., et al. 2020, MNRAS 491, 2864 % >8 authors
\bibitem[]{westmeier08}Westmeier T., Br\"uns C., Kerp J. 2008, MNRAS, 390, 1691
\bibitem[]{wk08}Westmeier T., Koribalski B.S. 2008, MNRAS, 388, L29
\bibitem[]{westmeier12}Westmeier T., Popping A., Serra P. 2012, PASA, 29, 276
\bibitem[]{westmeier14}Westmeier T., Jurek R., Obreschkow D., Koribalski B.S., Staveley-Smith L. 2014, MNRAS 438, 1176 % Busy Function
\bibitem[]{westmeier15}Westmeier T., Staveley-Smith L., Calabretta M., Jurek R., Koribalski B.S., Meyer M., Popping A., Wong O.I. 2015, MNRAS 453, 338 
\bibitem[]{westmeier17}Westmeier T., et al. 2017, MNRAS 472, 4832 % >8 authors 
\bibitem[]{westmeier18}Westmeier T. 2018, MNRAS 474, 289 
\bibitem[]{sofia-2018}Westmeier T., Giese N., van der Hulst T., Jurek R., Popping A., Serra P. 2018, Proceedings of {\em FAST--MeerKAT and SKA Pathfinder Synergies}, eds. B. Peng, C. Carignan \& M. Zhu, submitted 
\bibitem[]{wh12}Whiting M., Humphreys B. 2012, PASA 29, 371 
\bibitem[]{whiting17}Whiting M., Voronkov M., Mitchell D., et al. 2017, ASPC 512, 431 
% \bibitem[]{whiting20}Whiting M., et al. 2020, in prep. % ASKAPsoft pipeline
\bibitem[]{winkel10}Winkel B., Kalberla P.M.W., Kerp J., Fl\"oer L. 2010, ApJS 188, 488
\bibitem[]{winkel16}Winkel B., Kerp J., Fl\"oer L., Kalberla P.M.W., Ben Bekhti N., Keller R., Lenz D. 2016, A\&A 585, 41 
\bibitem[]{wolfinger13}Wolfinger K., Kilborn V.A., Koribalski B.S., et al. 2013, MNRAS 428, 1790 %  8 authors !
\bibitem[]{wolfinger16}Wolfinger K., Kilborn V.A., Ryan-Weber, E.V., Koribalski B.S. 2016, PASA 33, 38 
\bibitem[]{wolf18}Wolf C., et al. 2018, PASA 35, 10 
\bibitem[]{wong06}Wong O.I., et al. 2006, MNRAS 371, 1855 
\bibitem[]{wong09}Wong O.I., et al. 2009, MNRAS 399, 2264 
\bibitem[]{wong14}Wong O.I., Kenney J.D.P., Murphy E.J., Helou G. 2014, ApJ 783, 109 
\bibitem[]{wong16}Wong O.I., et al. 2016, MNRAS 460, 1106 
\bibitem[]{wong17}Wong O.I., et al. 2017, MNRAS 466, 574 
\bibitem[]{wright10}Wright E.L.., et al. 2010, AJ 140, 1868 % >8 authors
\bibitem[]{yang05}Yang X., Mo H.J., Jing Y.P., van den Bosch F.C. 2005, MNRAS 358, 217 
\bibitem[]{yang07}Yang X., et al. 2007, ApJ 671, 153 
\bibitem[]{yang08}Yang X., Mo H.J., van den Bosch F.C. 2008, ApJ 676, 248 
\bibitem[]{yoon17}Yoon H., Chung A., Smith R., Jaff\'e Y.L. 2017, ApJ 838, 81 
\bibitem[]{young07}Young L.M., Skillman E.D., Weisz D.R., Dolphin A.E. 2007, ApJ 659, 331
\bibitem[]{yun93}Yun M.S., Ho P.T.P., Lo K.Y 1993, ApJ 411, L17
\bibitem[]{yun94}Yun M.S., Ho P.T.P., Lo K.Y 1994, Nature 372, 530
\bibitem[]{zhang19}Zhang K., et al. 2019, SCPMA 62, 959506 % >8 authors
\bibitem[]{zoldan17}Zoldan A., De Lucia G., Xie L., Fontanot F., Hirschmann M. 2017, MNRAS 465, 2236
\bibitem[]{zwaan97}Zwaan M.A., et al. 1997, ApJ 490, 173
\bibitem[]{zwaan01}Zwaan M.A., van Dokkum P.G., Verheijen M.A.W. 2001, Science 293, 1800
\bibitem[]{zwaan03}Zwaan M.A., et al. 2003, AJ 125, 2842 
\bibitem[]{zwaan04}Zwaan M.A., et al. 2004, MNRAS 350, 1210
\bibitem[]{zwaan05}Zwaan M.A., Meyer M.J., Staveley-Smith L., Webster R.L. 2005, MNRAS 359, L30 
\end{thebibliography}
\end{document}